\documentclass{aa}  

\usepackage{xcolor}
\usepackage{soul}

\usepackage{graphicx}
\usepackage{caption}
\usepackage{txfonts}
\usepackage{gensymb}
\usepackage{float}
\usepackage[colorlinks=true, allcolors=blue]{hyperref}

\begin{document}

   \title{The HOSTS survey: Suspected variable dust emission and constraints on companions around $\theta$\,Boo}

    \titlerunning{The HOSTS survey: Variable dust emission and constraints on companions around $\theta$\,Boo}

   \author{G. Garreau\inst{1}
        \and D. Defrère\inst{1}
        \and S. Ertel\inst{2,3}
        \and V. Faramaz-Gorka\inst{2}
        \and G. Bryden\inst{4}
        \and M. Sommer\inst{5}
        \and D. Mesa\inst{6}
        \and K. Wagner\inst{2}
        \and T. De Prins\inst{1}
        \and R. Laugier\inst{1}
        \and A. Weinberger\inst{7}
          \and J. Farinato\inst{6}
          \and C. Haniff\inst{8}
          \and P. M. Hinz\inst{9}
          \and J. W. Isbell\inst{2,3}
          \and G. M. Kennedy\inst{10}
          \and A. Lorenzetto\inst{6}
          \and E. R. Maier\inst{11}
          \and L. Marafatto\inst{6}
          \and S. Marino\inst{12}
          \and M. A. Martinod\inst{1}
          \and B. Mennesson\inst{4}
          \and H. Rousseau\inst{2,3}
          \and E. Spalding\inst{13}
          \and D. Vassallo\inst{6}
          \and M. C. Wyatt\inst{5}
          }

   \institute{Institute of Astronomy, KU Leuven,
            Celestijnenlaan 200D bus 2401, 3001 Leuven, Belgium
            \and 
            Department of Astronomy and Steward Observatory, The University of Arizona, 933 N Cherry Ave, Tucson, AZ 85719, USA
            \and
            Large Binocular Telescope Observatory, The University of Arizona, 933 N Cherry Ave, Tucson, AZ 85719, USA
            \and
            Jet Propulsion Laboratory, California Institute of Technology, Pasadena, CA 91109, USA
            \and
            Institute of Astronomy, University of Cambridge, Madingley Road, Cambridge, CB3 0HA, UK
            \and
            INAF-Osservatorio Astronomico di Padova, Vicolo dell’Osservatorio 5, Padova, Italy, 35122-I
            \and
            Earth and Planets Laboratory, Carnegie Institution for Science, Washington, DC 20015, USA
            \and
            Cavendish Laboratory, University of Cambridge, J J Thomson Avenue, Cambridge, CB3 0HE, UK
            \and
            Department of Astronomy \& Astrophysics, University of California, Santa Cruz, CA 95064, USA
            \and
            Department of Physics and Centre for Exoplanets and Habitability, University of Warwick, Gibbet Hill Road, Coventry CV4 7AL, UK
            \and
            Lowell Observatory, 1400 W. Mars Hill Rd, Flagstaff, AZ 86001, USA
            \and
            Department of Physics and Astronomy, University of Exeter, Stocker Road, Exeter EX4 4QL, UK
            \and
            Sydney Institute for Astronomy, School of Physics, University of Sydney, Sydney NSW 2006, Australia\\
             }

   \date{Received October 18, 2024; accepted May 7, 2025}

  \abstract
   {During the Hunt for Observable Signatures of Terrestrial Systems (HOSTS) survey by the Large Binocular Telescope Interferometer (LBTI), an excess emission from the main-sequence star, $\theta$\,Boo (F7V spectral type, 14.5\,pc distance) was observed. This excess indicates the presence of exozodiacal dust near the star's habitable zone (HZ). Previous observations with \textit{Spitzer} and \textit{Herschel} show no evidence of outer cold dust within their respective detection limits. Because exozodiacal dust is generally believed to originate from material located farther out in the system, its source around $\theta$\,Boo remains unclear.}
{We conducted additional nulling and high-contrast adaptive optics (AO) observations to spatially constrain the dust distribution, search for variability, and directly image potential companions in the system. This study presents the results of these observations and provides an interpretation of the inner system's architecture.}
   {We observed the star using the LBTI's N'-band nulling mode during three epochs in 2017, 2018, and 2023. For each epoch, we modeled and constrained the dust distribution using the standard LBTI nulling pipeline, assuming a vertically thin disk with a face-on inclination. We also performed high-contrast AO observations in the L'-band and H-band to constrain the presence of substellar companions around the star.}
{We find several solutions for the dust distribution for each epoch. However, the LBTI nulling observations are unable to discriminate between them.
   Using upper limits from previous observations, we constrain the representative size of the dust grains to approximately 3-5\,µm.
   We also measured a tentative increase in dust brightness at the Earth-equivalent insolation distance between 2017 and 2023. This increase corresponds to the injection of $4\times10^{-8}-4\times10^{-7}\,M_\oplus$ of new material into the disk.
   We consider several options to explain the origin of the observed dust and its variability, but no clear sources are identified from the current observations, partly because our high-contrast AO observations could only constrain the presence of companions only down to $11\,M_\text{Jup}$ at 1.3\arcsec separation.}
   {}


   \maketitle
%

\section{Introduction}

     Direct imaging of exoplanets is a challenging type of observation in modern astronomy due to the high star-planet contrast and small angular separation. Although only  $\sim$1\,\% of currently known exoplanets have been directly imaged, these observations permit detailed characterization of their atmospheres.
     Present and upcoming improvements in high-angular-resolution capabilities, including 30\,m-class telescopes \citep{Quanz2015,Bowens2021} and interferometry (ground-based: \citeauthor{Defrere2018} \citeyear{Defrere2018}; \citeauthor{Wagner2021} \citeyear{Wagner2021}, \citeyear{Wagner2021SPIE}; \citeauthor{Ertel2022} \citeyear{Ertel2022}; or space-based: \citeauthor{Kammerer2022} \citeyear{Kammerer2022}), open up the Habitable Zone (HZ) of nearby main-sequence stars for direct imaging. The HZ commonly refers to the circumstellar region where planetary surfaces may sustain liquid water given an adequate atmosphere \citep[][]{Kopparapu2013}.
     This region is relevant for the search of potential exoplanets located at the Earth-equivalent insolation distance (EEID) -- the orbital distance where objects receive the same irradiance as Earth -- and the assessment of their habitability.
     However, this region can also be rich in debris material, similar to the zodiacal cloud of the Solar System, known as exozodiacal dust. Like other debris disks found at larger separations, exozodiacal disks are expected to appear at the end of planetary formation, during the disk-clearing phase \citep[between 10\,Myr and a few Gyr age,][]{HinzExozodiChapter, Ercolano2017}. Collisions and evaporation of planetesimals serve as sources of exozodiacal dust 
     \citep{Kennedy2015_PR_drag, Marboeuf2016, Faramaz2017, Rigley2022} while other phenomena tend to deplete it (e.g., Poynting-Robertson (P-R) drag: \citeauthor{Wyatt1950} \citeyear{Wyatt1950}; \citeauthor{Burns1979} \citeyear{Burns1979}; planetary sweeping: \citeauthor{Bonsor2018} \citeyear{Bonsor2018}; or radiation pressure: \citeauthor{Burns1979} \citeyear{Burns1979}).
     The study of exozodiacal disks around nearby main-sequence stars serves several main purposes. It enables investigation of the properties and dynamics of the dust \citep[][]{Mennesson2014,Defrere_2015,Lebreton2016}, provides insights into the architecture and formation mechanisms of planetary systems \citep[][]{Stark2008}, and allows evaluation of cometary bombardment and its impact on the habitability of inner planetary systems \citep[][]{Chyba1990,OBrien2014,Rotelli2016,Kral2018,Wyatt2020,Ritson2020}. Studying these disks is also important for estimating their impact on rocky planet imaging, both from the shot noise they generate in the infrared \citep[][]{Defrere2010,Stark2014,Stark2019}, and from their scattering of stellar light in the visible \citep[][]{Beichman2006,Defrere2012,Roberge2012}.
     
\begin{figure}
        \centering
        \includegraphics[width=\linewidth]{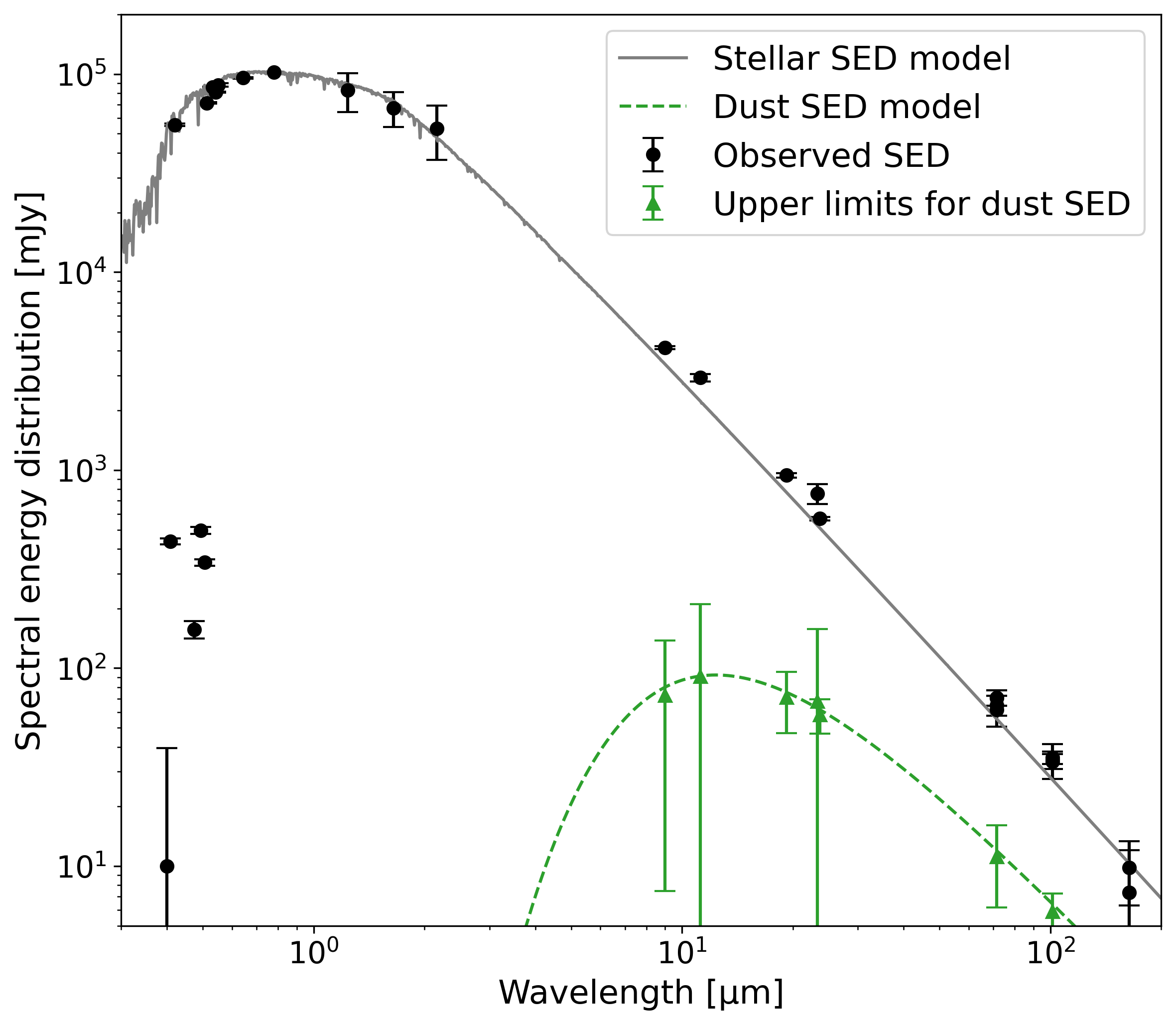}
        \caption{Observed spectral energy distribution (SED) of $\theta$\,Boo. The black points indicate the observed photometric data. The gray line represents the best fit stellar model, and the green dashed line shows the dust model. The green points indicate the upper limits of the dust emission.}
        \label{fig:SED_theta_boo}
    \end{figure}

     A survey of exozodiacal disks was carried out by the Hunt for Observable Signatures of Terrestrial Systems  \citep[HOSTS:][]{Ertel2018,Ertel_2020} which quantified the presence of warm dust ($\sim$300\,K, i.e., located in the HZ) and its surface density around nearby ($<$30\,pc) main-sequence stars. The survey was performed with the nulling mode of the Large Binocular Telescope Interferometer \citep[LBTI:][]{Defrere2015,Hinz2016,Ertel2020SPIE}.
     As part of the survey, the F7V type star $\theta$\,Boo was observed. This star, located at 14.5\,pc, showed signs of HZ dust emission estimated at $148.2\pm27.7$\,zodi \citep[1\,zodi being the dust surface density at the EEID of the Solar System, i.e., 1\,au; see definition in][]{Weinberger2015,Kennedy_2015}.
     Table\,\ref{tab:Theta Boo parameters} summarizes the stellar parameters of $\theta$\,Boo used in this study. Among these, the stellar luminosity L$_\star$ is used to calculate the local EEID following the method presented in  \cite{Weinberger2015}, where the radius $r_\text{EEID}$ is proportional to $\sqrt{\text{L}_\star}$. For $\theta$\,Boo, we find $r_\text{EEID}\simeq 2\,au$, which corresponds to a separation of $\sim $138\,mas.
     The star was previously observed by the space telescopes \textit{Spitzer} (2004) and \textit{Herschel} (2010) at 24, 70, and 100\,µm wavelength to search for warm ($\sim$600\,K) and cold dust ($\sim$120\,K) emission. These surveys did not show significant emission excess within their detection limits \citep{Bryden_2006,Trilling_2008,Gaspar_2013,Montesinos_2016}.
     Figure\,\ref{fig:SED_theta_boo} shows the observed spectral energy distribution (SED) of $\theta$\,Boo, and the upper limits for the dust emission obtained from space surveys\footnote{The photometric data, SED models, and upper limits are taken from \url{https://cygnus.astro.warwick.ac.uk/phsxfz/sdb/seds/masters/sdb-v2-142511.80+515102.7/public/index.html}}. The upper limits are estimated following the method used in \cite{Yelverton2019}, in which the observed data are fitted by a model with both a stellar and a dust component. The stellar component is a PHOENIX BT-Settl model \citep[][]{Allard2012}, and the dust component is a modified blackbody. 
     The error bars of the observed data are then propagated to the results of the dust model.
     
    \renewcommand{\arraystretch}{1.2}
    \begin{table}
        \centering
        \caption{Stellar parameters of $\theta$\,Boo used in this study. }
        \begin{tabular}{c c c c} \hline \hline
           Parameter  &  Value & Reference\\ \hline
           Distance d [pc]  &  14.5$\pm$0.04 & (1) \\ 
           T$_{\text{eff}}$ [K] & 6315$\pm$10 & (2) \\ 
           Age [Myr] & 500$\pm$250 & (3) \\
           L$_{\star}$ [L$_{\odot}$] & 4.01$\pm$0.02 & (4) \\
           M$_{\star}$ [M$_{\odot}$] & 1.31$\pm$0.07 & (4) \\
           R$_\star$ [R$_\odot$] & 1.73$\pm$0.01 & (4) \\
           ($B-V$) & 0.50 & (5)  \\
           $\nu \sin(i)$ [km/s] & 31.8$\pm$2 & (6) \\
           $\log R'_\text{HK}$ & -4.5$\pm$0.1 & (7) \\\hline
        \end{tabular}\\
        \raggedright
        \textbf{Notes.} T$_{\text{eff}}$ is the effective temperature; L$_{\star}$, the luminosity; M$_{\star}$, the mass; R$_\star$, the radius; and $\nu \sin(i)$, the projected rotational velocity. $R'_\text{HK}$ is an activity indicator defined as $F'_\text{HK}/\sigma T^4_{\text{eff}}$, where $F'_\text{HK}$ is the chromospheric flux in the H and K lines of Ca II. The estimated age of the star is, however, uncertain as it depends on the method used (see discussion in Sect.\,\ref{sec:Constrain on Planet}).
        \label{tab:Theta Boo parameters}\\
        \textbf{References.} (1) \cite{Gaia2016,Gaia2023}; (2) \cite{Montesinos_2016}; (3) \cite{Gaspar_2013}; (4) \cite{Boyajian2013}; \cite{Johnson1966}; (6) \cite{Rachford2009}; (7) \cite{Pace2013}
    \end{table}
    \renewcommand{\arraystretch}{1.0}
    
     This raises the question of the origin of the exozodiacal dust in the system. The presence of such warm dust is strongly correlated with the presence of a cold debris disk \citep[][]{Ertel_2020}. Statistical results from the HOSTS survey support this correlation, finding a detection rate of HZ dust of $78^{+8}_{-18}$\,\% in systems with a known cold debris disk. Only 3 of 28 stars have known HZ dust without cold debris, with $\theta$\,Boo being one of them.
     To investigate the nature of the exozodiacal disk of $\theta$\,Boo, a new observation in nulling mode was performed by the LBTI in 2023 that covered a wider range of parallactic angles than those obtained during the HOSTS survey. This work is part of an ongoing effort to study individual targets of the HOSTS survey ($\beta$\,Leo: \citeauthor{Defrere_2021} \citeyear{Defrere_2021}; $\alpha$\,Lyr: Faramaz et al. in prep.; 110\,Her: Rousseau et al. in prep.; $\epsilon$\,Eri: Weinberger et al. in prep.; $\beta$ Uma: Bryden et al. in prep.), and to investigate the nature of these disks.

     In this paper, we present a descriptive model for the spatial distribution of dust around $\theta$\,Boo. We also constrain the presence of companions using observations from the LBTI/L- and M-band InfraRed Camera (LMIRCam) and the System for coronagraphy with High-order Adaptive optics from R to K band (SHARK)-Near InfraRed (NIR) at the Large Binocular Telescope (LBT). Section\,\ref{sec:LBTO observations} describes the different observations of $\theta$\,Boo used in this study. Section\,\ref{sec:data reduction} explains the data reduction processes for these observations and their results. The constraints on the presence of possible companions in the system are presented in Sect.\,\ref{sec:Constrain on Planet}. The models of the dust distribution and its variability at the EEID are then shown in Sect.\,\ref{sec:Dust Variability}. Finally, we discuss the constraints on the dust distribution, its possible origins, and the source of its variability in Sect.\,\ref{sec:discussion}.


\section{Observations of $\theta$\,Boo}\label{sec:LBTO observations}

    \subsection{Instrument description}

        The nulling observations of $\theta$\,Boo were conducted at the LBT on UT 2017 April 11 and UT 2018 May 23 with the LBTI/Nulling Optimized Mid-Infrared Camera (NOMIC) in the N'-band (9.81-12.41\,µm) as part of the HOSTS survey. An additional nulling observation with NOMIC was obtained on UT 2023 May 25. High-contrast adaptive optics (AO) imaging with LMIRCam in the L'-band (3.41-3.99\,µm) and with SHARK-NIR in the H-band (1.38 to 1.82\,µm) was performed on UT 2024 February 24. Table\,\ref{tab:Observations properties} gives an overview of these observations.

        \begin{table*}
        \centering
        \caption{Overview of the $\theta$\,Boo observations.}
        \begin{tabular}{c c c c c c} \hline \hline
            Night [UT] & Spectral Band & Instrument & Observing mode & Time\,[UT] & PA\,[deg] \\ \hline
            2017-04-11 & N'& LBTI/NOMIC & Nulling &  07:51-09:06 & (-159,+155) \\
            2018-05-23 & N'& LBTI/NOMIC & Nulling &  05:05-05:43 & (-157,+179) \\
            2023-05-25 & N'& LBTI/NOMIC & Nulling & 03:58-07:35 & (-128,+119) \\
            2024-02-24 & L'& LBTI/LMIRCam & Imaging & 10:18-01:55 & (-140,+116) \\
            2024-02-24 & H & LBT/SHARK-NIR & Imaging & 10:18-01:55 & (-140,+116) \\ \hline
        \end{tabular}
        \label{tab:Observations properties}\\
        \raggedright
        \textbf{Notes.} UT stands for Universal Time, and PA stands for parallactic angle.
    \end{table*}
        
        \cite{Hill2012} describe the LBT, which consists of two 8.4\,m diameter apertures, each corrected with AO. The AO system achieves Strehl ratios of 80\%, 95\%, and 99\% at 1.6\,µm, 3.8\,µm, and 10\,µm, respectively, using Pyramid Wave Front Sensing (WFS) with the visible portion of the light \cite[][]{Bailey2010,Bailey2014,Pinna2016}. The infrared portion of the light is directed toward the Nulling and Imaging Camera (NIC), where it is split and focused onto different cameras as explained in \cite{Hinz2008}.

\begin{figure}
            \centering
            \includegraphics[width=\linewidth]{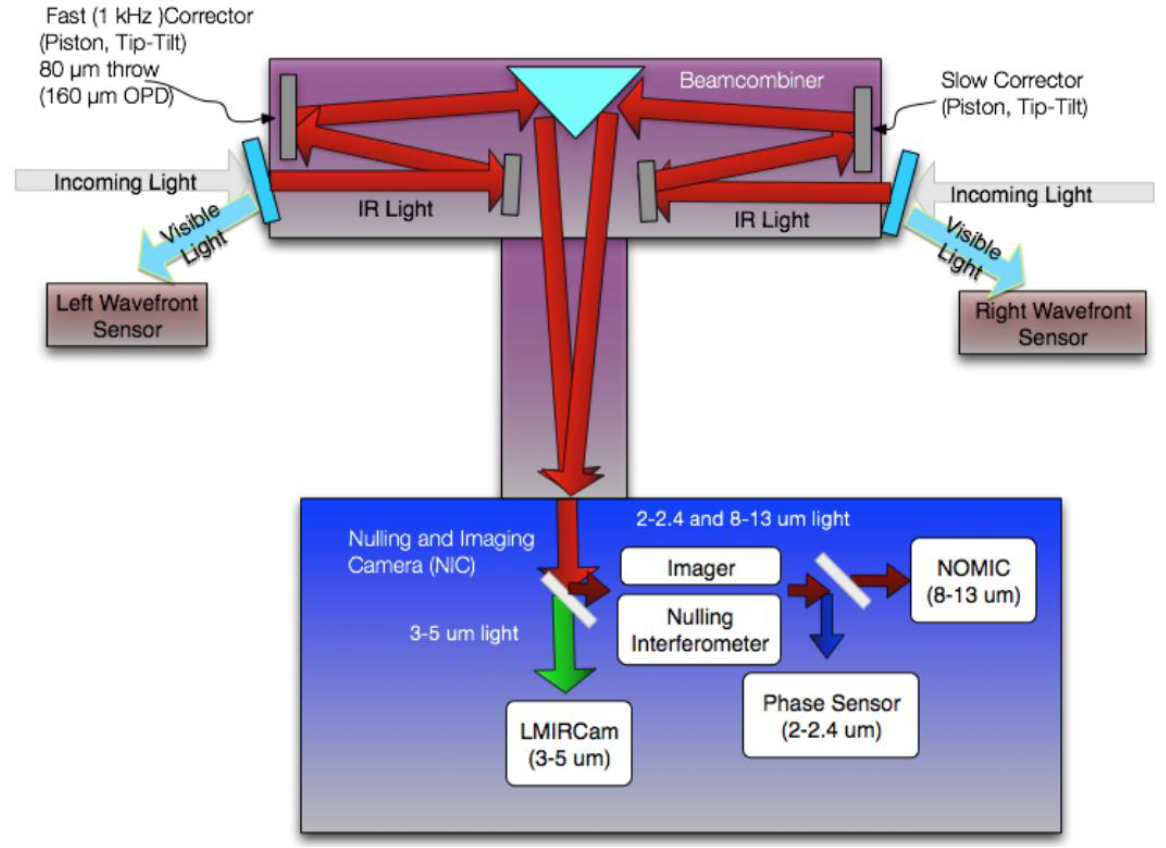}
            \caption{Block diagram of the LBTI architecture. Light from the two apertures enters from the top left and right and passes through the beam combiner subsystem (purple) and the NIC cryostat (blue). Both the beam combiner and the NIC are maintained at cryo-temperatures. At the entrance of the beam combiner subsystem, visible light is reflected by dichroics and directed to the wavefront sensing (WFS) units. Infrared light is transmitted to the beam combiner where two mirrors adjust the tip-tilt and the optical path difference (OPD) of the beams for nulling interferometry. After recombination of the two LBTI pupils by the roof mirror, the light is split in the NIC. The L-band (3-5\,µm) is directed to LMIRCam for high-contrast imaging of exoplanets, whereas the K-band (1.5-2.5\,µm) is directed to Phasecam to monitor the differential tip-tilt and phase between the two apertures. Finally, the N-band (8-13\,µm) light is directed to NOMIC which can be used in either imaging interferometry or nulling interferometry modes.
            }
            \label{fig:LBTI overview}
        \end{figure}

        The LBTI is the instrument that combines the light from the two LBT apertures to perform interferometric observations \citep[][]{Hinz2016}. LMIRCam and NOMIC can be used in this configuration by imaging both beams separately or by overlapping them. Figure\,\ref{fig:LBTI overview} provides an overview of the LBTI architecture. A K-band fast readout camera, called Phasecam, monitors the phase difference between the two apertures \citep[][]{Defrere2014}. It measures the differential piston and corrects it using two piston correctors located upstream on the tip-tilt mirrors.
        
        The SHARK-NIR instrument is located near the Gregorian focus of the left aperture and performs observations in the near-infrared \citep[][]{Farinato_2022,Marafatto_2022}. The LMIRCam instrument covers wavelengths from the J- to M-bands, while NOMIC operates in the N-band.

        The only relevant change made at the LBT between the initial observations of 2017 and 2018 and those in 2024 was the AO upgrade \citep[][]{Pinna2016}. This upgrade is not expected to have a major impact on the nulling observations, as the previous AO system already achieved a similar Strehl ratio in the N-band on the HOSTS targets. Moreover, any change in AO performance would have resulted in changes in instrumental null or phase jitter, which are considered in the data reduction pipeline and the error estimations.

    \subsection{N'-band imaging observations}\label{sec:nulling observations}

        In the N'-band, the NOMIC camera was used in a configuration designed for nulling observations. In this mode, the phase difference between the two apertures is locked at $\pi$.
        When combined in the pupil plane, light from a point source at the center of the field of view (FOV) destructively interferes and is not observed when reimaged at the detector plane. However, off-axis sources add an extra phase difference between the apertures, depending on their on-sky position. This results in emission from off-axis sources and any sufficiently extended stellar components (termed ``stellar leakage'' or ``instrumental null depth'') being imaged on the NOMIC camera.
        
        The throughput of off-axis sources as a function of their on-sky position is described by the ``transmission map'' $T$. For the LBTI, which has two apertures separated by a baseline $B=$14.4\,m, $T=\sin^2(\pi \phi B/\lambda)$, where $\phi$ is the on-sky position projected onto the baseline vector and $\lambda$ is the wavelength \citep[][]{Kennedy_2015}. The transmission of an object on the camera therefore depends on its on-sky position and the parallactic angle rotation during the night (i.e., the baseline rotation). The coherent FOV $\theta_c$ -- the on-sky region where light sources can interfere -- is defined as
        \begin{equation}\label{eq:coherent FOV}
           \theta_c \equiv \frac{\lambda^2}{B \times \Delta \lambda},
        \end{equation}
        where $\Delta \lambda$ is the spectral bandwidth \citep[][]{Thompson1986}. For nulling in the N'-band (9.81-12.41\,µm), the coherent FOV is $\theta_c\simeq670$\,mas.
        
        The amount of instrumental null depth can be estimated using a set of calibrator stars (labeled ``CAL'') observed before and after the science target $\theta$\,Boo.
        These calibrators were selected following \cite{Mennesson2014}, using the two catalogs from \cite{Borde2002} and \cite{Merand2005}, with additional stars from the JSDC catalog and the SearchCal tool from \cite{Chelli2016}.  Table\,\ref{tab:Properties of Theta Boo and Calibrators} lists the calibrators used in this study and their properties.
        
        Observations consisted of several observation blocks (OBs), each comprising 2000$\times$42.7\,ms frames. Between nulling OBs, the telescope was offset by $\pm$2.3\,\arcsec at the detector in a vertical nod movement. The instrument uses an Aquarius detector, with a FOV of 18\arcsec$\times$18\arcsec and a pixel scale of 17.9\,mas.
        This allows for measurement of the mid-infrared thermal background noise at pixels that may have received a signal in previous observations.
        Each observation includes multiple nulling OBs (with overlapping beams), one photometric OB (with separated beams), and one background OB (with beams nodded off the detector).
        The N'-band observations on 2017 April 11, 2018 May 23, and 2023 May 25 targeted exozodiacal dust at a separation range of 40-700\,mas (i.e., 0.6-10.4\,au for $\theta$\,Boo). Table\,\ref{tab:Observations properties} summarizes the three observing nights.
        The observation sequences for each night were as follows: CAL1-$\theta$\,Boo-CAL2 (2017 April 11); CAL1-$\theta$\,Boo-CAL3 (2018 May 23); and CAL1-$\theta$\,Boo-CAL4-$\theta$\,Boo-CAL5 (2023 May 25). Each sequence began with CAL1, enabling comparison of results across the three nights.

        \begin{table*}
        \centering
        \caption{Basic properties of $\theta$\,Boo and its calibrators used for the nulling observations on 2017 April 11, 2018 May 23, and 2023 May 25.}
        \begin{tabular}{c c c c c c c c c c} \hline\hline
           ID  & HD & R.A. & Dec. & Spectral Type & $m_V$ & $m_K$ & $\theta_{LD}$$\pm$1$\sigma$\,[mas] & References \\ \hline
            $\theta$\,Boo & 126660 & 14 25 11.8 & +51 51 02.7 & F7V & 4.04$\pm$0.002 & 2.80$\pm$0.09 & 1.14$\pm$0.05 & [Du02], [Kh09]\\
           CAL 1 & 128902 & 14 38 12.6 & +43 38 31.7 & K4III & 5.72$\pm$0.004 & 2.32$\pm$0.259 & 1.83$\pm$0.25 & [Kh09] \\
           CAL 2 & 138265 & 15 27 51.4 & +60 40 12.8 & K5III & 5.91$\pm$0.004 & 1.59$\pm$0.238 & 2.83$\pm$0.35 & [Kh09] \\ 
           CAL 3 & 131507 & 14 51 26.4 & +59 17 38.4 & K4III & 5.48$\pm$0.003 & 2.21$\pm$0.233 & 1.97$\pm$0.24 & [Kh09] \\
           CAL 4 & 128000 & 14 32 30.9 & +55 23 52.8 & K5III & 5.72$\pm$0.004 & 2.14$\pm$0.195 & 2.1$\pm$0.21 & [Kh09] \\
           CAL 5 & 138265 & 15 27 51.4 & +60 40 12.8 & K5III & 5.91$\pm$0.004 & 1.59$\pm$0.238 & 2.83$\pm$0.35 & [Kh09] \\ \hline
        \end{tabular}
        \label{tab:Properties of Theta Boo and Calibrators}
        \raggedright
        \textbf{References.} Coordinates and spectral types from SIMBAD; V\&K magnitudes and error bars from [Du02] \cite{Ducati2002} and [Kh09] \cite{Kharchenko2009}; Limb-darkened angular diameters and their 1$\sigma$ uncertainties from \cite{Chelli2016}.
    \end{table*}

    \subsection{L'-band imaging observations}\label{sec:LMIRCam observation}

       We obtained L'-band observations with LMIRCam on the LBT right arm (with one 8.4\,m aperture) using Director’s Discretionary Time (DDT). 
        These observations, conducted on 2024 February 24, targeted a potential companion at a separation range of 0.3-4\arcsec\,(i.e., 4-58\,au for $\theta$\,Boo, see Sect.\,\ref{sec:LMIRCam reduction}). They consisted of 28 sequences of exposures, each with 2000 images with 55\,ms integration time, totaling 3080\,s ($\sim$51\,minutes).
       We did not use a coronograph and centering was achieved using a rotational-based algorithm following \cite{Morzinski2015}.
        The Teledyne H2RG detector provided a FOV of 20\arcsec$\times$20\arcsec and a pixel scale of 10.9\,mas.
        We acquired observations in pupil-stabilized mode that allowed the rotation of the FOV to perform angular differential imaging \citep[ADI:][]{Marois_2006}. During observations, the parallactic angle rotated by 104$\degree$, and the seeing fluctuated between 1.05\arcsec and 2\arcsec.
        To monitor the thermal sky background and detector drifts, the stellar image was nodded up and down on the detector by an offset angle of 4.5\arcsec\,between every observing block. 

    \subsection{NIR imaging observations}\label{sec:SHARK observation}

    In parallel with LMIRCam, we conducted simultaneous observations with SHARK-NIR operating on the LBT left arm. Using a broadband H filter (central wavelength 1.6~µm, bandwidth 0.218~µm) and a Gaussian Lyot coronagraph with an inner working angle (IWA) of 150~mas, we acquired data with a Teledyne H2RG detector (FOV: 18$\arcsec\times$18$\arcsec$; pixel scale: 14.5\,mas).
    To avoid detector saturation near the star and minimize exposure time, we selected an observing mode where only a central region of dimensions 2048 x 200 pixels was read out.
    This limits the usable region for companion detection, and for the definition of contrast, to a separation of 1.3\arcsec\,(see Sect.\,\ref{sec:SHARK reduction}).
Each frame had a 0.84~s exposure time,  and we acquired 6855 frames. This gave a total exposure time of 5758.2~s ($\sim$96~minutes) on the target. The observations were also taken to allow ADI during post-processing.
    To correctly estimate the obtained contrast, images of the stellar point spread function (PSF) without the coronagraph were taken at the start and end of the observations.
    These frames were taken using an appropriate neutral density filter (ND3) to avoid detector saturation.

\section{Data reduction}\label{sec:data reduction}

    \subsection{N'-band nulling data}\label{sec:data reduction nulling}
        We used the LBTI nulling pipeline from \cite{Defrere2016} for the N'-band nulling data reduction and calibration.
        The pipeline first corrects the raw images by removing bad pixels and subtracting the mid-infrared thermal background. Because the background cannot be measured at the star position when the star is present, it is estimated from the series of OBs with an opposite nod position. The background is then subtracted using either the mean value of the series or a principle component analysis (PCA) approach \cite[][]{2024Rousseau}. In this work, we used the mean background subtraction. The pipeline then computes the flux contained within an aperture radius using the background-subtracted images.
        The PSF of the telescope at 11\,µm ($\lambda/D$), where D is the diameter of an individual LBT aperture, has a full width at half maximum (FWHM) of $\sim$ 286\,mas, which is oversampled by 16\,pixels (each 17.9\,mas).
        We therefore selected a standard photometric aperture radius of $\sim$$0.5\lambda/D$ (i.e., 8\,pixels, or 143\,mas), which maximizes the signal-to-noise ratio (S/N) for a point-like source's measured flux.
        
\begin{figure*}
            \centering
            \includegraphics[width=0.45\linewidth]{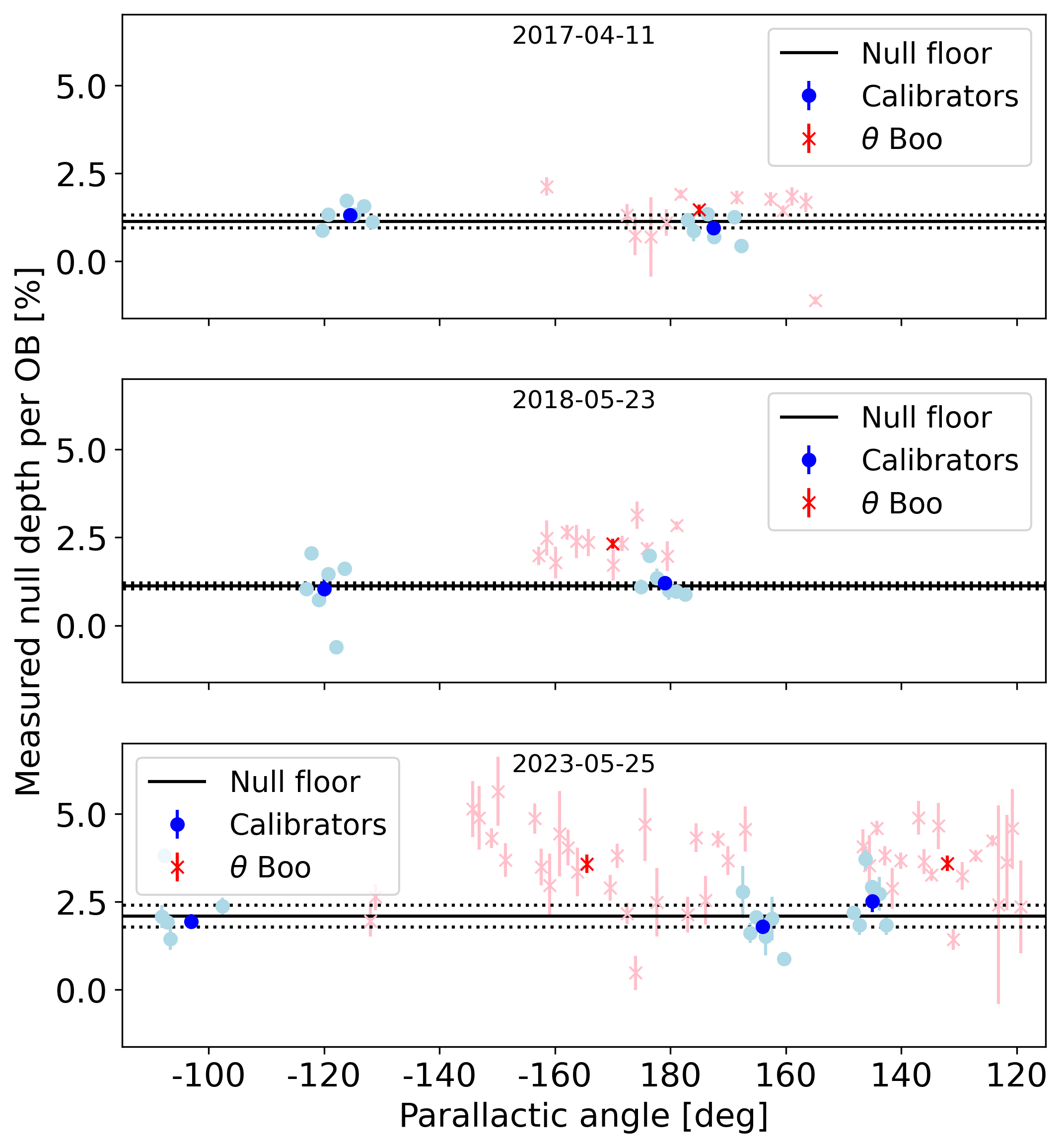}
            \includegraphics[width=0.45\linewidth]{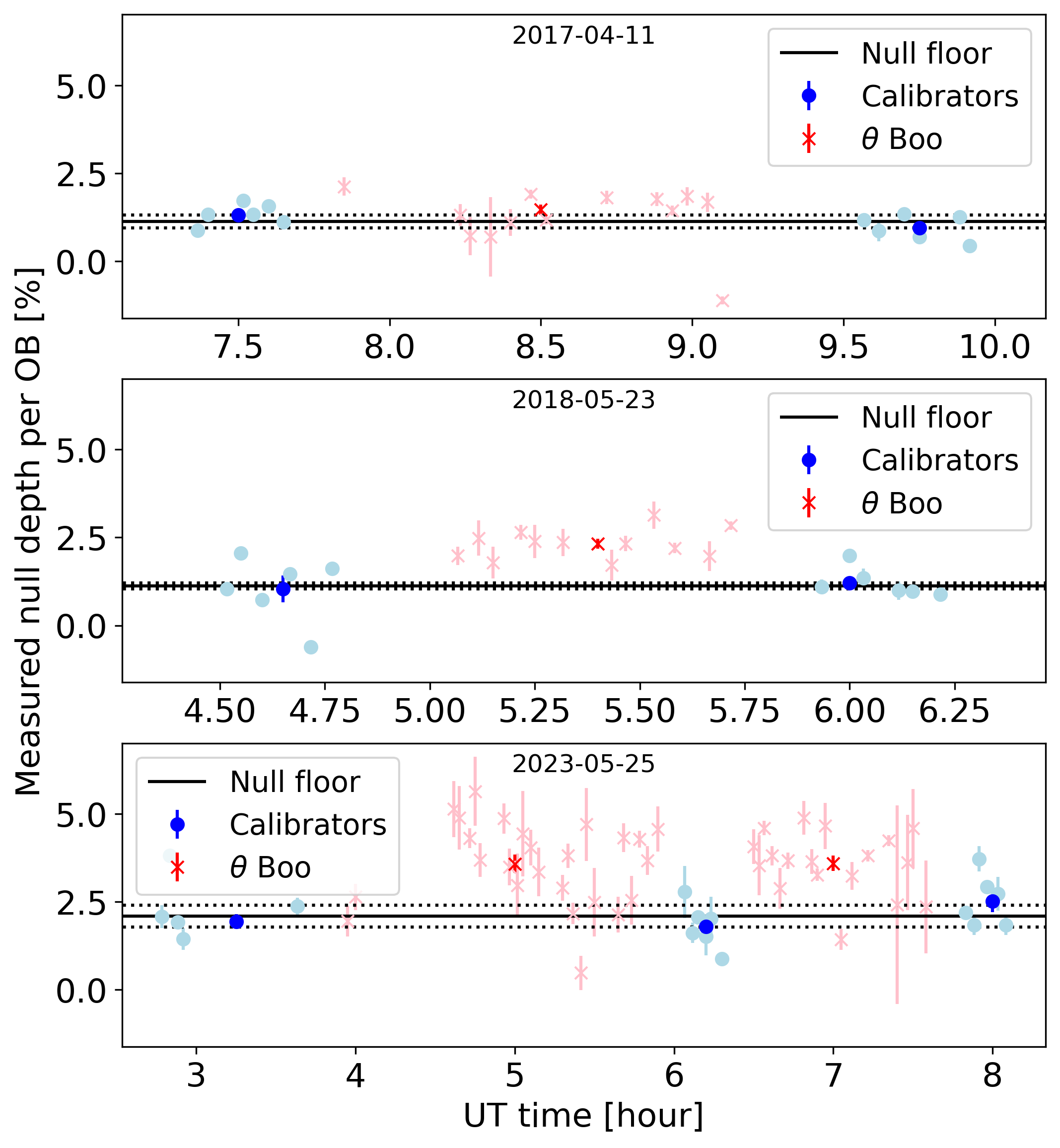}
            \caption{Null depths obtained after the data reduction pipeline described in Sect.\,\ref{sec:data reduction nulling}. These points are computed using the reduction pipeline with an aperture radius of 24\,pixels (i.e., 430\,mas). The OBs are plotted as a function of their parallactic angle (left) and UT of observation (right), for each of the three nights: 2017 April 11 (top), 2018 May 23 (middle), and 2023 May 25 (bottom). In each panel, light blue points and light red crosses show the measured null depth of the OBs for the calibrators and for $\theta$\,Boo, respectively. Dark blue points and dark red crosses indicate the null depth values for each pointing, corresponding to the unweighted mean values of the OBs included in the pointing. The instrumental null depths measured from the calibrators are averaged to calculate the null floor, which represents the reference level below which any excess cannot be reliably detected. The null floor and its error bar are shown by the solid and dotted black lines, respectively. The error bar corresponds to the 1$\sigma$ standard deviation.}
            \label{fig:example TF}
        \end{figure*}

        To probe the dust spatial distribution, we performed photometric measurements using several aperture radii: 8, 16, 24, 32, and 40\,pixels (i.e., 143, 286, 430, 573, and 716\,mas).
        After computing the flux for each radius, the pipeline calculates and calibrates the null for each OB using the nulling self-calibration (NSC) approach, originally developed for the Palomar Fiber Nuller \citep{Hanot2011,Mennesson2011} and adapted for the LBTI \citep{Defrere2016,Mennesson2016}. This approach removes errors in the nulling setpoint between the science star and its calibrators. The computed null is also corrected for the instrumental null depth generated by the science target and calibrators (see Sect.\,\ref{sec:nulling observations}). The limb-darkened angular diameters $\theta_{LD}$ and their 1$\sigma$ uncertainties are listed in Table\,\ref{tab:Properties of Theta Boo and Calibrators}.
        The null depths obtained for the calibrators and $\theta$\,Boo are shown in Fig.\,\ref{fig:example TF}. The instrumental null depths from the calibrators define a ``null floor'', corresponding to their mean value. The difference between the $\theta$\,Boo nulls and the null floor is attributed to dust emission passing through the transmission pattern $T$ (see Sect.\,\ref{sec:nulling observations}). This difference is referred to as the ``calibrated null depth.''
        Small sample statistics \citep[as described in][]{Mawet2014} are not used in the NSC for small apertures. The statistical noise is estimated from the time series of images and the flux measurement in each image. With approximately 2000 images per OB and several OBs per observation, the noise estimated by the NSC is not affected by small sample statistics.

\begin{figure*}
            \centering
            \includegraphics[width=0.49\linewidth]{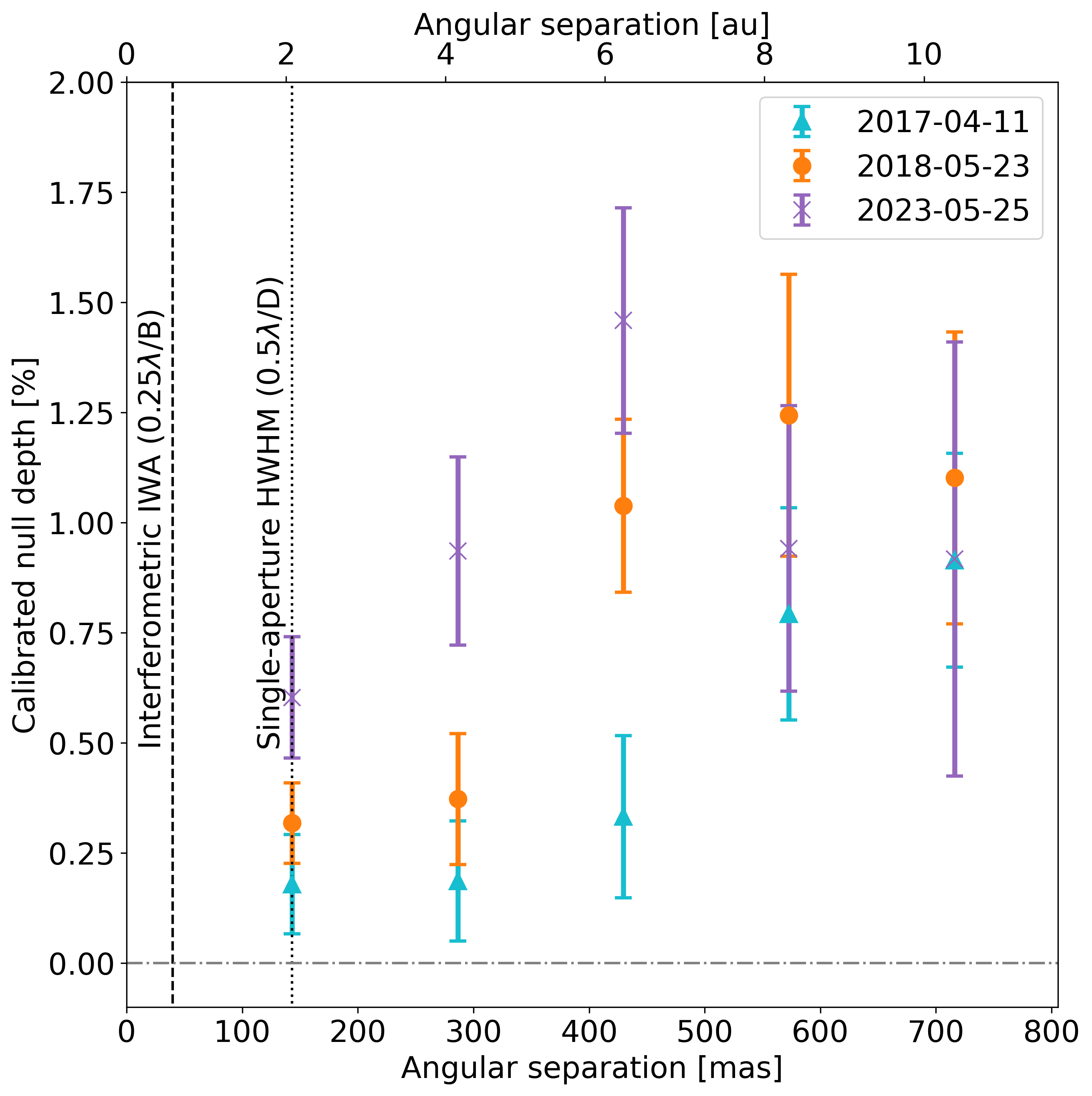}
            \includegraphics[width=0.49\linewidth]{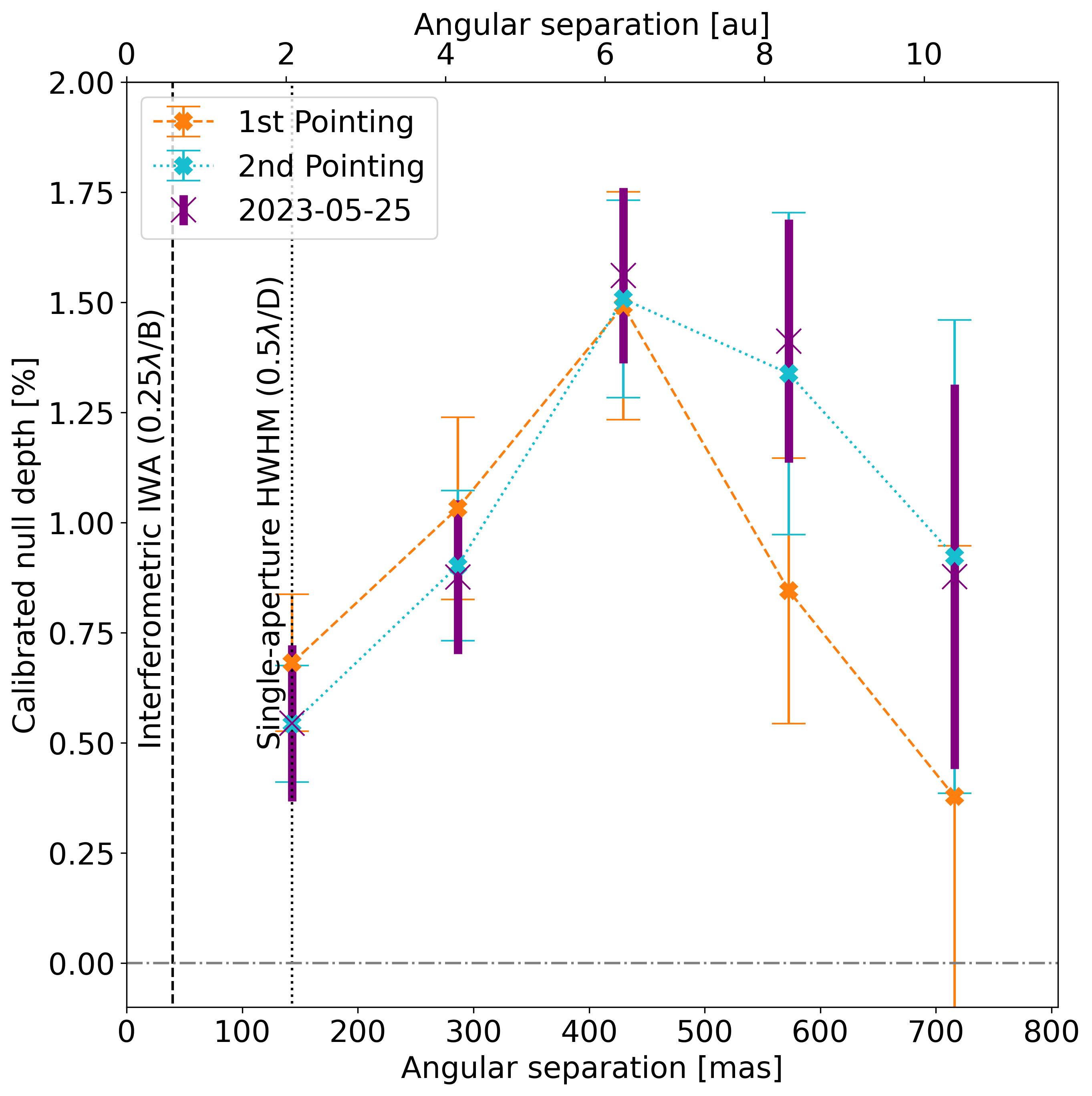}
            \caption{Left: Calibrated null depths for the different aperture radii and for the three nights of observation with NOMIC. The aperture radii are 8, 16, 24, 32, and 40\,pixels (i.e., 143, 286, 430, 573, and 716\,mas, respectively). Right: Calibrated null depths for the 2023 May 25 observation as a function of aperture radius. Results for the two individual pointings are shown as dotted and dashed lines, and the combined results are shown as crosses. The first pointing corresponds to the OBs taken from 04:00 to 06:00\,UT and the second pointing is from 06:30 to 07:30\,UT.}
            \label{fig:Null vs Radius}
        \end{figure*}

        After calculating the calibrated null depth for each OB of the science target with a given aperture radius, we take their mean value as the calibrated null depth for that radius.
        The calibrated null depths for all apertures jointly probe the excess flux from the dust and its radial distribution. The left panel of  Fig.\,\ref{fig:Null vs Radius} shows the data reduction results for the three nulling observation nights. The null depths of the OBs associated with each point are listed in Appendix\,\ref{ap:all TF}. The difference between each aperture radius is 8 pixels (143\,mas), corresponding to $0.5\lambda/D$. The different calibrated nulls are correlated due to the PSF size, so a model is required to determine  the dust location in the system.
       Modeling results for these observations are presented in Sect.\,\ref{sec:Dust Variability}.
        
        To assess whether these results depend on the parallactic angle, we analyzed the two pointings from the 2023 May 25 observation individually. Figure\,\ref{fig:example TF} shows the parallactic angle ranges for the two pointings, and the right panel of Fig.\,\ref{fig:Null vs Radius} plots the calibrated null excess for the pointings individually and for the combined data. The results show no significant differences in calibrated nulls compared to the combined results, indicating that the excess can be assumed to be independent of the on-sky parallactic angle.
        This suggests that the exozodiacal disk is likely close to face-on if the disk is vertically thin, which is useful for modeling the dust distribution in Sect.\,\ref{sec:Dust Variability}.
        Additionally, this allows us to compare results from the three nights of observation despite their different parallactic angle ranges.

        We used the calibrated nulls from Fig.\,\ref{fig:Null vs Radius} to estimate the total infrared excess from the dust measured by the LBTI at 11\,µm.
        Estimating this excess without prior assumptions on the dust angular distribution is challenging due to the transmission map $T$. Since $T$ follows a $\sin^2$ function with the on-sky position parallel to the baseline vector, we assume that approximately 50\,\% of the excess generated by the dust in the FOV is reimaged on the camera \citep[][]{Rigley2020}. To estimate the total excess for each observing night, we take the aperture radius with the highest calibrated null, and double it. This yields total emission excesses at 11\,µm of 1.83$\pm$0.49\,\%, 2.49$\pm$0.66\,\%, and 2.92$\pm$0.51\,\% of the stellar emission for the 2017, 2018, and 2023 observing nights, respectively. The theoretical blackbody emission of $\theta$\,Boo is $\sim$2315\,mJy, which gives an excess of 42$\pm$11, 58$\pm$15, and 68$\pm$12\,mJy from the dust for the 2017, 2018, and 2023 observing nights, respectively. These estimates are consistent with the upper limit shown in Fig.\,\ref{fig:SED_theta_boo}. An increase in the total excess by $\sim$2$\sigma$ is observed between 2017 and 2023. The mean excess value is 56$\pm$7.4\,mJy.

    \subsection{L'-band direct imaging data}\label{sec:LMIRCam reduction}
    
    We processed the LMIRCam data using the LBTI Exozodi Exoplanet Common Hunt (LEECH) survey pipeline \citep[][]{Stone_2018}. The pipeline first replaces bad pixels with the median of the eight closest pixels. 
    As with the N'-band observations, the thermal background is subtracted using the median value of the opposite nod images from the closest OB. The images are then coarsely corrected for distortion using the dewarp coefficients of \cite{Maire2015} which remain sufficiently precise, as astrometric precision is not required. The telescope PSF has a FWHM of 95\,mas at 3.7\,µm. The detector pixels have an on-sky size of 10.7\,mas, so the PSF is oversampled by a factor of about 4. Each image is therefore binned 2$\times$2 to remove cosmic rays or any remaining bad pixels.
    
    The pipeline then fits and removes the diffracted light from the star using PCA \cite[][]{Amara_Quanz_2012,Soummer_2012,Gomez_Gonzalez_2017} before derotating and stacking the images.
    To optimize high contrast capability, the PCA method was performed annulus by annulus with each annulus having a width of 9 pixels ($\sim$$2\lambda/D$) and a subtraction region annulus of 1 pixel width. Artificial planets were injected at different angular radii, and the number of PCA components was chosen to maximize their S/N. The best contrast was achieved after several iterations. The planet injection and contrast estimation procedures are detailed in \cite{Wagner2019}. The two nod positions were reduced separately, but their images were recombined using a weighted mean. Weights were set for each annulus to optimize the detection of the artificial planets. Regions of poor sensitivity due to dust diffraction within LMIRCam were down-weighted accordingly.

    \begin{figure*}
        \centering
        \includegraphics[width=0.4\linewidth]{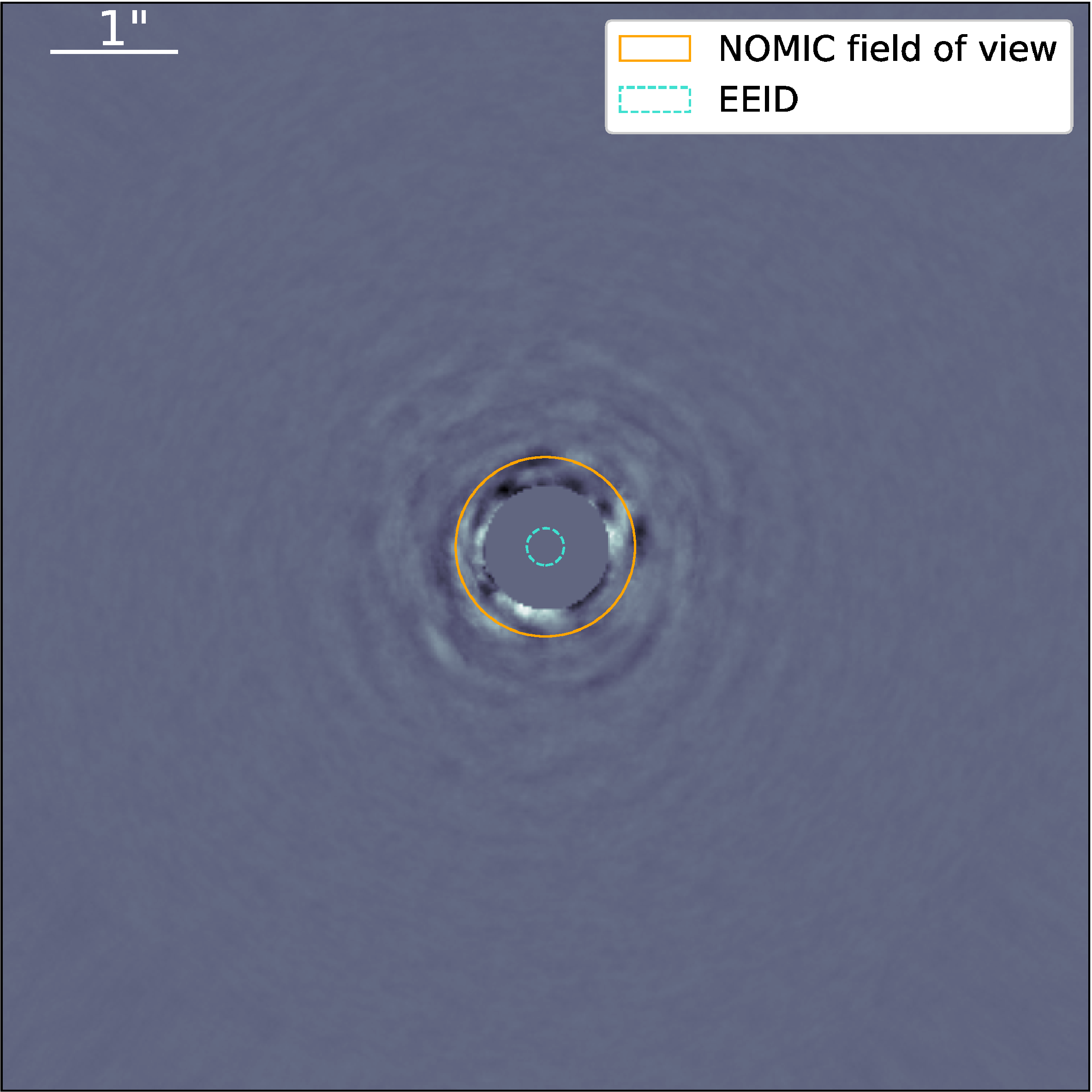}
        \includegraphics[width=0.4\linewidth]{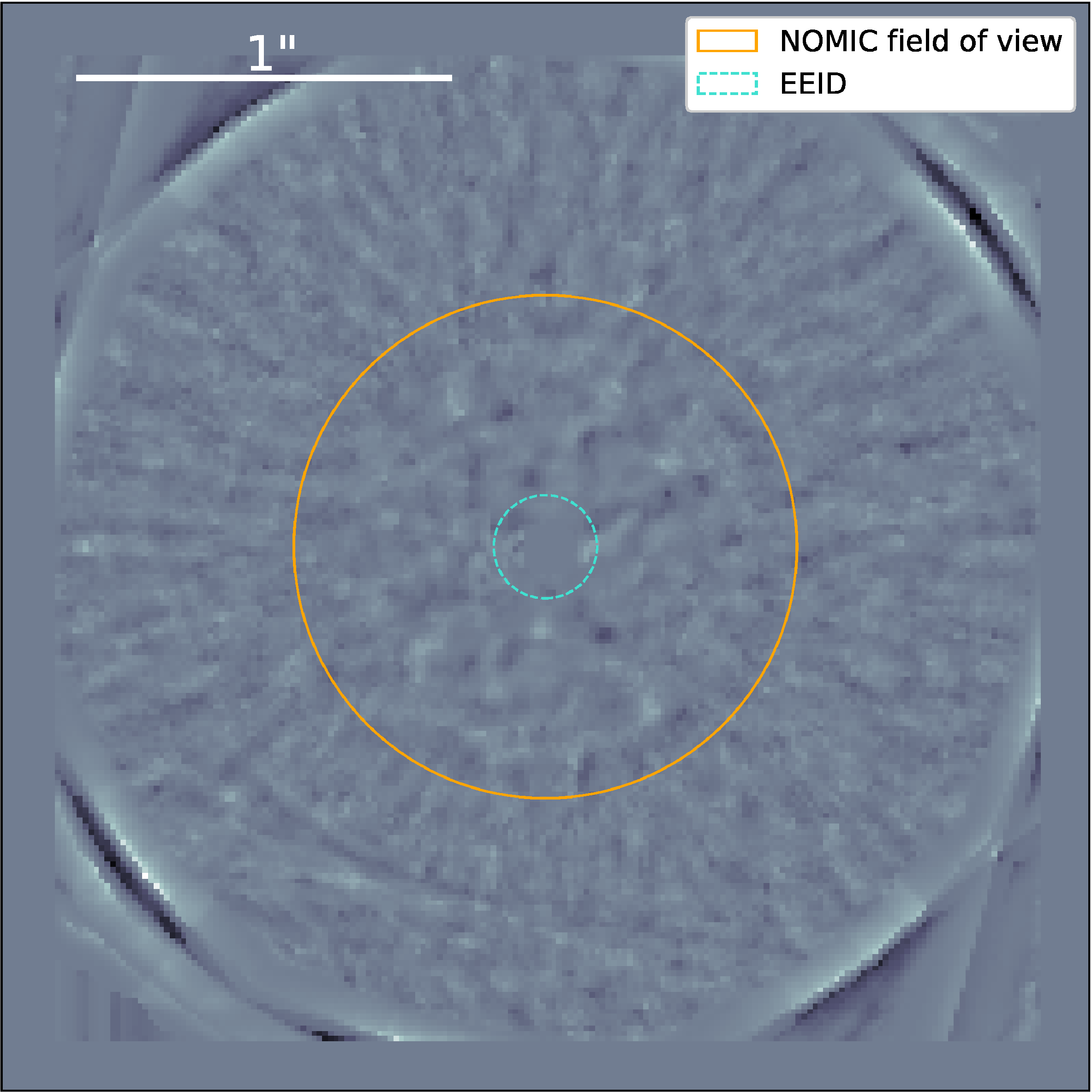}
        \caption{Left: ADI image of $\theta$\,Boo obtained with LMIRCam after processing with the LEECH-survey pipeline. The image has a FOV of 4\arcsec. A central mask of 5$\lambda/D$ radius is used to improve the visualization. No feature indicating the presence of a companion is detected between 0.3\arcsec\,and 4\arcsec. Right: ADI image of $\theta$\,Boo from SHARK-NIR after processing with the SHARP pipeline using five PCA components. The image has a FOV of 1.3\arcsec. 
        For both images, the outer circle indicates the coherent FOV of NOMIC in the N'-band ($\sim$$670$\,mas; see Eq.\,(\ref{eq:coherent FOV})), and the inner dashed circle represents the EEID ($\sim$$138$\,mas). No feature indicating the presence of a companion is detected between 0.15\arcsec\,and 1.3\arcsec.}
        \label{fig:LMIRCam image}
    \end{figure*}
    
    The left panel of Fig.\,\ref{fig:LMIRCam image} shows the reduced image of $\theta$\,Boo. No feature resembling a companion can be observed between 0.3\arcsec\,and 4\arcsec. The detection limits for this observation are presented in Sect.\,\ref{sec:Constrain on Planet}.
    To account for the reduced number of resolution elements near the star -- and its impact on the confidence level of the detection thresholds -- the detection limits were calculated including small sample statistics
    \citep[][]{Mawet2014}.

    \subsection{NIR direct imaging data}\label{sec:SHARK reduction}
    The SHARK-NIR data were reduced using the Python pipeline called SHARP, currently under development. The pipeline first subtracts the dark and divides the image by the flat. The internal deformable mirror generates four satellite spots that are images of the star with lower brightness and symmetric with respect to the position of the target, arranged in a cross configuration that can be used to determine the position of the star behind the coronagraph. Fitting two lines through two opposite spots, their intersection accurately defines the central position of the stellar PSF. Frames of poor quality, for example those not adequately masked by the coronagraph due to bad weather conditions, were removed. After this procedure, 6422 frames remained. To improve the reduction time, the frames were grouped in sets of four and averaged, forming a data cube of 1605 frames. Speckle subtraction was then performed on this cube, using the PCA algorithm, followed by image derotation and stacking. The PCA was run with 1, 5, 10, 15, 20, and 25 components. A reduced image was produced for each case; however, no significant differences were observed between the different images.

    The right panel of Fig.\,\ref{fig:LMIRCam image} shows the resulting SHARK-NIR image using five PCA components. No feature resembling a companion is observed between 0.15\arcsec and 1.3\arcsec. We then calculate the brightness contrast by measuring the standard deviation in one-pixel-wide rings and dividing by the normalization factor obtained using the non-coronagraphic images described in Sect.\,\ref{sec:SHARK observation}.
    The detection limits corresponding to this observation are presented in Sect.\,\ref{sec:Constrain on Planet}. As with the LMIRCam data, the detection limits were calculated accounting for small sample statistics.


    \section{Detection limits in the L'-band and NIR}\label{sec:Constrain on Planet}

\begin{figure*}
        \centering
        \includegraphics[width=0.49\linewidth]{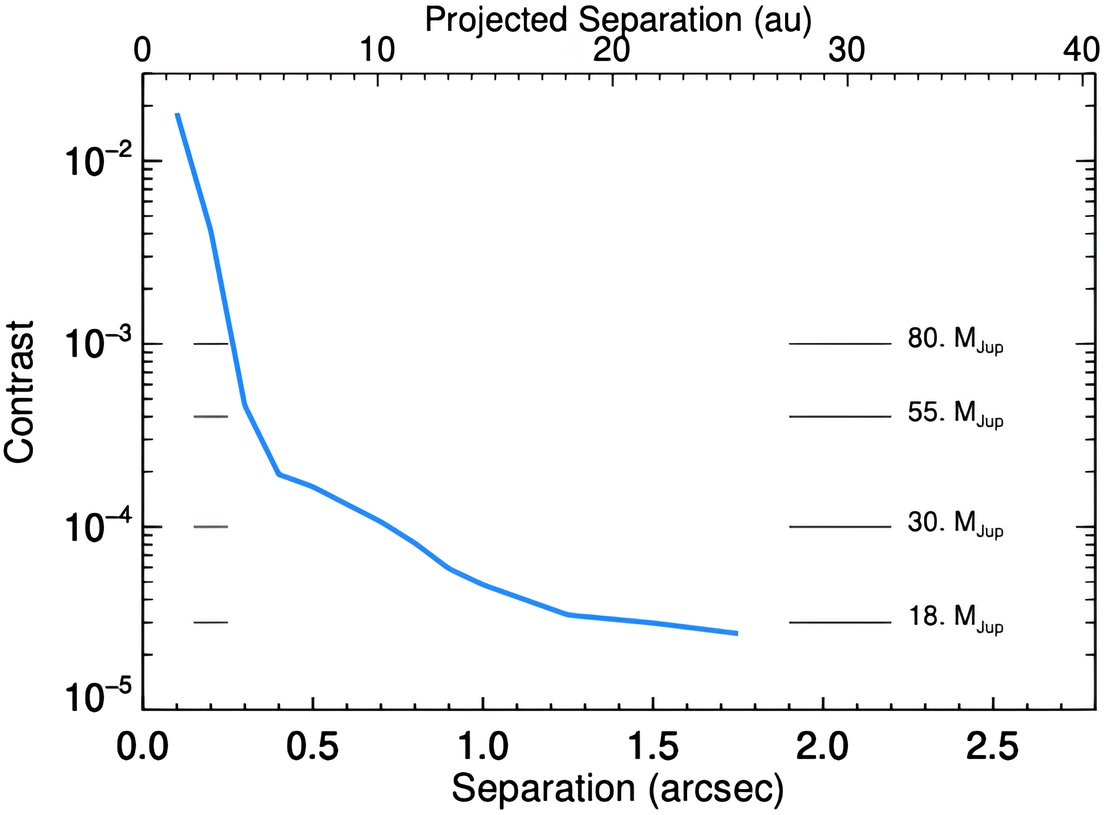}
        \includegraphics[width=0.49\linewidth]{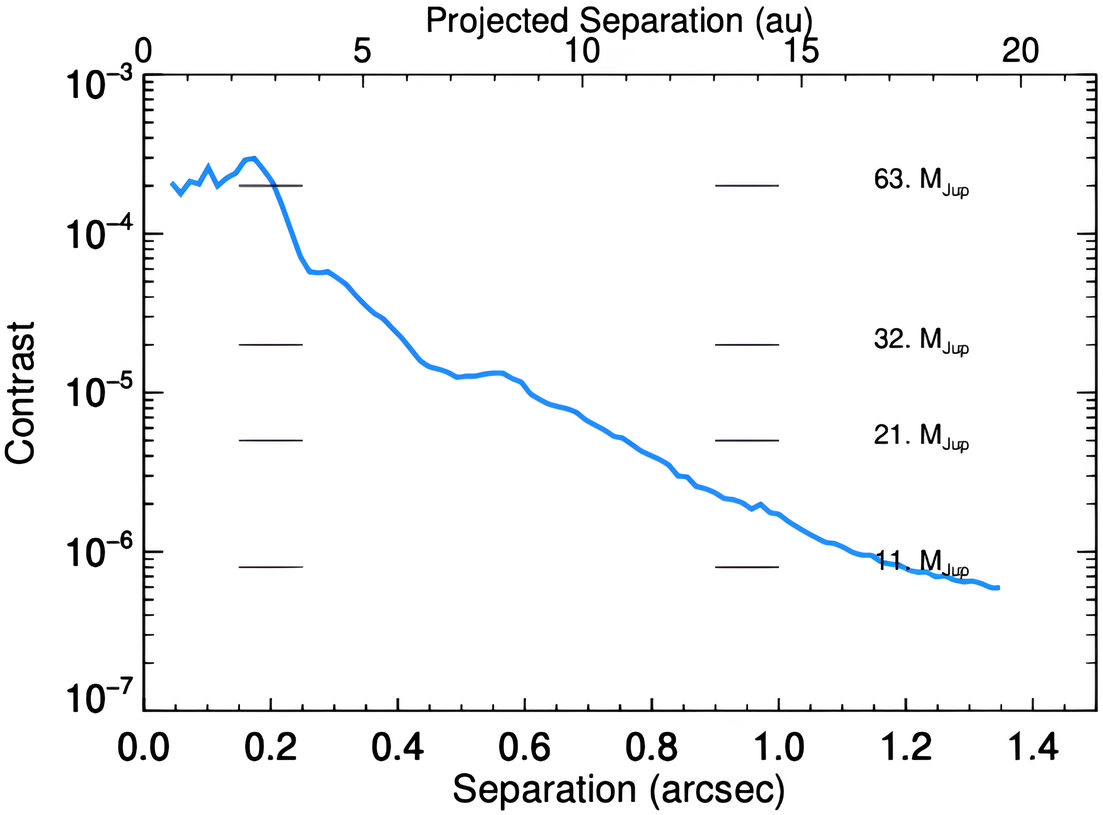}
        \caption{Detection limits for companions around $\theta$\,Boo in the L'-band with LMIRCam (left) and in the H-band with SHARK-NIR (right). The blue line shows the contrast as a function of angular separation. For four contrast levels, the corresponding detection limit is given in terms of mass (in units of Jupiter mass). The best mass constraint is obtained in the H-band around 1\arcsec, with a limit of 11\,$M_{\text{Jup}}$. The system is assumed to be 500\,Myr for these mass constraints.}
        \label{fig:LMIRCam contrast curves}
    \end{figure*}

    The images obtained from LMIRCam and SHARK-NIR after post-processing (Fig.\,\ref{fig:LMIRCam image}) show no apparent point-like feature, but they can still be used to compute companion detection limits. Figure\,\ref{fig:LMIRCam contrast curves} presents these detection limits (corresponding to S/N=5) as contrast curves relative to angular separation from the star.
    Constraints on planetary masses are derived using the COND evolutionary model described in \cite{Baraffe2003}. The model assumes a ``hot start'' formation scenario, in which gravitational instabilities in the protoplanetary disk cause dust and gas to collapse and form a planet \citep[][]{Spiegel2012}. The collapsing gas conserves its entropy, resulting in a high initial entropy for the planet.
    The COND model neglects dust opacity by assuming that atmospheric dust immediately falls below the photosphere after formation. Given the contrast curves and stellar parameters, the model provides planetary masses corresponding to various contrast levels.
    The best detection limit in the NIR is achieved by SHARK-NIR with a limit of 11\,$M_\text{Jup}$ around 1.3\arcsec. The LMIRCam is unable to probe the planetary mass regime, with a limit of 18\,$M_\text{Jup}$ for separations $>$1.5\arcsec.

    The mass constraints for these observations are calculated assuming an age of 500\,Myr for $\theta$\,Boo from \cite{Gaspar_2013}. However, the age of the system is uncertain, with different methods yielding a range of estimates. Table\,\ref{tab:Age Theta Boo} summarizes the current estimates for each method, which span from 400\,Myr to several Gyrs.
    \cite{Rachford2009} reported that $\theta$\,Boo is approximately twice as luminous as a zero-age main-sequence star of the same color. This implies that the star is either at the end of its main-sequence phase or is a binary with equal components. Using the Stefan-Boltzmann law with the stellar parameters listed in Table\,\ref{tab:Theta Boo parameters} yields a stellar radius of 1.77\,R$_\odot$.
    Interferometric observations at the Center for High Angular Resolution Astronomy (CHARA) measured a stellar radius of $1.73\pm0.01$\,R$_\odot$ \citep[][]{Boyajian2013}. The theoretical and measured stellar radii are about 30\,\% larger than that of a zero-age F7V star \citep[i.e., 1.324\,R$_\odot$,][]{Pecaut2013}. This indicates that $\theta$\,Boo is likely at the end of its main-sequence phase and not a binary with equal components. Its age is therefore likely toward the higher end of the range in Table\,\ref{tab:Age Theta Boo}, between 3 and 4.7\,Gyr.
   For ages beyond 500\,Myr, the internal heat resulting from planet formation has already dissipated, and there should be no significant difference in the evolutionary model. The mass limits are therefore still relevant even for this age range.
    It should also be noted that $\theta$\, Boo has a binary companion detected at a separation of 70\arcsec \citep[$\sim$1000\,au,][]{Eggen1956}. However, this separation is too large to impact the presence of a debris disk in the system \citep[][]{Yelverton2019}.

    \begin{table}
        \centering
        \caption{Summary of age estimates of $\theta$\,Boo from different independent methods.}
        \begin{tabular}{c c c} \hline \hline
        Method & Estimated age [Gyr] & References \\ \hline 
        Isochrones & 2.9-4.7 & (1) \\
        Chromospheric activity & 0.4-0.9 & (2) \\ 
        X-ray emission & 0.4 & (3)\\ \hline
        \end{tabular}\\
        \label{tab:Age Theta Boo}
        \raggedright 
        \textbf{References.}
        (1) \cite{Marsakov1995,Lachaume1999,Holmberg2009,Boyajian2013,Luck2017};
        (2) \cite{Barry1988,Vican2012,Gaspar_2013,Stanford-Moore2020};
        (3) \cite{Vican2012}.
    \end{table}

\section{Dust modeling}\label{sec:Dust Variability}

In this section, we constrain the spatial distribution of the exozodiacal dust. We first describe the dust emission models used to represent exozodi emission. These models are then combined with an MCMC fitting procedure, which allows us to recover information about the shape of the exozodiacal dust disk around $\theta$\,Boo at different epochs, and thus explore its temporal evolution.

    \subsection{Model description}\label{sec:model description}
    
    Here we describe how we generate exozodiacal dust emission models for comparison with the LBTI observations shown in Fig.\,\ref{fig:Null vs Radius}. Our models are based on the framework presented in \cite{Kennedy_2015} -- which serves as the baseline for the HOSTS survey at the LBTI. 
    
    In this framework, exozodiacal dust disks are assumed to be both optically and vertically thin. The face-on surface density of the dust, $\Sigma$, at a distance $r$ from the star is assumed to follow a power-law distribution:
    \begin{equation}\label{eq:dust distribution}
        \Sigma (r) = \Sigma_0\times r^{\alpha},
    \end{equation}
    between inner and outer radii $R_{\text{in}}$ and $R_{\text{out}}$, respectively, with $r$ expressed in au.
The disk surface density of cross-sectional area is represented by $\Sigma$, and hence is expressed in au$^2$/au$^2$, and is analogous to optical depth. The surface density at 1\,au is denoted by $\Sigma_0$, and $\alpha$ is the surface density slope of the disk.
    The dust surface brightness $S$ at wavelength $\lambda$ is then given by 
    \begin{equation}\label{eq:dust brightness}
        S(r)=2.35\times 10^{-11}\Sigma(r)B_{\nu}(\lambda,T_{\text{d}}(r)),
    \end{equation}
    with $B_{\nu}$ being the dust grain emission at temperature $T_{\text{d}}$. The numerical factor of $2.35\times 10^{-11}$ converts the surface brightness to units of Jy.arcsec$^{-2}$.
    The temperature $T_{\text{d}}$ is that of a blackbody and is defined by
    \begin{equation}\label{eq:dust temp}
        T_{\text{d}} = 278.3L_{\star}^{0.25}r^{-0.5}\,K.
    \end{equation}
    We summarize the parameters describing the dust spatial distribution in Table\,\ref{tab:Model parameters}.
    The stellar parameters used to compute the dust emission can be found in Table\,\ref{tab:Theta Boo parameters}.

    \paragraph{Geometry of the system}
    The face-on dust distributions in the model are limited to centrosymmetric geometries with the following parameters: the inner radius $R_{\text{in}}$, the outer radius $R_{\text{out}}$, the dust density at 1\,au, $\Sigma_0$, and the dust density slope, $\alpha$. We define the mean radius of the belt as $R=(R_{\text{in}}+R_{\text{out}})/2$. Three different geometries are considered when modeling the observations. The first is a ``wide ring'' model with free $R_{\text{in}}$ and $R_{\text{out}}$ and fixed $\alpha=0$. The second is a ``thin ring'' model, which is a sub-model of the wide ring. It has a free $R_{\text{in}}$, a fixed $(R_{\text{out}}-R_{\text{in}})/R=0.1$\footnote{Debris disks are considered narrow for $\Delta R/R \lesssim 0.5$, see \cite{Hughes2018}}, and fixed $\alpha=0$. This model is useful to constrain the region where most of the dust emission is located. 
    The third geometry is a ``disk'' model where $R_{\text{in}}$ is fixed at the sublimation radius of silicates (corresponding to $T_\text{BB}=1500$\,K), $R_{\text{out}}$ fixed at an arbitrarily large distance (typically for $T_\text{BB}=88$\,K\footnote{\cite{Kennedy_2015} demonstrate that this distance is sufficiently large to not affect the model at 11\,µm.}), and $\alpha$ is free. For $\theta$\,Boo, $R_{\text{in}}$ and $R_{\text{out}}$ are calculated to be 0.07 and 20\,au respectively (i.e., 4.7 and 1380\,mas).

  The parameter $\Sigma_0$ is systematically included as a free parameter for each geometry and is constrained in the range [0, $10^{-2}$]. $R_{\text{out}}$ is constrained in the range [0, 20\arcsec], and $R_{\text{in}}$ in the range [0, $R_{\text{out}}$[. The range of $\alpha$ is [-10, 10]. For the ``wide ring'' and ``thin ring'' models, we set $\alpha=0$  to simulate a dust belt with a uniform optical depth radial distribution. This assumption is used to study the collisional evolution of cold debris disks, as in \cite{Wyatt2011} (see the P-R drag model description in Sect.\,\ref{sec:discussion origin}).

    \paragraph{Orientation of the system} The orientation of the disk is defined by two parameters: the disk inclination, $i$, and the disk position angle, PA.
    As mentioned in Sect.\,\ref{sec:data reduction nulling}, the 2023 null depths show no visible dependence on the parallactic angles -- and hence on the transmission map orientation -- which indicates that the disk inclination is expected to be close to face-on ($i=0\degree$), under the assumption of a vertically thin disk.
    Several preliminary test runs established that the MCMC procedure did not converge within reasonable computational times when fitting the geometry ($R_{\text{in}}$, $R_{\text{out}}$, $\Sigma_0$, $\alpha$) and orientation parameters ($i$, PA) simultaneously.
    We therefore assume the disk to be face-on, and perform the MCMC procedure with $i$ systematically set to $0\degree$ for the different datasets. The value for PA is then irrelevant.

    \begin{table}
        \centering
        \caption{Dust model parameters and their reference values for the zodiacal disk.}
        \begin{tabular}{c c c c} \hline \hline
           Symbol & Unit & Parameter & Reference\\ \hline
           $R_{\text{in}}$ & au & Inner disk radius & 0.034 \\ 
           $R_{\text{out}}$ & au & Outer disk radius & 10 \\ 
           $R$ & au & $(R_{\text{in}}+R_{\text{out}})/2$ & 5 \\
           $\Sigma_0$ & au$^{2}$/au$^{2}$ & Surface density at 1\,au & $7.12\times 10^{-8}$ \\ 
           $z$ & zodi & Surface density at EEID & 1 \\
           $\alpha$ & - & Surface density slope & -0.34 \\
           $i$ & \degree & Disk inclination & -\\ 
           PA & \degree & Disk position angle & - \\ \hline
        \end{tabular}\\
        \raggedright
        \textbf{Notes.} Reference values taken from \cite{Kennedy_2015}.
        \label{tab:Model parameters}
    \end{table}

    \subsection{Modeling strategy}
\begin{figure*}
        \centering
        \includegraphics[width=0.74\linewidth]{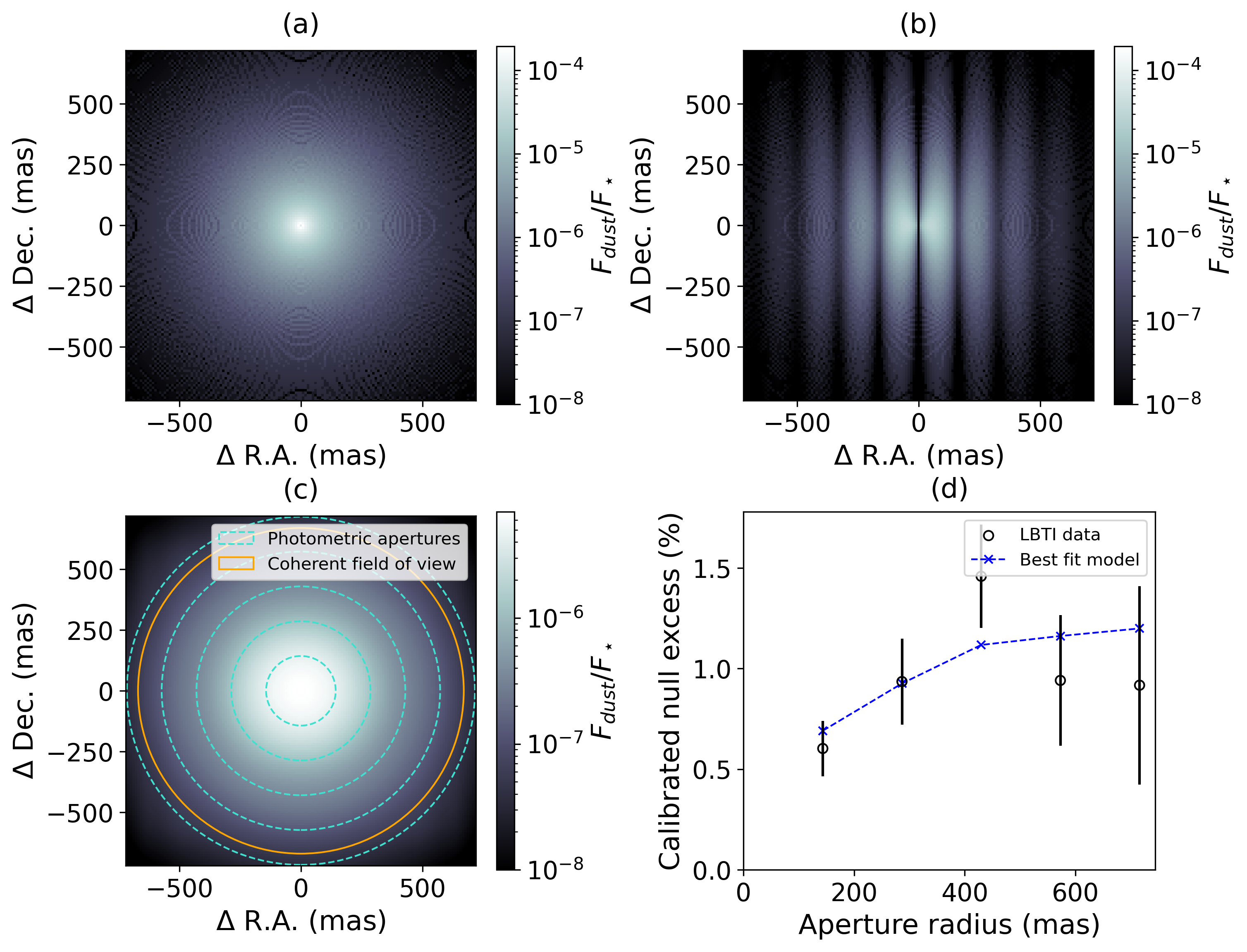}
        \caption{Results of the 2023 May 25 night modeling, showing the different steps of the procedure. The geometry used is a disk with $\Sigma_0=3\times 10^{-5}$ and $\alpha=0$, the orientation is face-on. From (a) to (d) show: the angular distribution of $F_\text{dust}/F_\star$, the impact of the LBTI transmission map, the convolution with the PSF of the LBT, and the comparison between the observed excess and the modeled excess.
      Panel (c) indicates the photometric apertures used in the reduction with the dashed circles. The coherent FOV is indicated with the solid orange circle.
        The maps have a pixel size of 10\,mas.}
        \label{fig:diskmodel2018outer}
    \end{figure*}

    Our goal is to model the dust emission as measured with NOMIC. The next step is therefore to compute the dust raw radiance, $F_\text{dust}$, by integrating Eq.\,(\ref{eq:dust brightness}) over the passband of NOMIC (N'-band: 9.81-12.41\,µm), taking into account the transmission of the NOMIC filter. The value of $F_\text{dust}$ is then normalized by the stellar radiance, $F_\star$, calculated from its blackbody emission. Finally, the model accounts for the impact of the PSF and the transmission map on the detected flux: the raw radiance map is first multiplied by the LBTI transmission pattern (Sect.\,\ref{sec:nulling observations}) and then convolved with the modeled PSF of the LBT.
    The measured excess can then be estimated by integrating the pixels over the different apertures defined in Sect.\,\ref{sec:data reduction nulling}, and compared to the results from Fig.\,\ref{fig:Null vs Radius}.
    This comparison uses an MCMC approach: we generate 30 walkers with flat priors across all parameters, each allowed to perform 500 sampling steps (after a burn-in phase of 1000 models). 
    For each model, we calculate its likelihood function with the LBTI observation and its corresponding $\chi^2$. The MCMC then explores the parameter space to minimize the $\chi^2$ value of the models. The model with the minimum $\chi^2$ is finally returned as the best-fit model.
    Figure\,\ref{fig:diskmodel2018outer} shows this modeling procedure as applied to the 2023 May 25 observation. The different steps are detailed, from the computation of raw radiance to the retrieval of the excess signature on the detector. The various photometric apertures used to constrain the dust distribution are also shown, with the coherent FOV calculated from Eq.\,(\ref{eq:coherent FOV}).

    \renewcommand{\arraystretch}{1.75}
    \begin{table*}
        \centering
        \caption{Results of modeling for the combined and individual  2017, 2018, and 2023 datasets using the disk, wide ring, and thin ring geometries.}
        \begin{tabular}{c c c c c c c c c c} \hline \hline
            Observing night(s) & Geometry & $\Sigma_0/10^{-5}$\,[au$^2$/au$^2$] & $R_\text{in}$\,[au] & $R_\text{out}$\,[au] & $\alpha$ & $z$ & $\chi^2$ & $\chi^2_\text{red}$ \\ \hline
            All nights & Disk & $\boldsymbol{0.86^{+0.32}_{-0.28}}$ & 0.07 & 20 & $\boldsymbol{1.38^{+0.38}_{-0.43}}$ & $305^{+34}_{-44}$ & $34.4^{+2.78}_{-1.11}$ & $2.65^{+0.21}_{-0.09}$\\ 
                & Wide ring & $\boldsymbol{7.30^{+2.75}_{-2.12}}$ & $\boldsymbol{2.05^{+0.42}_{-0.35}}$ & \textbf{NC} & 0 & $1025^{+386}_{-298}$ & $36.0^{+2.62}_{-1.22}$ & $3.00^{+0.22}_{-0.10}$ \\
                    & Thin ring & $\boldsymbol{54.5^{+6.54}_{-5.58}}$ & $\boldsymbol{2.95^{+0.53}_{-0.36}}$ & $3.25^{+0.58}_{-0.40}$ & 0 & 0 & $38.9^{+3.40}_{-1.12}$ & $2.99^{+0.26}_{-0.09}$ \\ \hline
            2017-04-11 & Disk & $\boldsymbol{0.38^{+0.59}_{-0.26}}$ & 0.07 & 20 & $\boldsymbol{1.73^{+0.92}_{-1.16}}$ & $158^{+62}_{-67}$ & $4.31^{+3.93}_{-2.16}$ & $1.44^{+1.31}_{-0.72}$ \\ 
                & Wide ring & $\boldsymbol{63^{+180}_{-37}}$ & $\boldsymbol{6.17^{+2.02}_{-1.57}}$ & \textbf{NC} & 0 & 0 & $4.97^{+4.35}_{-1.80}$ & $2.49^{+2.20}_{-0.90}$ \\
                    & Thin ring & $\boldsymbol{186^{+175}_{-87}}$ & $\boldsymbol{6.69^{+1.22}_{-1.35}}$ & $7.36^{+1.34}_{-1.48}$ & 0 & 0 & $4.77^{+3.22}_{-1.34}$ & $1.59^{+1.10}_{-0.45}$ \\ \hline
            2018-05-23 & Disk & $\boldsymbol{0.78^{+0.46}_{-0.32}}$ & 0.07 & 20 & $\boldsymbol{1.6^{+0.47}_{-0.54}}$ & $322^{+53}_{-64}$ & $4.08^{+2.70}_{-1.20}$ & $1.36^{+0.90}_{-0.40}$ \\
                & Wide ring & $\boldsymbol{14^{+8.1}_{-4.8}}$ & $\boldsymbol{2.81^{+0.74}_{-0.62}}$ & \textbf{NC} & 0 & 0 & $4.81^{+2.69}_{-1.29}$ & $2.40^{+1.35}_{-0.65}$ \\ 
                    & Thin ring & $\boldsymbol{68.3^{+51}_{-12}}$ & $\boldsymbol{3.63^{+0.85}_{-0.69}}$ & $4.00^{+0.94}_{-0.76}$ & 0 & 0 & $6.34^{+3.39}_{-1.34}$ & $2.11^{+1.13}_{-0.45}$ \\ \hline 
            2023-05-25 & Disk & $\boldsymbol{2.98^{+0.60}_{-0.83}}$ & 0.07 & 20 & $\boldsymbol{0.05^{+0.69}_{-0.86}}$ & $443^{+98}_{-170}$ & $4.49^{+2.16}_{-1.23}$ & $1.50^{+0.72}_{-0.41}$ \\
                & Wide ring & $\boldsymbol{5.52^{+3.67}_{-1.97}}$ & $\boldsymbol{1.29^{+0.49}_{-0.75}}$ & \textbf{NC} & 0 & $775^{+515}_{-277}$ & $3.76^{+2.37}_{-0.89}$ & $1.88^{+1.20}_{-0.45}$\\ 
                    & Thin ring & $\boldsymbol{77.9^{+18.7}_{-29.2}}$ & $\boldsymbol{2.42^{+0.54}_{-0.46}}$ & $2.66^{+0.60}_{-0.51}$ & 0 & 0 & $3.44^{+3.48}_{-1.17}$ & $1.15^{+1.16}_{-0.39}$ \\ 
            \hline
        \end{tabular}\\
        \raggedright 
        \textbf{Notes.} NC stands for not constrained. The parameters fitted by the MCMC are written in bold text with the median values and standard deviations taken from their posteriors, given in Appendix\,\ref{ap:corner plots}.
        Other parameters are either fixed or derived from the fitted ones.
        The zodi level, $z$, is estimated by calculating the ratio between $\Sigma$(r=EEID) and the zodiacal dust density at 1\,au (i.e., $7.12\times 10^{-8}$). Values of $z$=0 indicate that $R_\text{in}$>2\,au, so no dust is expected at the EEID from the corresponding model.
        \label{tab:Final modeling parameters}
    \end{table*}
    \renewcommand{\arraystretch}{1.0}
    
    \subsection{Results}\label{sec:modeling results}

\begin{figure}
        \centering
        \includegraphics[width=\linewidth]{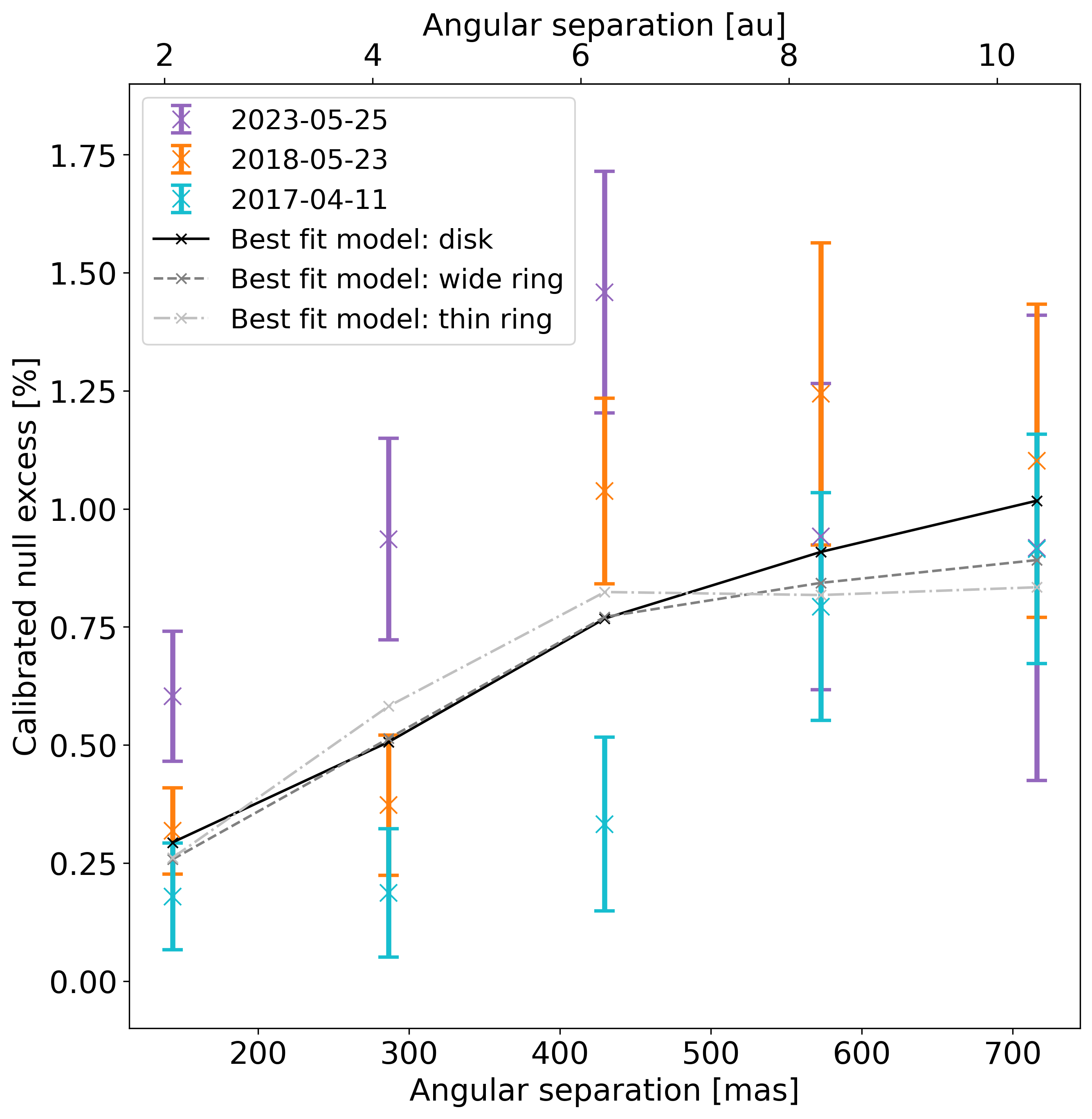}
        \caption{Best-fit models for the combined 2017, 2018, and 2023 datasets. Solid dark lines represent the best-fit models using a disk geometry, as defined in Sect.\,\ref{sec:model description}. Dashed gray lines represent the best-fit models using a wide ring geometry and dash-dotted silver lines represent the best-fit models using a thin ring geometry. Parameters of the models are listed in Table\,\ref{tab:Final modeling parameters}.}
        \label{fig:bestfitmodel all}
    \end{figure}

    First, assuming that the system is in a steady state, we use the three geometries considered to model the dust spatial distribution (thin ring, wide ring, and disk) to fit the combined 2017, 2018, and 2023 datasets. Figure\,\ref{fig:bestfitmodel all} shows the best-fit models, and Table\,\ref{tab:Final modeling parameters} summarizes the median values of the model parameters with their standard deviation, obtained from their MCMC posterior distributions. All MCMC posteriors are presented in Appendix\,\ref{ap:corner plots}.
    The median values of $\chi^2$ and the reduced $\chi^2$ ($\chi^2_\text{red}=\chi^2/\nu$, where $\nu$ is the degree of freedom) are also indicated.

\begin{figure*}
        \centering
        \includegraphics[width=0.33\linewidth]{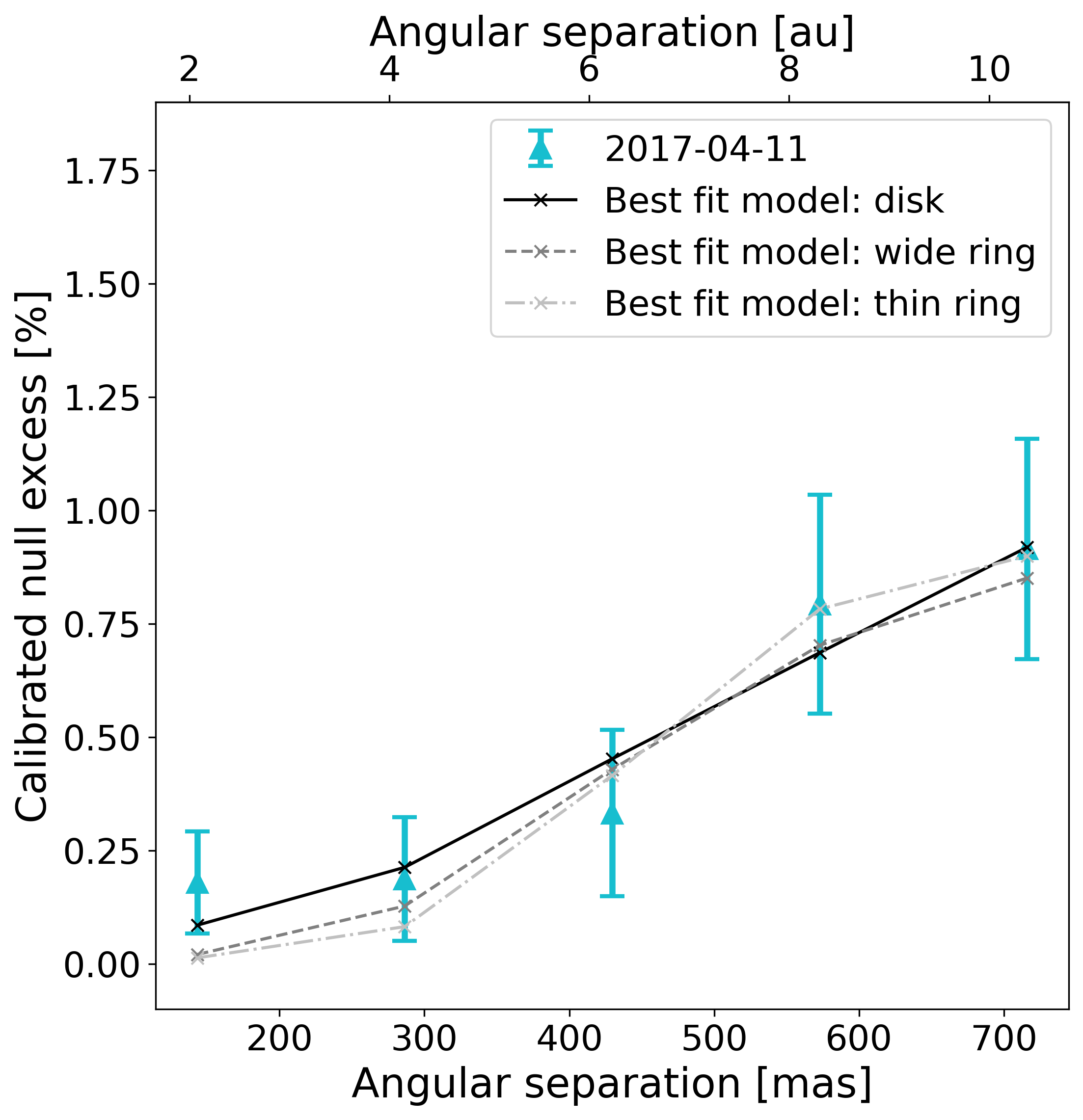}
        \includegraphics[width=0.33\linewidth]{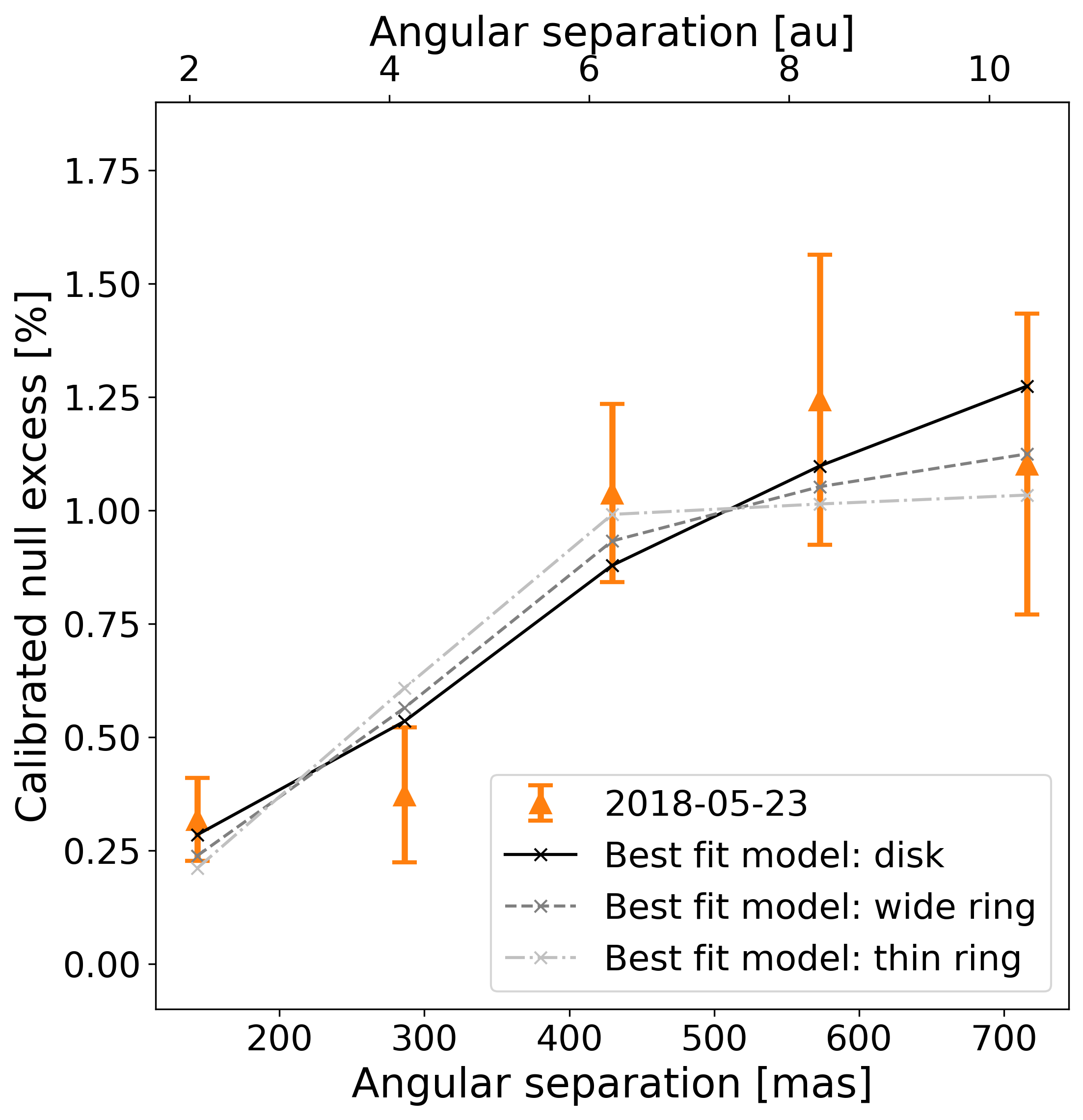}
        \includegraphics[width=0.33\linewidth]{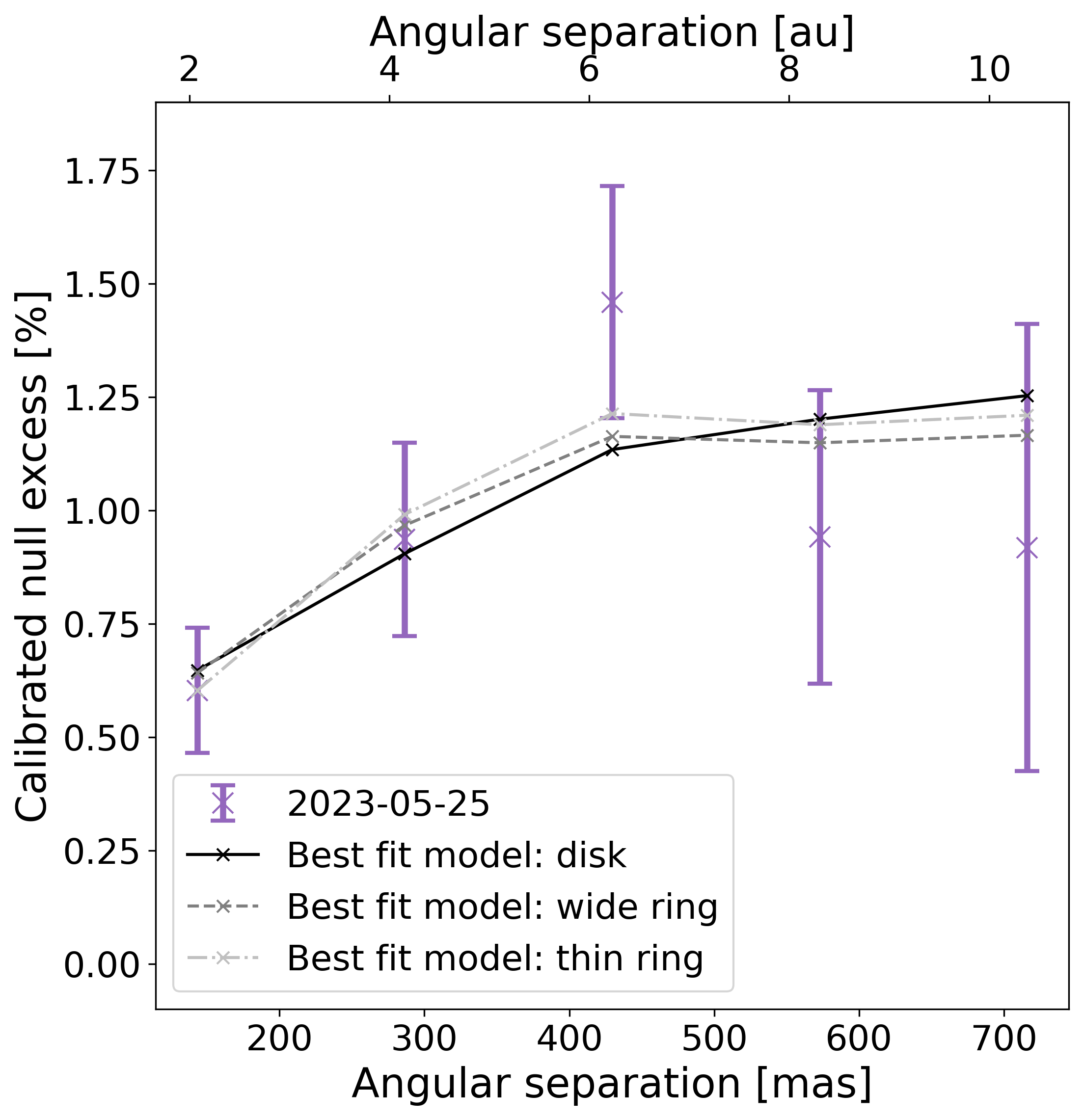}
        \caption{Best fit models for the 2017 (left), 2018 (middle), and 2023 (right) datasets. Solid dark lines represent the best-fit models using a disk geometry, as defined in Sect.\,\ref{sec:model description}. Dashed gray lines represent the best-fit models using a wide ring, and dash-dotted silver lines represent the best-fit models using a thin ring geometry. The parameters for the models are listed in Table\,\ref{tab:Final modeling parameters}.}
        \label{fig:bestfitmodel}
    \end{figure*}

    The different geometries are then used to fit the individual 2017, 2018, and 2023 datasets. Figure\,\ref{fig:bestfitmodel} shows the best-fit models.
    The best-fit values of $\chi^2_\text{red}$ are significantly worse when fitting the combined nights compared to the individual nights. The steady-state hypothesis for the system is therefore not favored, but cannot be fully rejected.
     It is worth noting that, although the thin ring is a submodel included within the wide ring model, their results are not systematically compatible. Adding $R_\text{out}$ as an additional free parameter allows exploration of more geometries and can yield models with better $\chi^2$ -- although not significantly better.
    For the individual fits, the $\chi^2$ values obtained for each geometry are consistent across the three observing nights. Therefore, our data do not distinguish between the three possible dust distributions, assuming a face-on, optically thin exozodi.
    
\begin{figure*}
        \centering
        \includegraphics[width=0.96\linewidth]{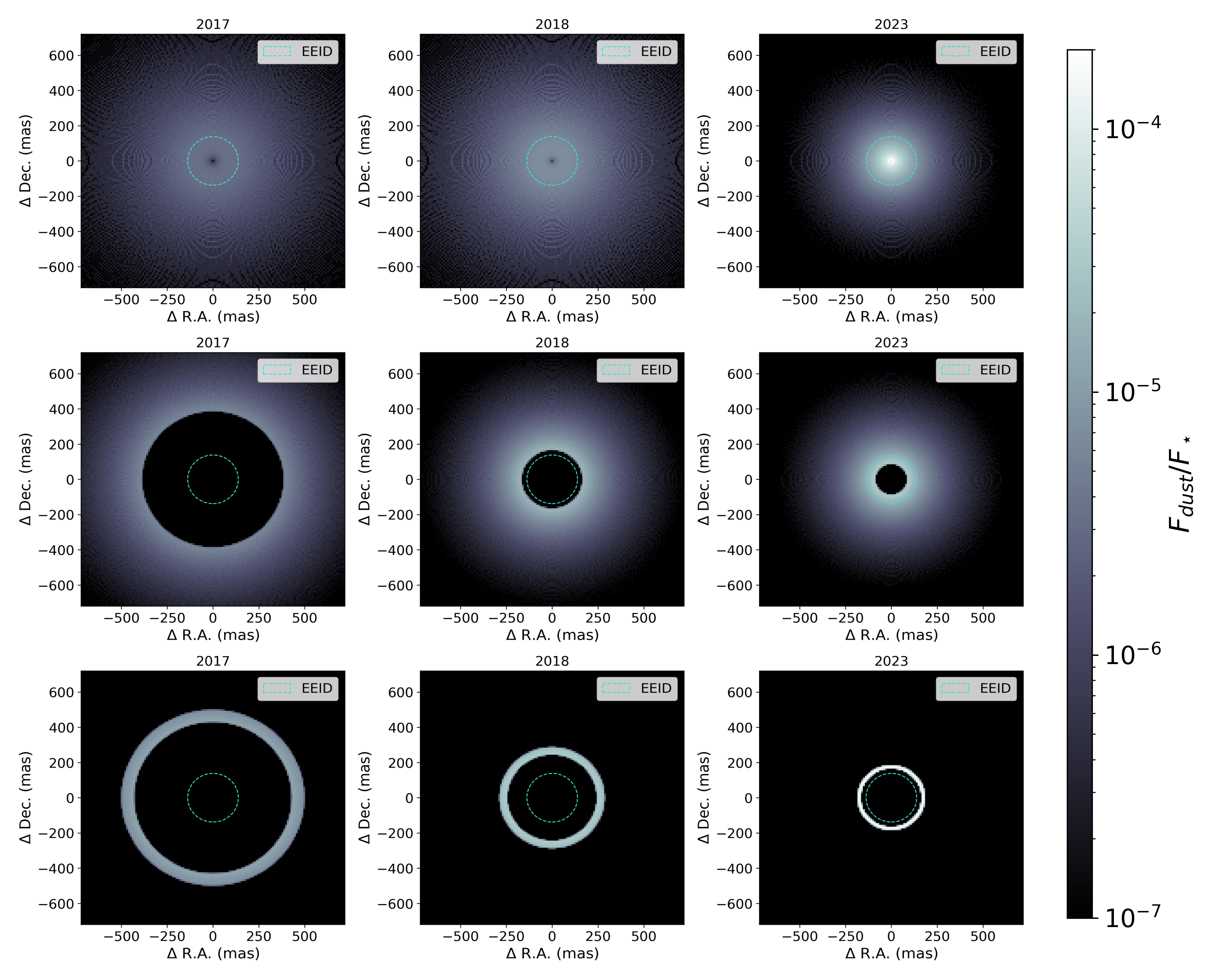}
        \caption{2D maps of the dust surface brightness ($F_\text{dust}$) for the models of 2017 (left), 2018 (middle), and 2023 (right). Maps correspond to the disk (top), wide ring (middle), and thin ring (bottom) geometries. $F_\text{dust}$ is normalized by the stellar radiance ($F_\star$). Both $F_\text{dust}$ and $F_\star$ are integrated over the NOMIC passband. The dashed circle indicates the EEID ($\sim$138\,mas). The maps have a pixel size of 10\,mas.}
        \label{fig:2D map dust brightness}
    \end{figure*}

    Figure\,\ref{fig:2D map dust brightness} presents the 2D maps of dust radiance for the individual models of 2017, 2018, and 2023, using the median values of the model parameters from Table\,\ref{tab:Final modeling parameters}.

    \subsection{Analysis of the dust models}\label{sec:analysis nulling models}

    \paragraph{Dust temperature}
    Equation\,(\ref{eq:dust brightness}) shows a degeneracy between the dust temperature and the dust density $\Sigma(r)$. For the dust brightness detected by NOMIC, this degeneracy means that the dust density cannot be precisely estimated without constraining the dust temperature -- and, by extension, the dust composition and grain size distribution. The values of $\Sigma_0$ in Table\,\ref{tab:Final modeling parameters} therefore depend on the assumption that the dust grains behave as ideal blackbodies and should be interpreted with caution. Since the nulling observations performed with the LBTI are photometric, they cannot constrain the dust temperature using spectral information. Nonetheless, using the same temperatures $T_\text{d}(r)$ between the 2017, 2018, and 2023 datasets allows for a comparison of their $\Sigma$ values and an estimate of dust variability. Estimates of the dust size distribution, and hence its equilibrium temperature profile, are discussed in Sect.\,\ref{sec:discussion}.

    \paragraph{Dust variability at the EEID} As a probe for the amount of exozodiacal dust predicted by the models in the HZ of $\theta$\,Boo, we consider the zodi level, $z$. This zodi level is defined as the dust surface density, $\Sigma$, at the EEID ($\sim$2\,au or 138\,mas for $\theta$\,Boo) divided by the dust surface density at 1\,au in the Solar System \citep[][]{Weinberger2015}. For a disk model, $\Sigma$ can be calculated self-consistently for each combination of parameters $\Sigma_0$ and $\alpha$ using Eq.\,(\ref{eq:dust distribution}). We use the MCMC chain samples to build posterior distributions of $z$ for the 2017, 2018, and 2023 disk models -- $z_\text{2017}$, $z_\text{2018}$, and $z_\text{2023}$, respectively. The resulting median values for the 2017, 2018, and 2023 datasets, together with their $1\sigma$ confidence intervals, are shown in Table\,\ref{tab:Final modeling parameters}.
    It should be noted that the zodi level obtained for the 2017 dataset is consistent with that calculated by the HOSTS survey (148.2$\pm$27.7\,zodis). 

\begin{figure}
        \centering
        \includegraphics[width=\linewidth]{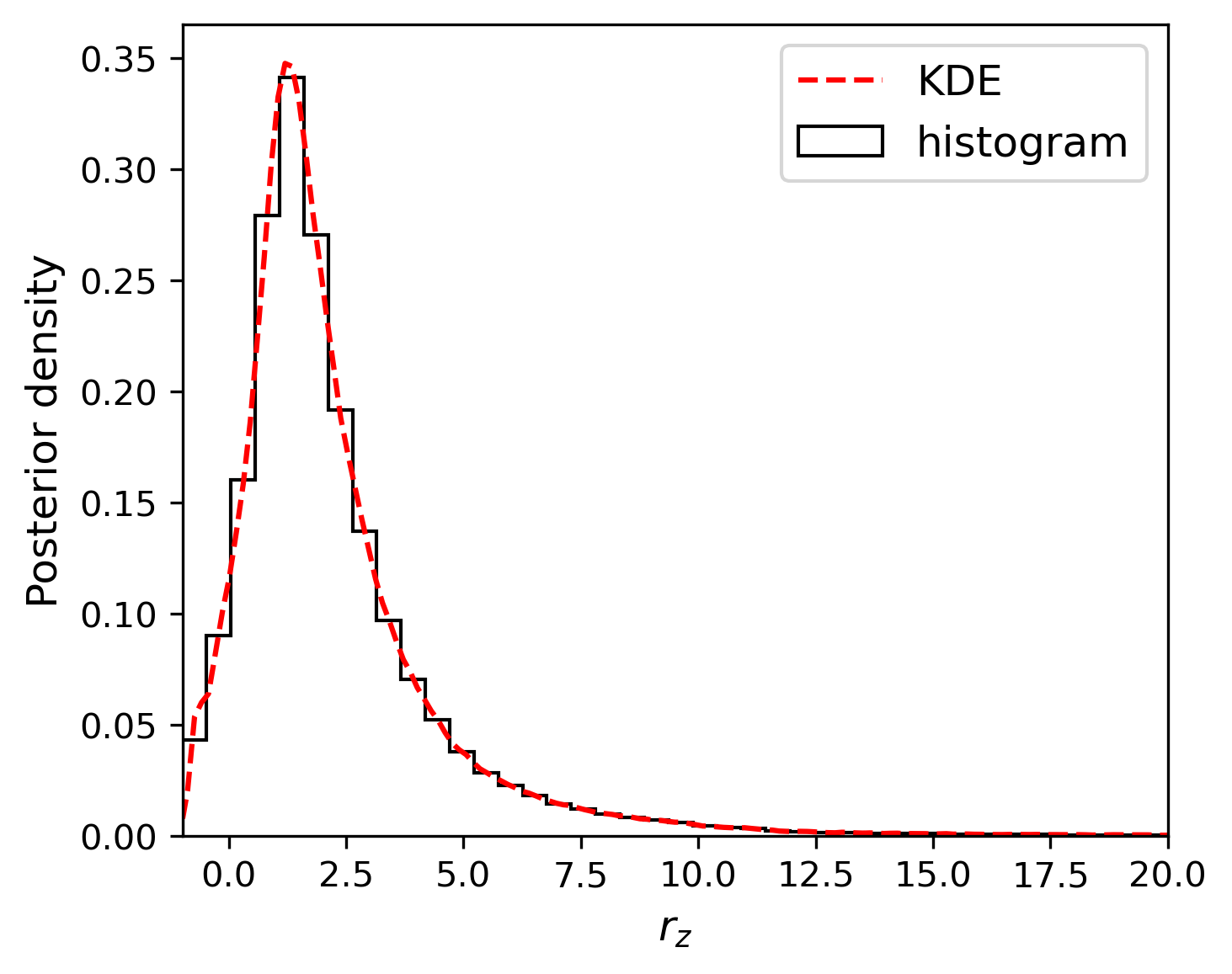}
        \caption{MCMC posterior distribution sample histogram of $r_z$, representing the relative difference in zodi level between the 2017 and 2023 models. A Gaussian kernel density estimate (KDE) is used to fit the sample and calculate the posterior probabilities of the dust variability.}
        \label{fig:zodi_factor_increase}
    \end{figure}

    To further quantify the increase in HZ dust brightness found in our models between 2017 and 2023, we calculate the posterior distribution of the ratio $r_z=(z_\text{2023} - z_\text{2017})/z_\text{2017}$. From the two MCMC runs for 2017 and 2023, we sample 800 models from each run and randomly pair MCMC models from the two epochs 640,000 times to construct the posterior histogram of the ratio. 
    The posterior can then be integrated to determine posterior probabilities for the zodi level increase between epochs. For example, the integral over $r_z > 2$ denotes the probability that the zodi level at least doubled between 2023 and 2017. To perform this integration efficiently and analytically, we approximate the posterior using a Gaussian kernel density estimate \citep[KDE:][]{Rosenblatt1956,Parzen1962}, which is shown together with the posterior histogram in Fig.\,\ref{fig:zodi_factor_increase}. The resulting posterior probability that there is an increase in zodi level is $93.7\,\%$ (\textasciitilde $1.86\sigma$).

    The posterior probability thus suggests a tentative increase in the HZ dust brightness between 2017 and 2023.
    We note that the probability calculations still make implicit assumptions about the correctness and quality of both the data acquisition (including calibration and error model) and the dust modeling, and depend slightly on the details of the adopted kernel density estimate. These results should therefore be interpreted with caution.
    The zodi levels calculated for 2017, 2018, and 2023 are also useful for providing an order-of-magnitude estimate of the absolute dust brightness in the HZ of $\theta$\,Boo. However, these values depend on the assumed dust temperature, which is not well constrained. Estimates of the zodi level for realistic dust grains with different sizes are given in Sect.\,\ref{sec:discussion}. 

\section{Discussion}\label{sec:discussion}

\subsection{Constraints from the models}\label{sec:discussion SED}

\begin{figure}
    \centering
    \includegraphics[width=\linewidth]{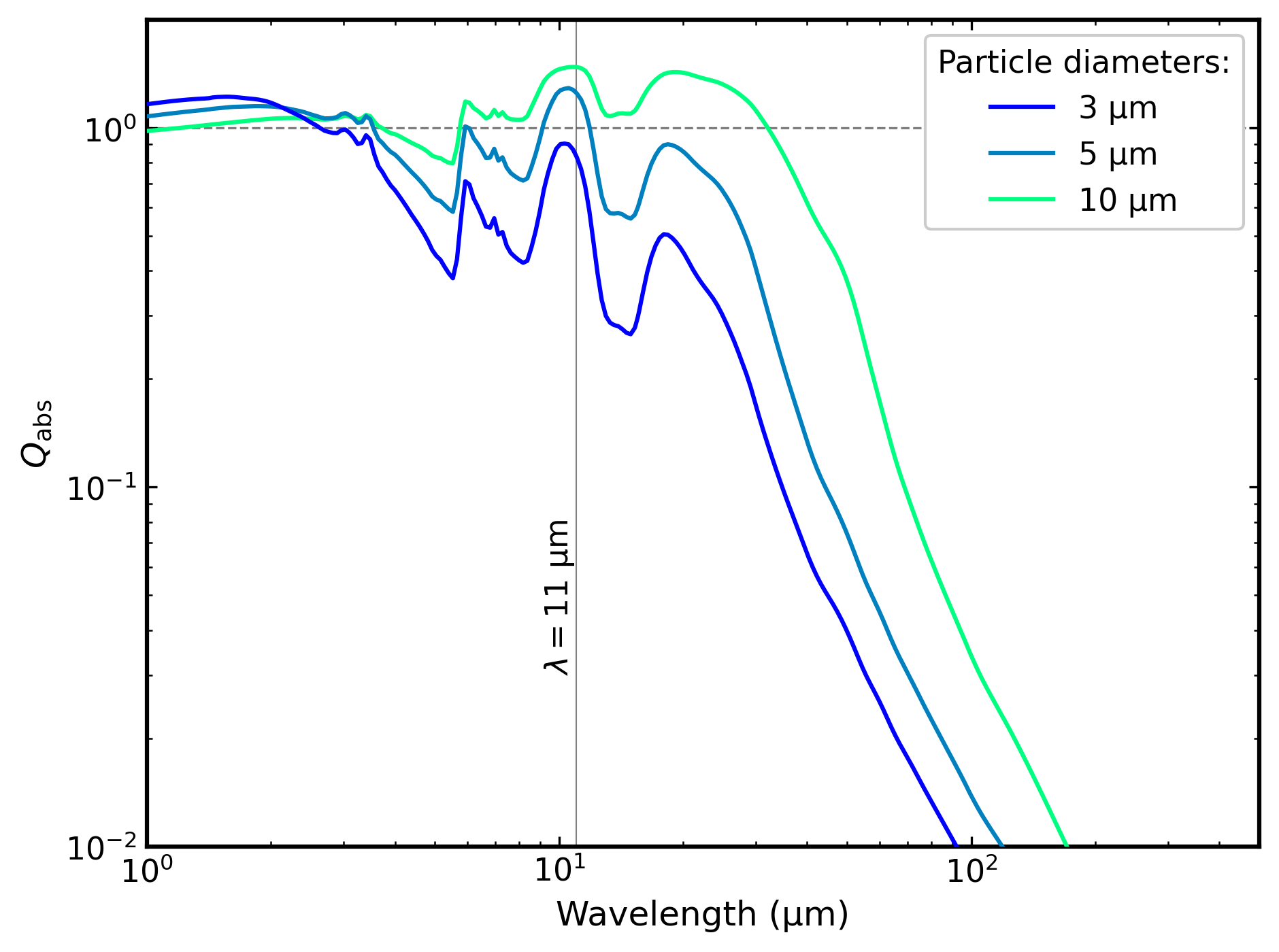}
    \caption{Simulated emissivities for dust grains of size 3, 5, and 10\,µm. The volume fraction of the grains is 1/3 amorphous silicates and 2/3 refractory organics.}
    \label{fig:Qabs}
\end{figure}

\paragraph{SED of the dust models}
The LBTI observations around 11\,µm cannot discriminate between the different dust distribution models obtained in Sect.\,\ref{sec:Dust Variability} or the properties of the dust grains. 
One way to further constrain their properties is to simulate the SED from each model and compare it with the upper limits from Fig.\,\ref{fig:SED_theta_boo}. 
Since the emitting dust may be considerably smaller than the wavelengths, we must account for the reduced emissivity $Q_\text{abs}$ of realistic grains compared to ideal blackbodies.
We assume compact spherical grains composed of amorphous silicates and refractory organics with volume fractions of 1/3 and 2/3, respectively \citep{Li1998}. We then compute $Q_\text{abs}$ and the grains' resulting equilibrium temperatures using the Mie-type optical properties code of \cite{Sommer2025}, which utilizes the optical constants for the constituent materials from \cite{Li1997}. Figure\,\ref{fig:Qabs} shows the resulting $Q_\text{abs}$. We consider grains with diameters of 3, 5, and 10\,µm. This choice is motivated by the assumption that the geometric optical depth is dominated by the smallest grains bound to the system, with an estimated blowout size of approximately 2.5\,µm (calculated using the same grain model).

We retrieve the radial profile of the dust surface brightness $S$ from the models shown  in Fig.\,\ref{fig:2D map dust brightness}.
For the three grain sizes considered, the radial profile of the dust's optical depth, $\Sigma$, is recomputed to match its surface brightness with $S$, using the relation
\begin{equation}\label{eq:optical depth recompute}
    \Sigma(r) = \frac{S(r)}{Q_\text{abs}(\lambda)\times B_\nu(\lambda,T_\text{d}(r))}.
\end{equation}
Since the surface brightness $S$ is constant between the models with ideal blackbodies (Sect.\,\ref{sec:Dust Variability}) and the models with realistic grains, all models should fit the data in a similar way.
Finally, we choose to truncate the optical depth profile of the dust at 10\,au (i.e., $\sim$700\,mas) for the SED calculation, since the dust located outside this distance is beyond our coherent FOV. This putative outer dust also has an infrared emission that does not comply with the upper limits, particularly for models with a slope $\alpha > 1$. We argue that if such quantities of outer dust were present at these distances in a static way, they would have been detected by previous far-infrared observations.

\begin{figure*}
    \centering
    \includegraphics[width=0.33\linewidth]{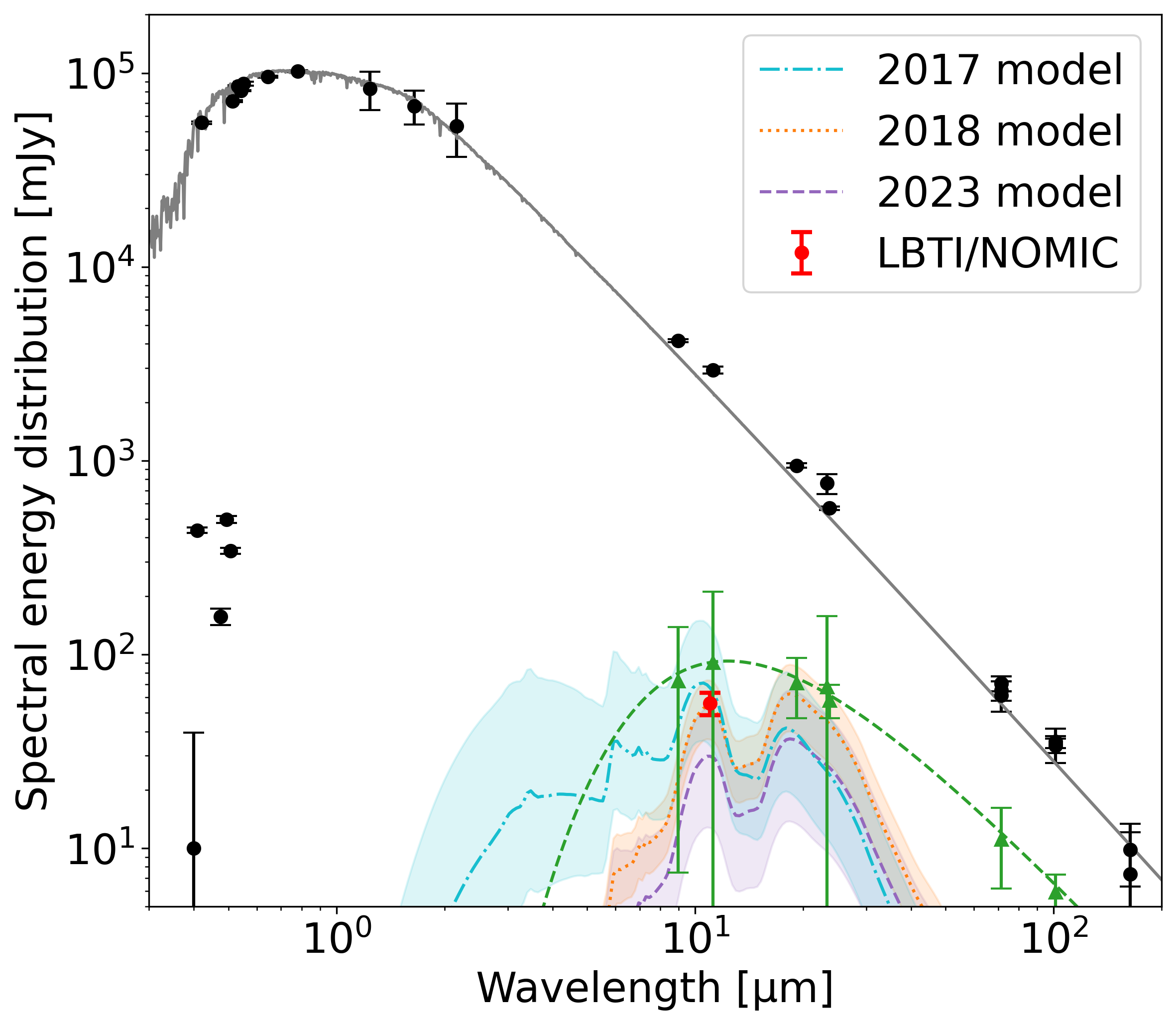}
    \includegraphics[width=0.33\linewidth]{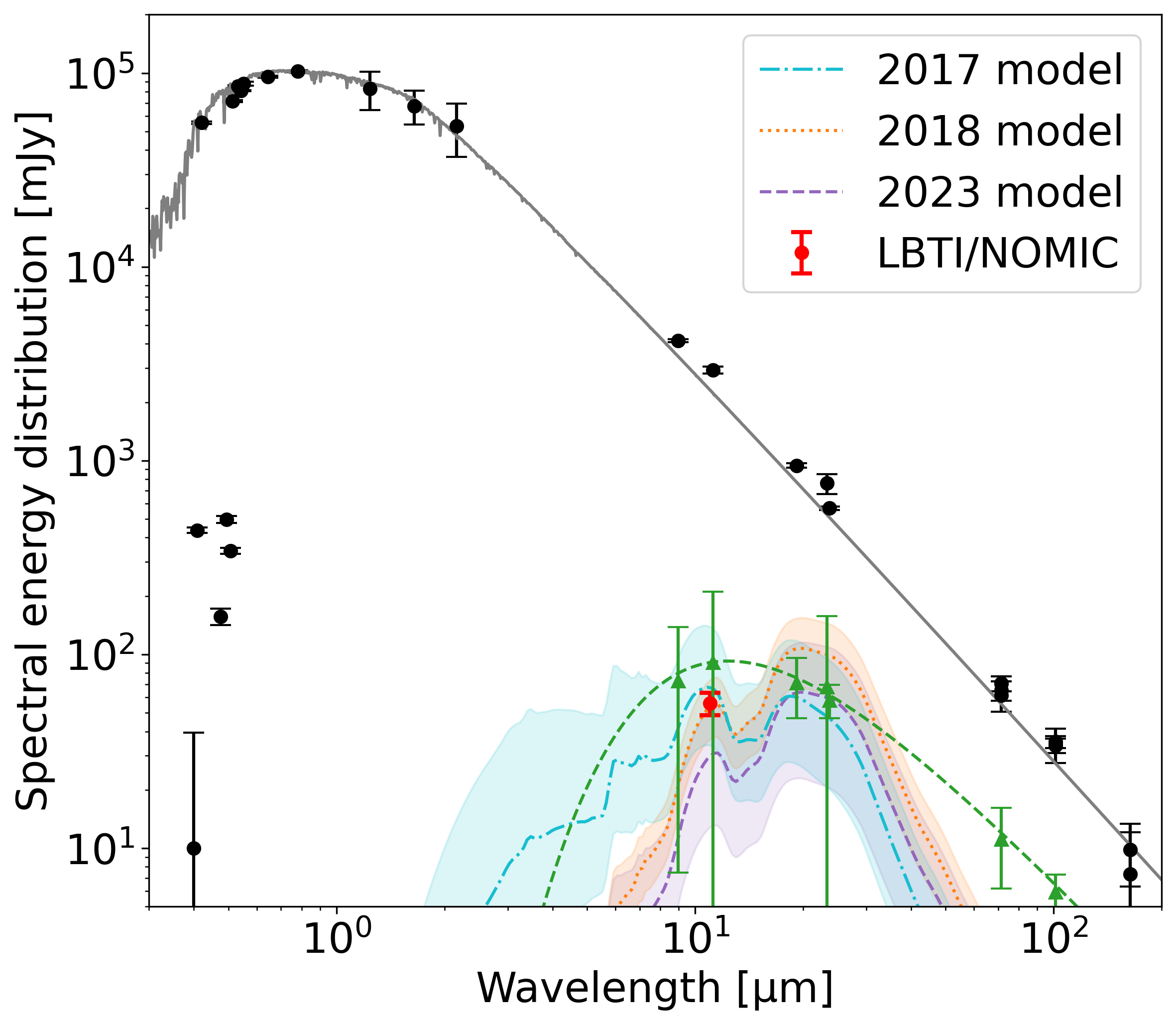}
    \includegraphics[width=0.33\linewidth]{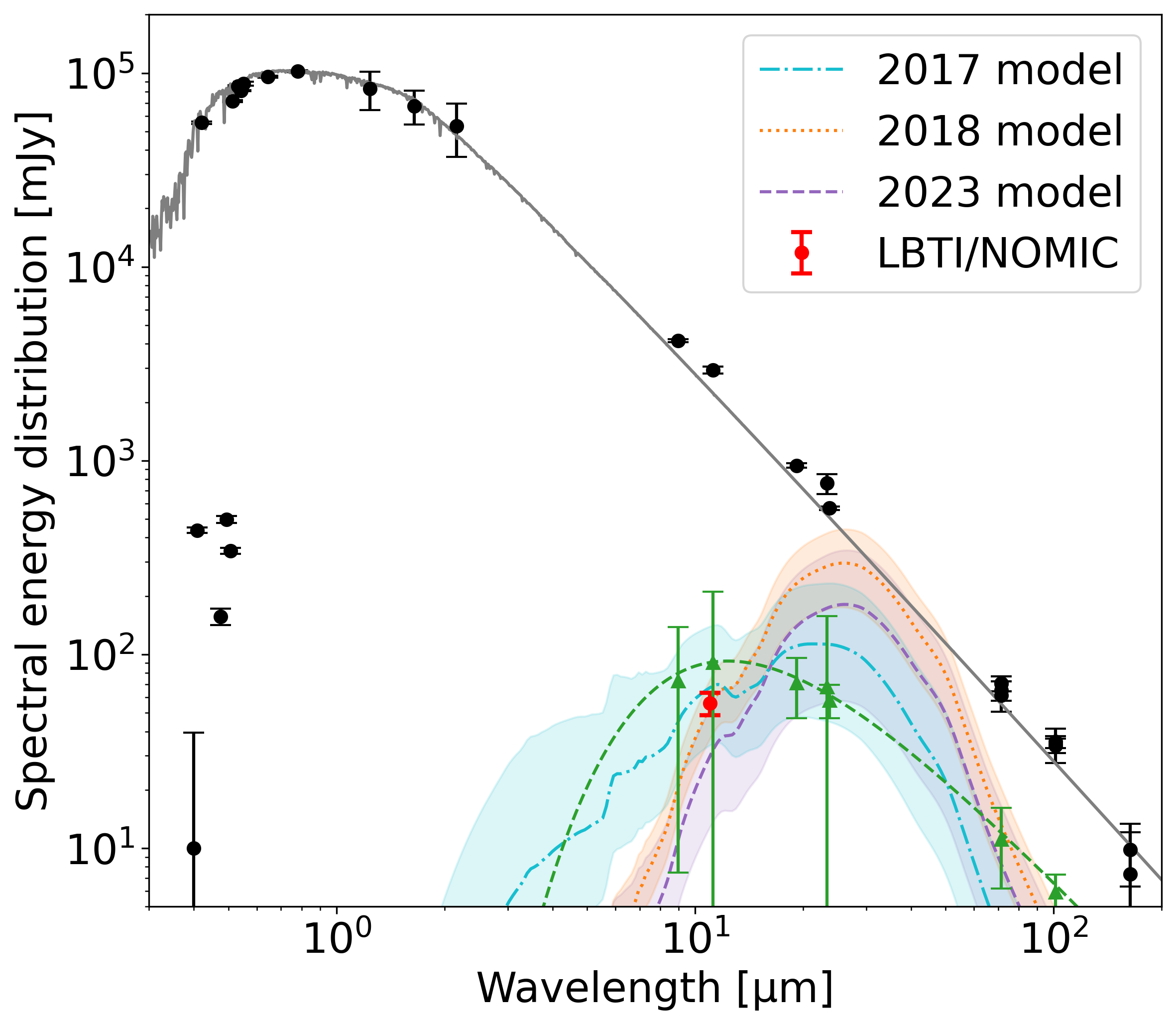}

    \includegraphics[width=0.33\linewidth]{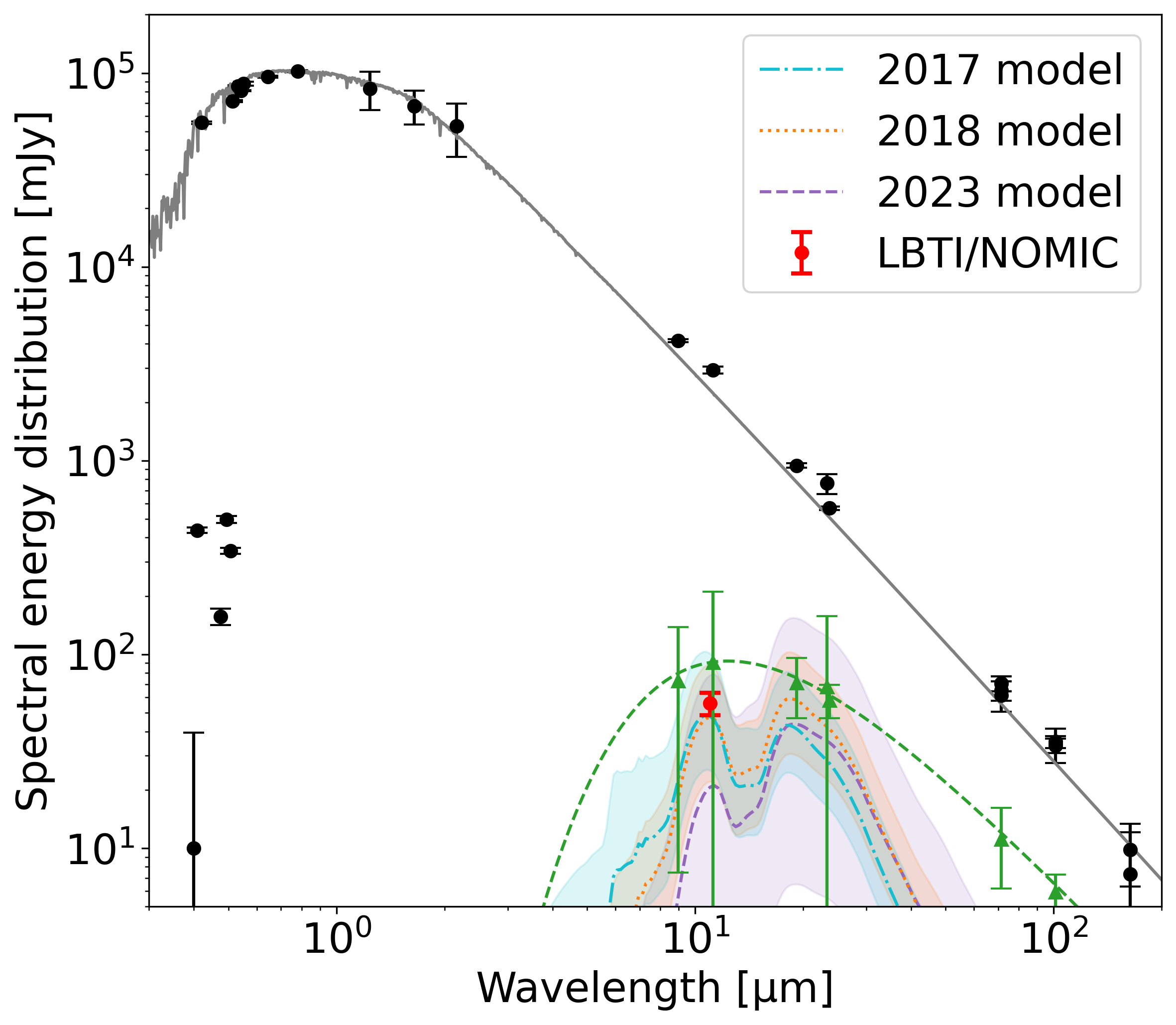}
    \includegraphics[width=0.33\linewidth]{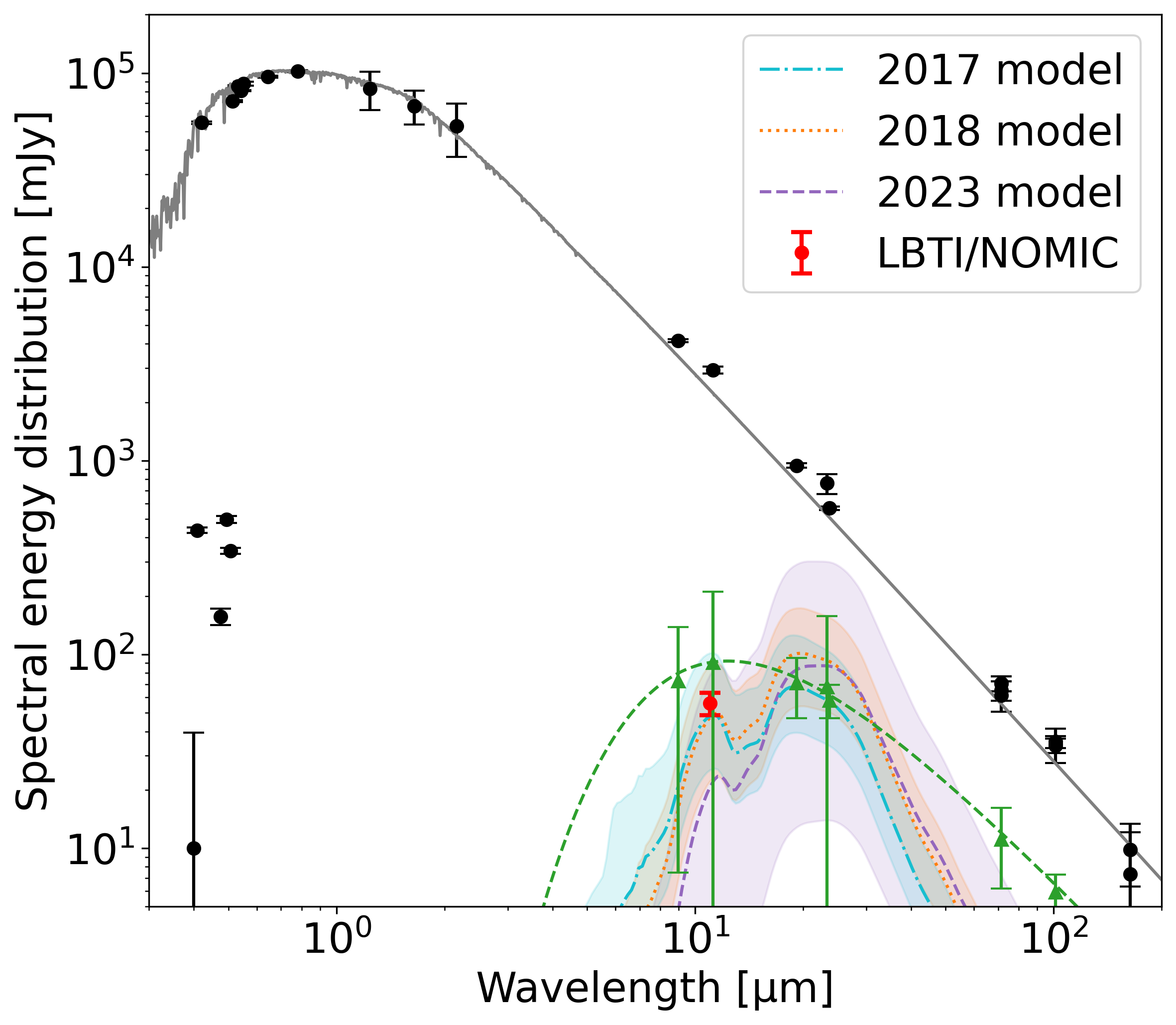}
    \includegraphics[width=0.33\linewidth]{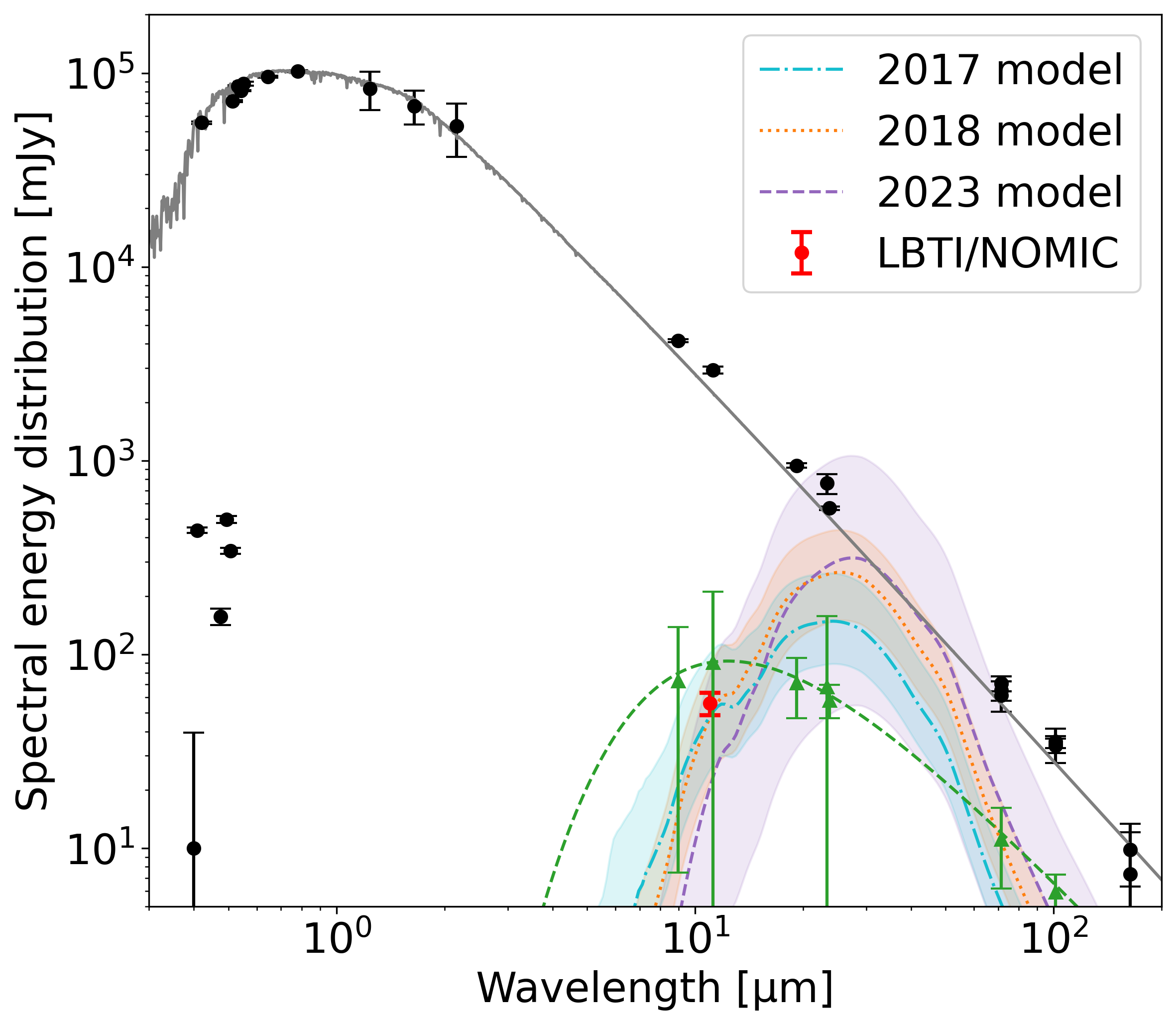}

    \caption{Simulated dust SEDs for realistic dust grains with sizes of 3\,µm (left), 5\,µm (middle), and 10\,µm (right). The SEDs correspond to the disk (top) and wide ring (bottom) models for 2017, 2018, and 2023.}
    \label{fig:SED}
\end{figure*}

Figure\,\ref{fig:SED} shows the SEDs corresponding to the recomputed optical depth profiles. Only the SEDs for the disk and wide ring models are presented for comparison, as the SEDs for the thin ring models yield  similar results.
The estimate of the mean infrared excess at 11\,µm from the current LBTI/NOMIC observations, calculated in Sect.\,\ref{sec:data reduction nulling} is also included. This data point serves as a consistency check, confirming that all models are compatible with our observations at 11\,µm. No increase is observed at 11\,µm between the 2017 and 2023 models. This outcome arises because the apparent increase in dust brightness is localized to the inner region of the system, while the dust emission integrated over the entire FOV remains constant for the three observations. This is evident in Fig.\,\ref{fig:Null vs Radius}, where the data points corresponding to the largest apertures are consistent.

The SED results indicate that for a grain size of 3\,µm, all models are consistent with the upper limits. For a grain size of 5\,µm, some models show a discrepancy with the upper limits near a wavelength of 20\,µm. For a grain size of 10\,µm, all models exhibit a significant infrared excess between 20\,µm and 50\,µm. Therefore, the simulated SEDs do not clearly discriminate between the different dust distribution models obtained, but favor a dust grain size distribution between 3 and 5\,µm in order to satisfy the various upper limits. Table\,\ref{tab:grain size to zodi} provides the updated estimates of zodi level from the recomputed optical depth profiles. The zodi levels change by a factor of 2.8 for dust grains between 3 and 10\,µm.

Further constraints on the dust SED could be obtained using observations from the \textit{James Webb} Space Telescope \citep[JWST:][]{Rigby2023,Gardner2023}, particularly with the Medium Resolution Spectrometer \citep[MRS:][]{Wells2015,Argyriou2023} of the Mid-InfraRed Instrument \citep[MIRI:][]{Wright2023}. The measured far-infrared emission could then be compared with our modeled SEDs to identify not only the dust distribution but also the dust properties, such as size, temperature, and composition.

\paragraph{Orientation of the system}
The results presented in Sect.\,\ref{sec:data reduction nulling} support the hypothesis that the planetary system of $\theta$\,Boo has a low inclination, possibly even a face-on orientation. This hypothesis is central for the modeling strategy described in Sect.\,\ref{sec:Dust Variability}; however,  no other independent observations of the planetary system are currently available to further constrain this orientation.
It is possible to constrain the stellar inclination from spectroscopic measurements of $\theta$\,Boo. The inclination of debris disks is often found to match the stellar inclination \citep[][]{Watson2011,Guilloteau2011}. To determine the stellar inclination, we follow the method used in \cite{Faramaz2014} for $\zeta^2$\,Reticuli. This method uses the color index ($B-V$), the stellar radius, $R_\star$, and the activity indicator, $R'_\text{HK}$, to estimate the rotational period at the equator of the star using the activity-rotation diagram from Fig.\,6b in \cite{Noyes1984}. This period is then compared with measurements of $\nu \sin(i)$, from which a value of $i$ can be deduced. The stellar properties of $\theta$\,Boo are summarized in Table \ref{tab:Theta Boo parameters}. We estimate the rotational period to be $P=3.22\pm0.83$\,days and the equatorial rotational velocity to be $\nu=27.2\pm7.1$\,km/s. This is consistent with observations giving $\nu\sin(i)=31.8\pm2$\,km/s, which suggests that the star is likely viewed edge-on (i.e., with $i\sim 90\degree$). If the exozodiacal disk shares the same inclination as the star, this is incompatible with our LBTI observations for a vertically thin disk. To be compatible, the disk would need to be sufficiently thick so that no significant changes in null depths would have been observed across the different parallactic angles in Fig.\,\ref{fig:example TF}. However, this hypothesis cannot currently be tested as no models are available for the LBTI to simulate observations of vertically thick disks. 

The disk may also be misaligned with the stellar rotation axis, forming a polar circumstellar disk. This configuration has proven to be stable over the stellar lifetime and can arise from interactions in a young stellar cluster \citep[][]{Kennedy2012}. The wide companion of $\theta$\,Boo could potentially have caused such a misalignment. The disk may also be tilted or diffused by giant planets orbiting with a planet-disk misalignment \citep[][]{Pearce2014,Brady2023}. Additional observations are therefore required to further constrain the orientation of the planetary system around $\theta$\,Boo.

\paragraph{Detection of giant planets}
If the orbits of potential giant planets in the system share an inclination close to face-on with the exozodi, the constraints from radial velocity and transit observations become less stringent.
Alternatively, the proper-motion anomaly measured by Gaia Early Data Release 3 (EDR3) could be analyzed to model what types of planets might generate such an anomaly \citep{Kervella2022}. Observations of $\theta$\,Boo with Gaia are showing a renormalized unit weight error (RUWE) of 3.38, which is sufficient to indicate the possible presence of a massive body in the system, but this anomaly might also be explained by the wide companion of $\theta$\,Boo.

\renewcommand{\arraystretch}{1.5}
\begin{table}
    \centering
    \caption{Zodi levels from the recomputed optical depth profiles of the disk models for realistic grain sizes of 3, 5, and 10\,µm.}
    \begin{tabular}{c c c c} \hline \hline
     &  \multicolumn{3}{c}{Observing night} \\ \cline{2-4} 
    Grain size [µm] & 2017-04-11 & 2018-05-23 & 2023-05-25 \\ \hline
    3 & $63^{+24}_{-26}$ & $128^{+21}_{-25}$ & $173^{+39}_{-67}$ \\
    5 & $85^{+33}_{-36}$ & $173^{+28}_{-34}$ & $235^{+53}_{-91}$ \\
    10 & $154^{+60}_{-65}$ & $313^{+52}_{-61}$ & $425^{+97}_{-165}$ \\\hline
    \end{tabular}\\
    \label{tab:grain size to zodi}
\end{table}
\renewcommand{\arraystretch}{1.0}

\subsection{Origin of the dust and its variability}\label{sec:discussion origin}

    \begin{figure*}
        \centering
        \includegraphics[width=0.49\linewidth]{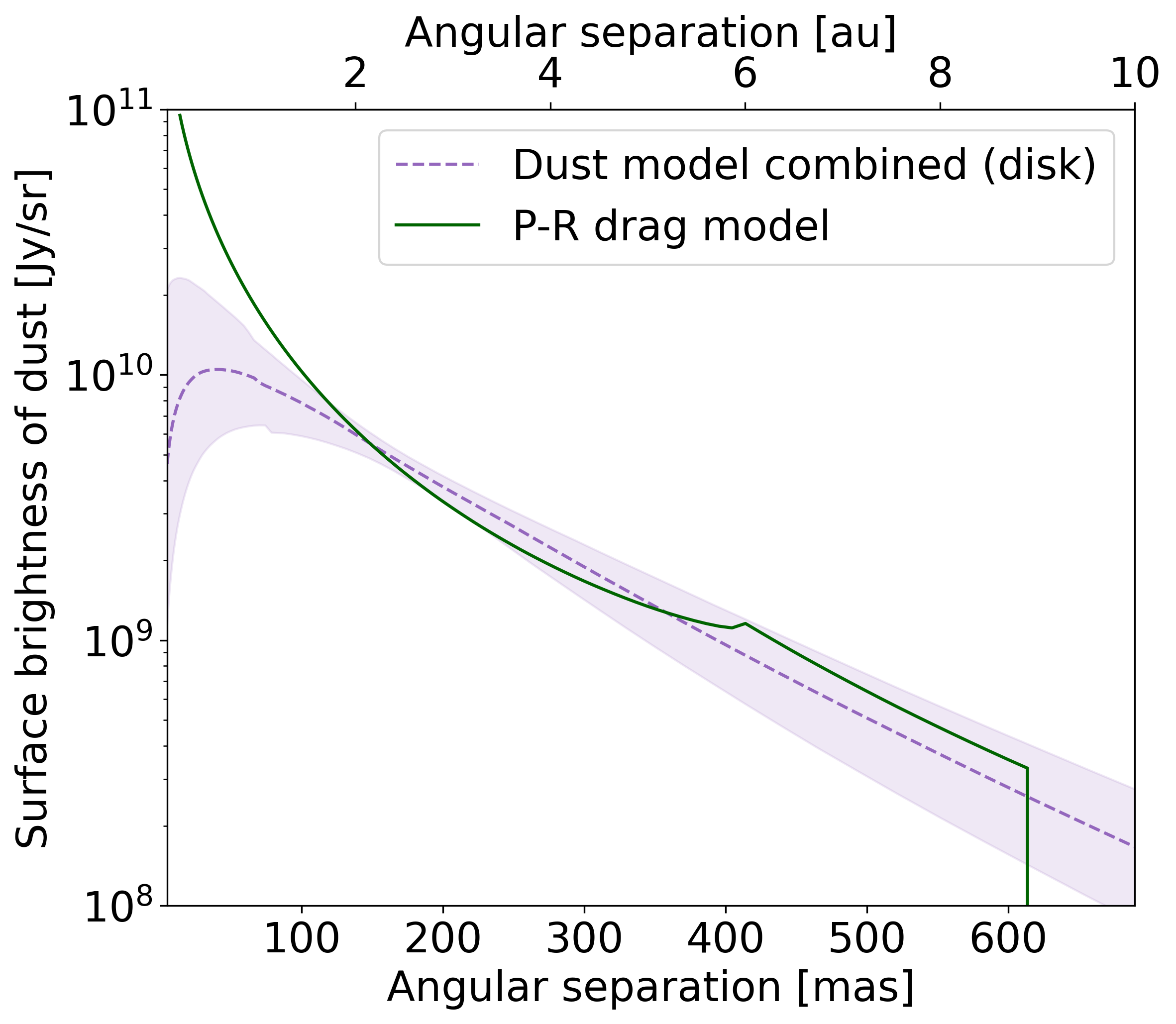}
        \includegraphics[width=0.49\linewidth]{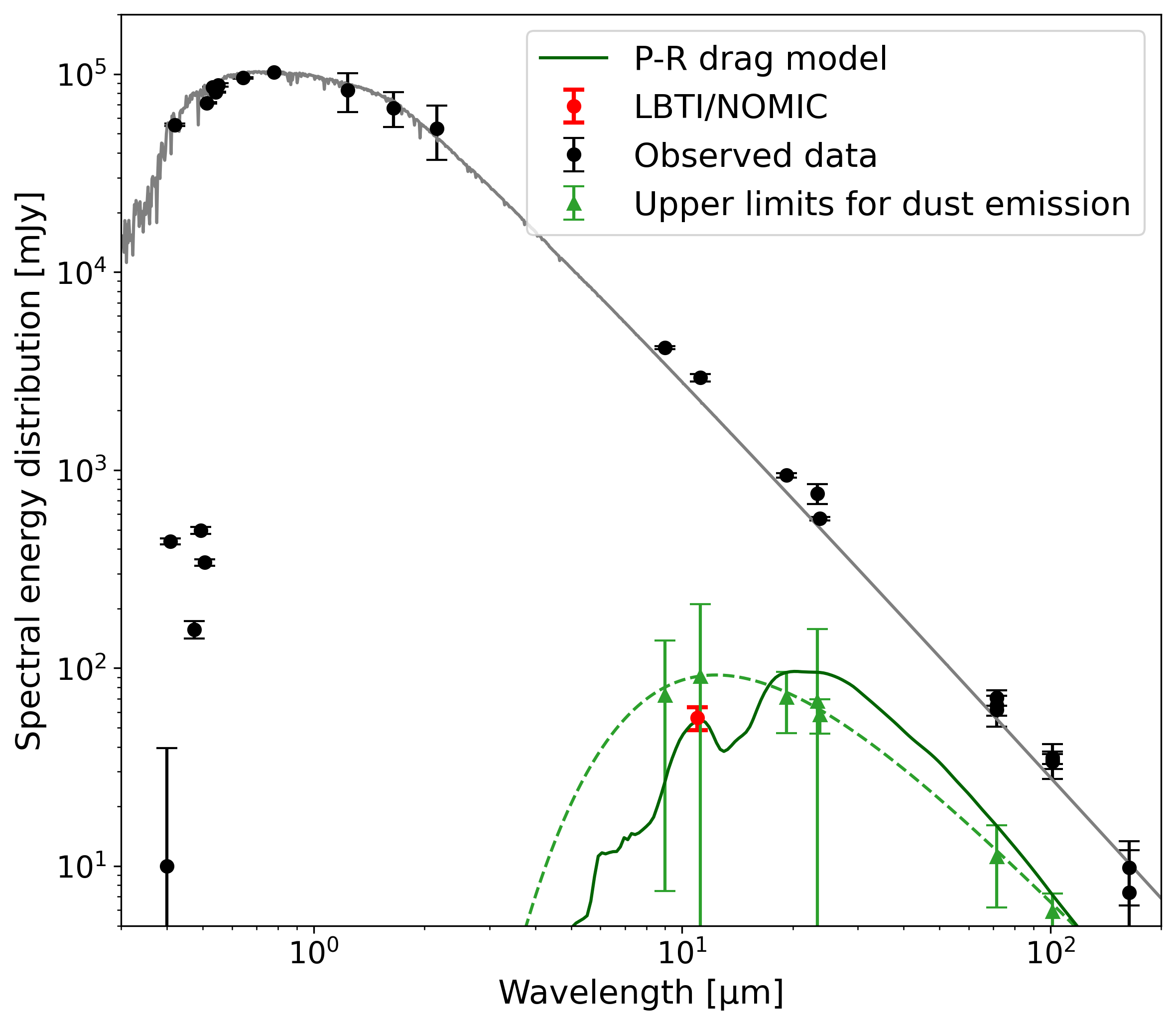}
        \caption{Left: Radial distribution of the dust surface brightness models for the combined 2017, 2018 and 2023 dataset. Models are calculated at a wavelength of 11\,µm, with 1$\sigma$ error bars derived from the MCMC posteriors. The solid line represents the predicted surface brightness from the P-R drag model for an outer belt extending from 6\,au to 9\,au, with a mass of $7\times 10^{-6}\,M_\oplus$. The dust composition is considered to be 1/3 silicate and 2/3 carbon. Right: SED of the dust from the P-R drag model.}
        \label{fig:PR drag model}
    \end{figure*}

\paragraph{Poynting–Robertson (P-R) drag hypothesis}
One of the main hypotheses to explain the presence of warm exozodiacal dust is the P-R drag mechanism. In this scenario, dust forms through planetesimal collisions in an outer belt located further out in the system and is dragged into the inner region by interaction with stellar radiation, eventually evaporating upon reaching sublimation temperature. 
To simulate this phenomenon, we applied the model presented in \cite{Rigley2020} to the $\theta$\,Boo system.
In this analytical model, the stellar parameters and the radius of the outer belt serve as input parameters to predict the size distribution of particles in the disk.
The size distribution is first simulated in the outer belt using the method described in \cite{Wyatt2011}.
The dust particles are then assumed to evolve independently inward until the system reaches a steady state where the optical depth distribution changes by less than 1\,\% in a logarithmic time step. The smallest grains that dominate the optical depth typically reach that state after $\sim$10\,Myr.
The radial profile for each particle is determined using the approach of \cite{Wyatt2005}. The model then calculates the 2D optical depth distribution as a function of particle size and distance from the star.
Finally, the model adjusts the width and mass of the outer belt to fit the surface brightness model from the observations.
To determine the emission from the grains in the modeled disk, we again assume the optical properties used to simulate their SEDs in Sect.\,\ref{sec:discussion SED}.

However, the simulations from this model are only applicable to a system in a steady state, on timescales of years. The time required for dust particles of blow-out size to drift from 8\,au to 2\,au is estimated to be  35,000\,yrs \citep[][]{Wyatt1950,Burns1979}.
We therefore compared the P-R drag model with the model for the combined observing nights of the LBTI (Fig.\,\ref{fig:bestfitmodel all}) using a disk geometry. The wide ring and thin ring geometries are not compatible with P-R drag alone and require the presence of a massive object (e.g., a giant exoplanet) to carve their edges. Figure\,\ref{fig:PR drag model} shows the result of the P-R drag model, including the best-fit surface brightness profile and its corresponding SED. The result is obtained for an outer belt extending from 6 to 9\,au, with a mass of $7\times10^{-6}\,M_\oplus$, considering all dust particles with sizes $\leq1$\,cm. We find that P-R drag can explain our model, with a discrepancy appearing only for $<1$\,au separations. However, the outer belt required to explain this distribution produces an infrared excess that exceeds the upper limits in Fig.\,\ref{fig:SED_theta_boo} around 25\,µm. This suggests that such an outer belt should have been detected by \textit{Spitzer}.

Since our models also do not favor the steady-state hypothesis (Sect.\,\ref{sec:modeling results}), we conclude that other phenomena are likely to influence the dust distribution and generate yearly variability.

\paragraph{Extreme debris disk hypothesis}
The models presented in  Sect.\,\ref{sec:Dust Variability} indicate possible variability of the exozodiacal disk around $\theta$\,Boo, both in its spatial distribution and brightness. Photometric variability of warm dust on monthly to yearly timescales has previously been reported for extreme debris disks \citep[EDDs,][]{Balog2009}. These disks are detected around FGK-type main-sequence stars with ages between $\sim$100 to $\sim$200\,Myr \citep[][]{Melis2012,Meng2012,Meng2015,Rieke2021,Moor2022}, and even $>1$\,Gyr \citep[][]{Thompson2019}. This variability is commonly attributed to major collisions of planetesimals during the terrestrial planet formation phase \citep[][]{Su2019}. \cite{Balog2009} also define EDDs as systems with total emission at least four times greater than the stellar photospheric output at 24\,µm.
For $\theta$\,Boo, the age of the system is estimated to be 3-4.7\,Gyr (Sect.\,\ref{sec:Constrain on Planet}), which is well after the terrestrial planet formation phase. Upper limits also indicate that no dust emission greater than one-tenth of the stellar photosphere has been detected at 24\,µm.
These findings cast doubt on the hypothesis of an EDD for $\theta$\,Boo, with massive planetesimal collisions as the source of the dust and its variations.

\paragraph{Catastrophic event hypothesis}
An alternative hypothesis to explain this variability is the occurrence of a catastrophic event such as a massive collision, fragmentation of a massive comet, or the passing of a massive object with highly eccentric orbit-that would have occurred in the inner region. After several years, such an event could generate new dust or disrupt the dust distribution, resulting in an observed brightness variability. 
For example, debris from a planetary collision occurring at 2\,au would have a circular orbit of 2.5\,years around $\theta$\,Boo. After five\,years, the debris would begin to form a belt in the HZ that could be detected by the LBTI.
Direct imaging observations in the H- and L-bands could not detect the presence of giant planets with masses lower than 11\,$M_\text{Jup}$. An eccentric planet could, however, disrupt a potential outer belt and periodically scatter exocomets inward \citep[][]{Faramaz2017}. 
From the increase in optical depth in our models, we estimate the additional mass of material needed to explain this variability. This material could originate from a massive collision or comet fragmentation. For a grain size of 3\,µm, the required mass ranges from $10^{-3}\,M_\text{Vesta}$ ($4\times10^{-8}\,M_\oplus$) for our disk models, to $3\times10^{-3}\,M_\text{Vesta}$ ($1.2\times10^{-7}\,M_\oplus$) for our wide ring models. For a grain size of 10\,µm, the estimated mass is tripled.

\section{Summary}
The HOSTS survey \citep[][]{Ertel_2020} demonstrated a correlation between the presence of cold outer dust and exozodiacal dust. $\theta$\,Boo is one of the three exceptions where exozodiacal dust has been detected by the LBTI without prior observations of significant mid- to far-infrared excess emission. In this study, we analyzed multi-epoch observations of the exozodiacal disk around $\theta$\,Boo using nulling interferometry with the LBTI in the N'-band. Three observations were obtained on UT 2017 April 11, UT 2018 May 23, and UT 2023 May 25, respectively. The data were reduced using the LBTI nulling pipeline and analyzed with a model dedicated to exozodiacal disk observations with the LBTI \citep[][]{Kennedy_2015}.

A prior assumption for our modeling procedure is that the planetary system of $\theta$\,Boo is close to face-on (i.e., $i\sim0\degree$). This is supported by our extended nulling observation in 2023, which shows no significant relation between the detected emission and the parallactic angle. However, calculations of the stellar inclination indicate that $\theta$\,Boo is likely to be edge-on (i.e., $i\sim90\degree$). For the exozodiacal disk to also be edge-on,  it would need to be sufficiently thick or diffuse. Because our modeling tool can only simulate a vertically thin disk, the face-on assumption is retained for this work. The results and conclusions presented here are therefore highly dependent on this assumption, and future observations are necessary to constrain the system's inclination with a higher confidence.

From the models obtained for the individual and combined 2017, 2018, and 2023 datasets, we propose several spatial distributions of the dust that can explain our observations.
The models for the combined nights fit the data less well than those for the individual nights, supporting the hypothesis of an unsteady state system.
The different dust distribution geometries fit each night similarly; therefore, we are unable to discriminate between them with the LBTI observations. Using the upper limits on dust excess emission in the mid- to far-infrared from the \textit{Spitzer} (2004) and \textit{Herschel} (2010) observations, we further constrain the dust properties. Although our simulated SEDs do not clearly discriminate between the different geometries, they favor a representative dust grain size distribution of 3-5\,µm.
Using our models for the dust spatial distribution of 2017 and 2023, we also estimate the zodi level and its variability at the EEID. Considering our disk models, which assume a broad dust distribution in the orbital plane and that dust grains behave as ideal blackbodies,
we estimate the dust brightness to be $158^{+62}_{-67}$ and $443^{+98}_{-170}$\,zodis for 2017 and 2023, respectively. Considering the probability distribution of the two values, we find a tentative increase in dust emission with 1.86$\sigma$ significance at the EEID. This result suggests possible brightness variability of exozodiacal disks in the HZ of main-sequence stars. Additional observations are needed to confirm this trend for $\theta$\,Boo.
However, the obtained zodi levels depend on the chosen dust temperature, and consequently on the size distribution of the dust grains. For realistic grains with a size of 3\,µm, for example, the obtained zodi levels are $63^{+24}_{-26}$ and $173^{+39}_{-67}$\,zodis for 2017 and 2023, respectively. Further constraints on the dust properties are therefore needed to estimate the absolute zodi level of the system.

Several hypotheses are considered for the origin of the exozodiacal dust and its variability: (1) P-R drag, (2) the EDD, and (3) a catastrophic event. 
The first hypothesis, P-R drag, is one of the main mechanisms explaining the presence of warm dust, continuously generated by a collisionally active asteroid belt analogue. 
However, nondetections of far-infrared excess limit the significance of such a belt. Using our steady-state model, with the combined 2017, 2018, and 2023 nights, we retrieve the dust distribution from the P-R drag for an outer parent belt extending from 6\,au to 9\,au with a mass of $7\times10^{-6}\,M_\oplus$. However, since our steady-state model is not favored, another mechanism is likely to influence the dust distribution besides P-R drag, and generate a yearly variability.

The second hypothesis, considering an EDD, could explain the variability, but is unlikely, given the age of $\theta$\,Boo and the level of excess infrared emission from the dust. The third hypothesis involves a catastrophic event, such as a massive collision, fragmentation of a massive comet, or passage of a massive object with a highly eccentric orbit. We estimate the amount of additional material injected into the system to explain this variability. Considering dust particles ranging from 3\,µm to 10\,µm, and for our different model geometries, we find an additional mass of $10^{-3}-10^{-2}\,M_\text{Vesta}$ ($M_\text{Vesta}\simeq4\times10^{-5}\,M_\oplus$).
The presence of a substellar companion that could disrupt the dust distribution has also been constrained using high-contrast AO observations in the H- and L'-bands with the SHARK-NIR and LMIRCam instruments, respectively. These observations were performed on UT 2024 February 24, but did not show any features indicating the presence of a point-like companion source in the system. Nonetheless, the contrast limits of the images allowed us to constrain the presence of giant planets down to 11\,$M_\text{Jup}$ at 1.3\arcsec\,angular separation, based on a ``hot start'' evolutionary model.

To overcome the current challenges in studying the exozodiacal disk of $\theta$\,Boo, additional observations will be required. In particular, detections at different wavelengths in the mid-infrared (e.g., using the JWST/MIRI instrument) would further constrain the dust properties and its temperature profile. More sensitive far-infrared observations would also search for emissions from a cold outer debris disk that could feed the warm dust, providing clues to the system's orientation. 
Additional LBTI observations with reduced noise would also enable a more precise modeling of the distribution and orientation of the warm dust, particularly by considering the dependence of the null measurements on the parallactic angle. The implementation of the PCA method for background subtraction \citep[][]{2024Rousseau}, along with the upcoming new detector for NOMIC, should reduce instrumental noise and improve future observations.

\begin{acknowledgements}
G.G., D.D., and M.A.M have received funding from the European Research Council (ERC) under the European Union's Horizon 2020 research and innovation program (grant agreement CoG - 866070).

S.E. is supported by the National Aeronautics and Space Administration through the Exoplanet Research Program (Grant No. 80NSSC21K0394) and the Astrophysics Decadal Survey Precursor Science program (Grant No. 80NSSC23K1473).

V.F. acknowledges funding from the National Aeronautics and Space Administration through the Exoplanet Research Program under Grants No. 80NSSC21K0394 and 80NSSC23K1473 (PI: S. Ertel), and Grant No 80NSSC23K0288 (PI: V. Faramaz). 

T.D.P. acknowledges support of the Research Foundation - Flanders (FWO) under grant 11P6I24N (Aspirant Fellowship).

The LBT is an international collaboration among institutions in the United States, Italy and Germany. LBT Corporation Members are: The University of Arizona on behalf of the Arizona Board of Regents; Istituto Nazionale di Astrofisica, Italy; LBT Beteiligungsgesellschaft, Germany, representing the Max-Planck Society, The Leibniz Institute for Astrophysics Potsdam, and Heidelberg University; The Ohio State University, representing OSU, University of Notre Dame, University of Minnesota and University of Virginia.

Part of the observing time for this program was granted as Director’s Discretionary Time by the LBT director.

We acknowledge the use of the Large Binocular Telescope Interferometer (LBTI) and the support from the LBTI team, specifically from Amali Vaz, Jordan Stone, Jenny Power, Phil Willems, Grant West, Jared Carlson, Andrew Cardwell, and Alexander Becker.

This work has made use of data from the European Space Agency (ESA) mission
{\it Gaia} (\url{https://www.cosmos.esa.int/gaia}), processed by the {\it Gaia} Data Processing and Analysis Consortium (DPAC,
\url{https://www.cosmos.esa.int/web/gaia/dpac/consortium}). Funding for the DPAC has been provided by national institutions, in particular the institutions participating in the {\it Gaia} Multilateral Agreement.

\end{acknowledgements}


\bibliographystyle{aa} 
\bibliography{report}

\newpage
\begin{appendix}
    \section{Null depths}\label{ap:all TF}
    \centering
    \vfill
    \begin{minipage}[H]{2\linewidth}
        \includegraphics[width=0.49\linewidth]{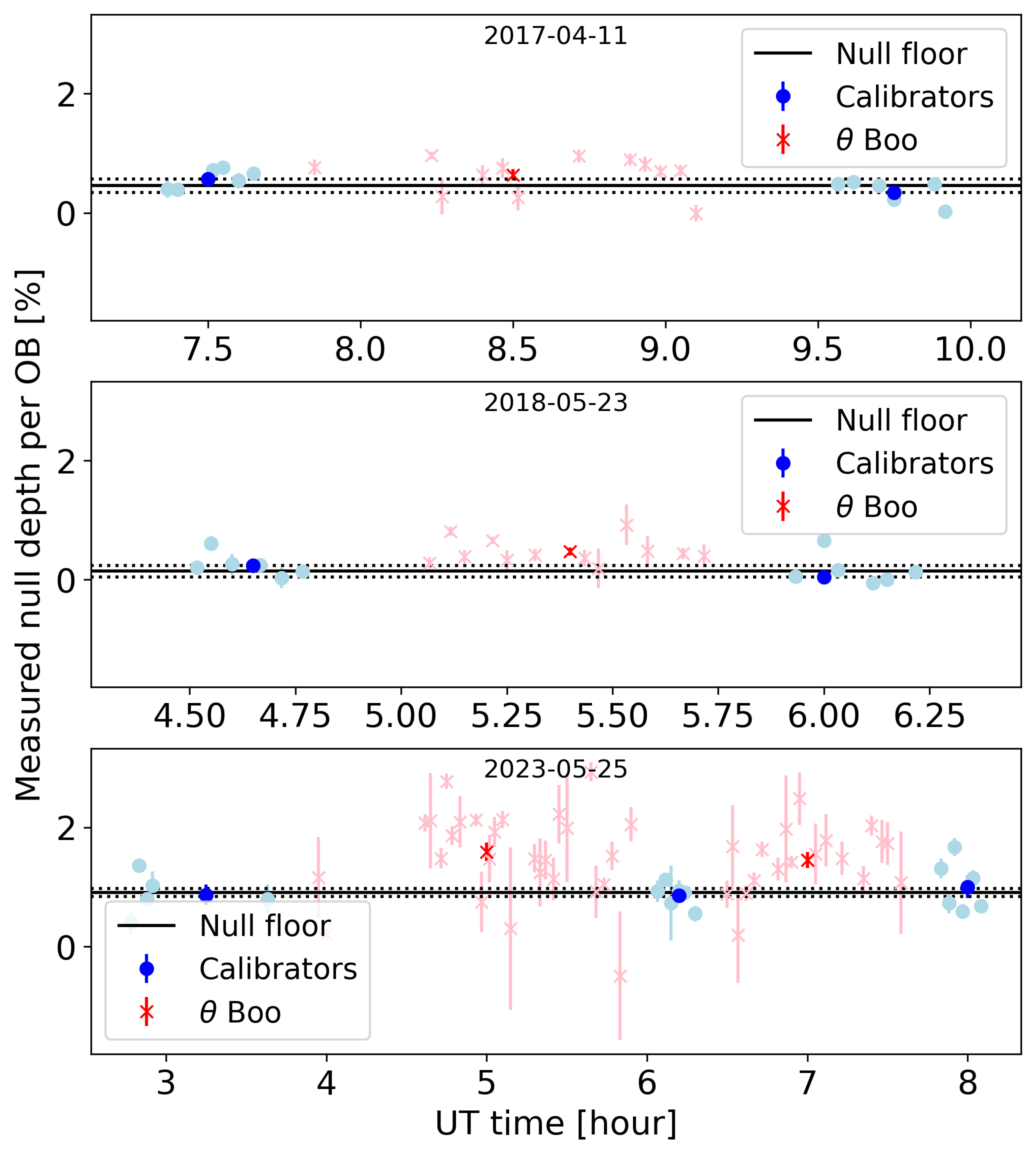}
        \includegraphics[width=0.49\linewidth]{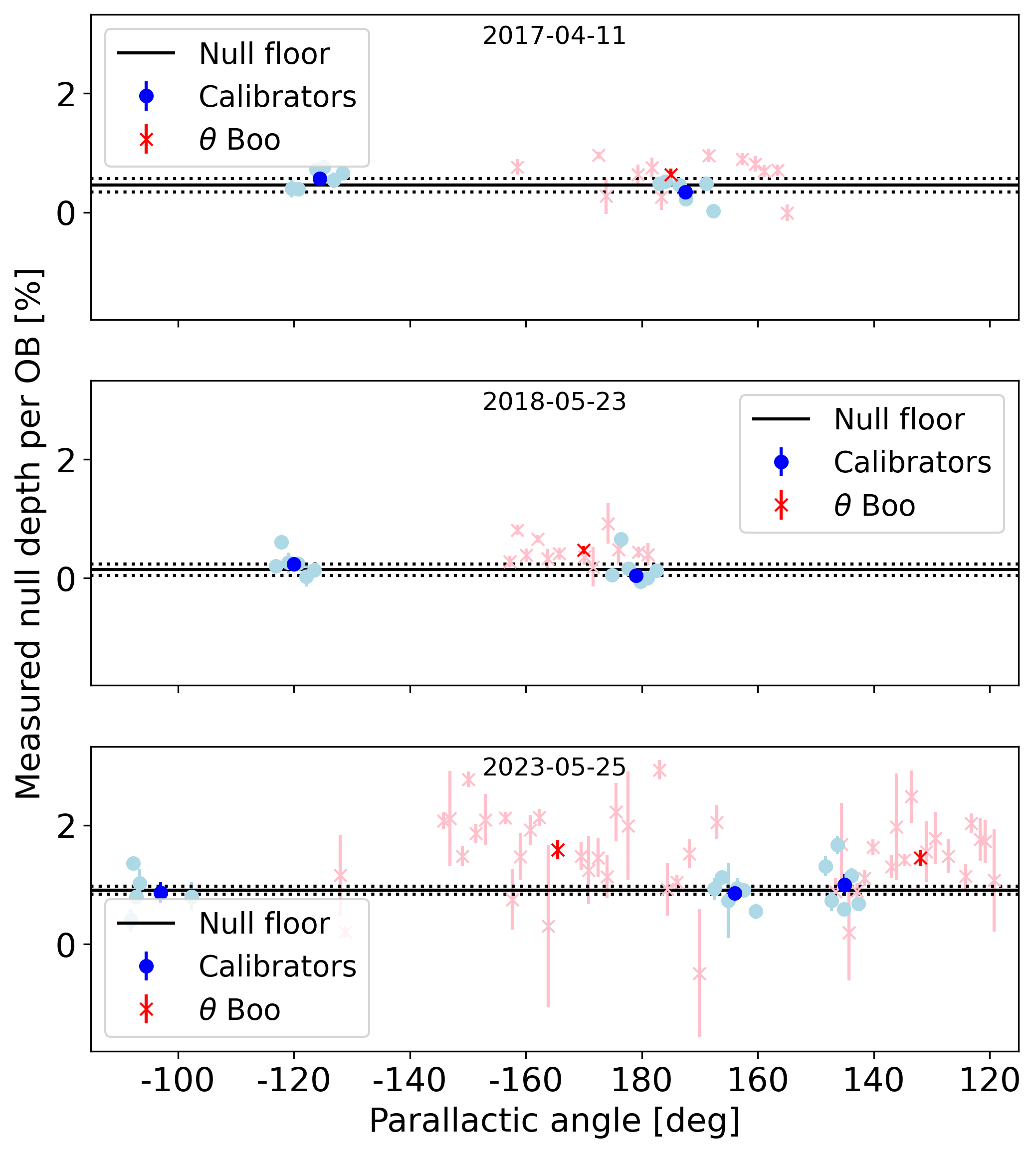}
        \captionof{figure}{Nulls with respect to UT (left) and  parallactic angle (right) for 8\,pixels aperture radius. (top) 2017 April 11, (middle) 2018 May 23, (bottom) 2023 May 25.}
    \end{minipage}
    \vfill
    \centering
    \begin{minipage}[H]{2\linewidth}
        \includegraphics[width=0.49\linewidth]{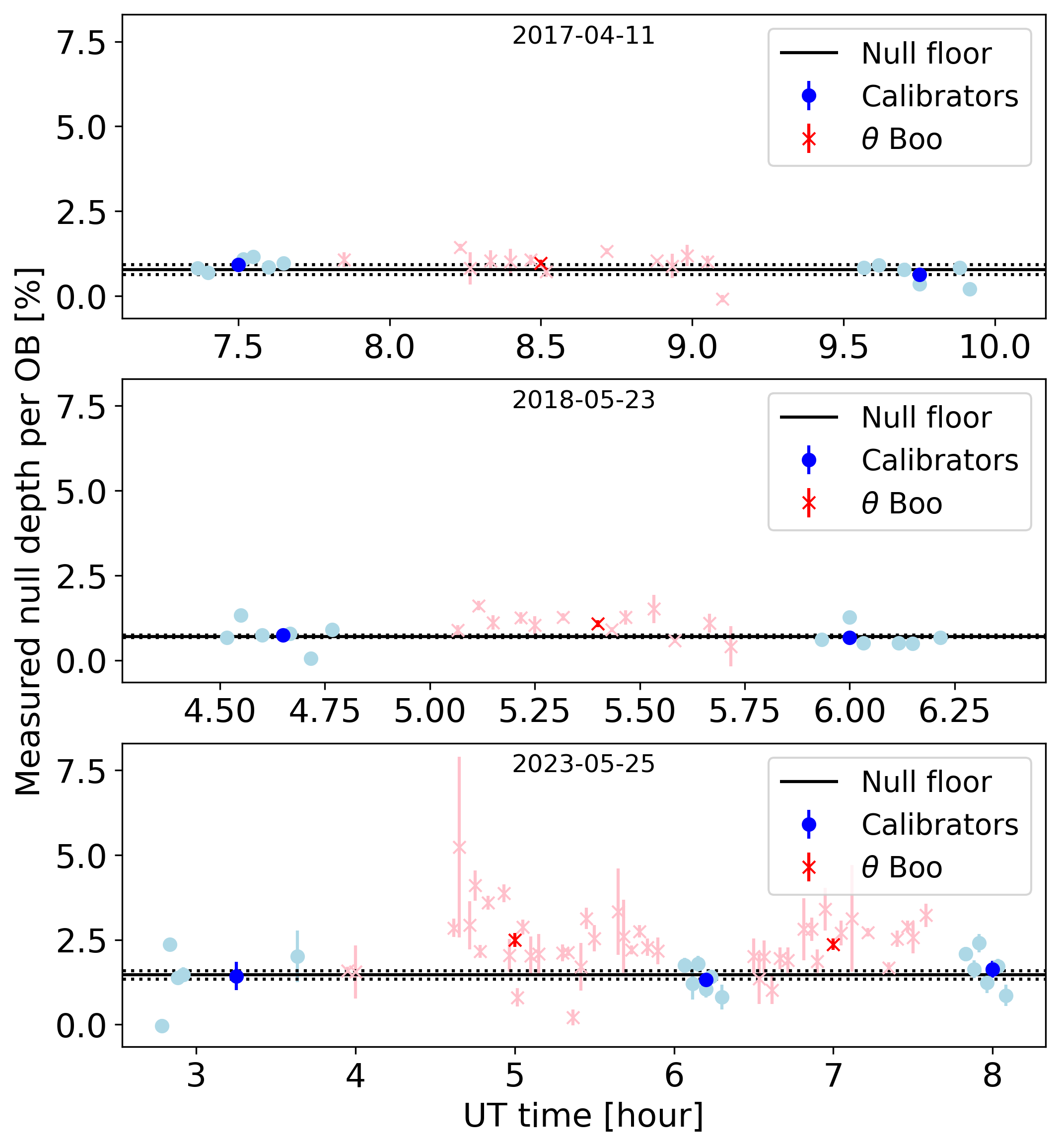}
        \includegraphics[width=0.49\linewidth]{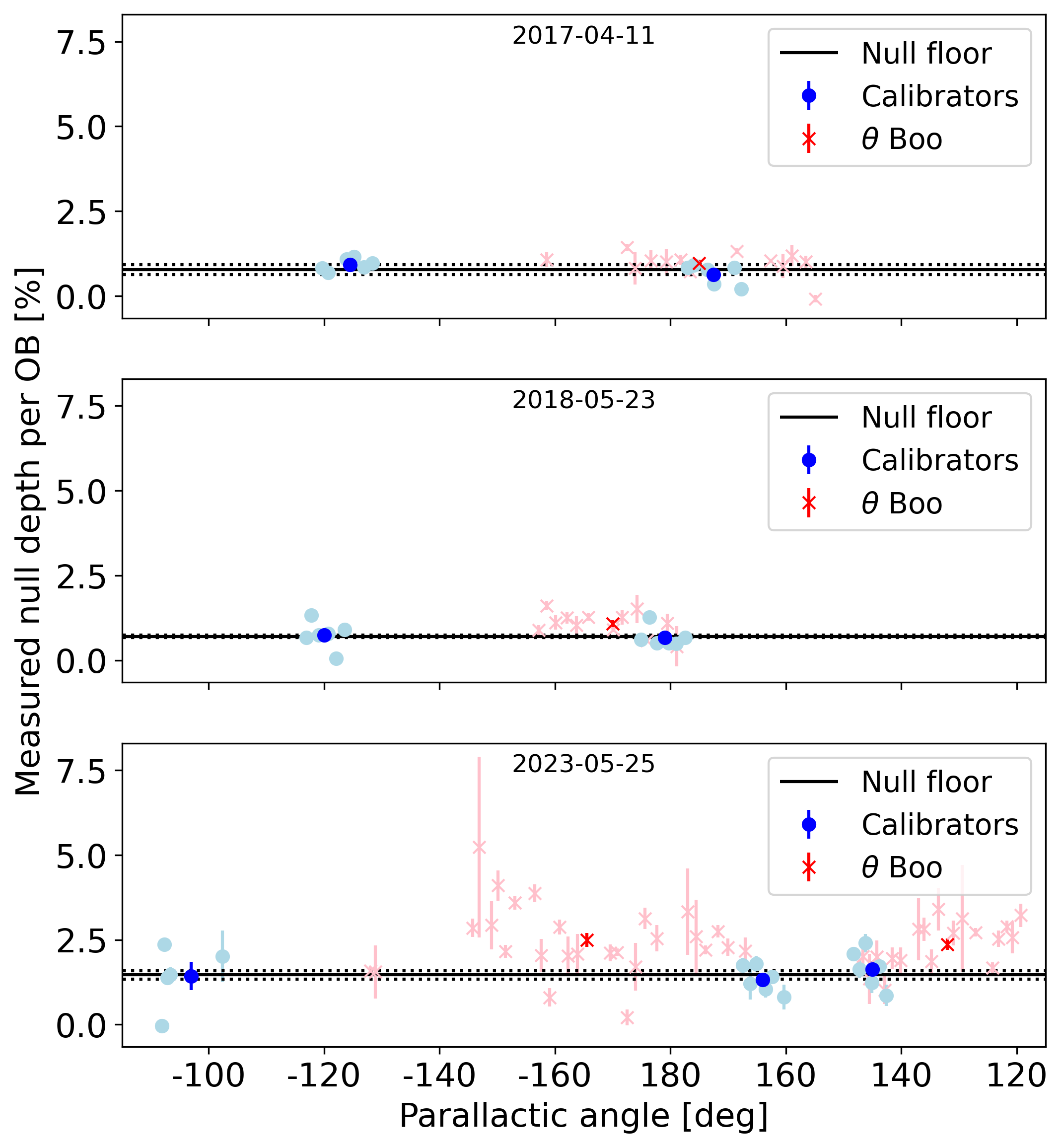}
        \captionof{figure}{Nulls with respect to UT (left) and  parallactic angle (right) for 16\,pixels aperture radius. (top) 2017 April 11, (middle) 2018 May 23, (bottom) 2023 May 25.}
    \end{minipage}
    \vfill

\newpage
\quad
\newpage
    \vfill
    \centering
    \begin{minipage}[H]{2\linewidth}
        \includegraphics[width=0.49\linewidth]{Tet_boo_TF_time_24.png}
        \includegraphics[width=0.49\linewidth]{Tet_boo_TF_PA_24.png}
        \captionof{figure}{Nulls with respect to UT (left) and  parallactic angle (right) for 24\,pixels aperture radius. (top) 2017 April 11, (middle) 2018 May 23, (bottom) 2023 May 25.}
    \end{minipage}
    \vfill
    \centering
    \begin{minipage}[H]{2\linewidth}
        \includegraphics[width=0.49\linewidth]{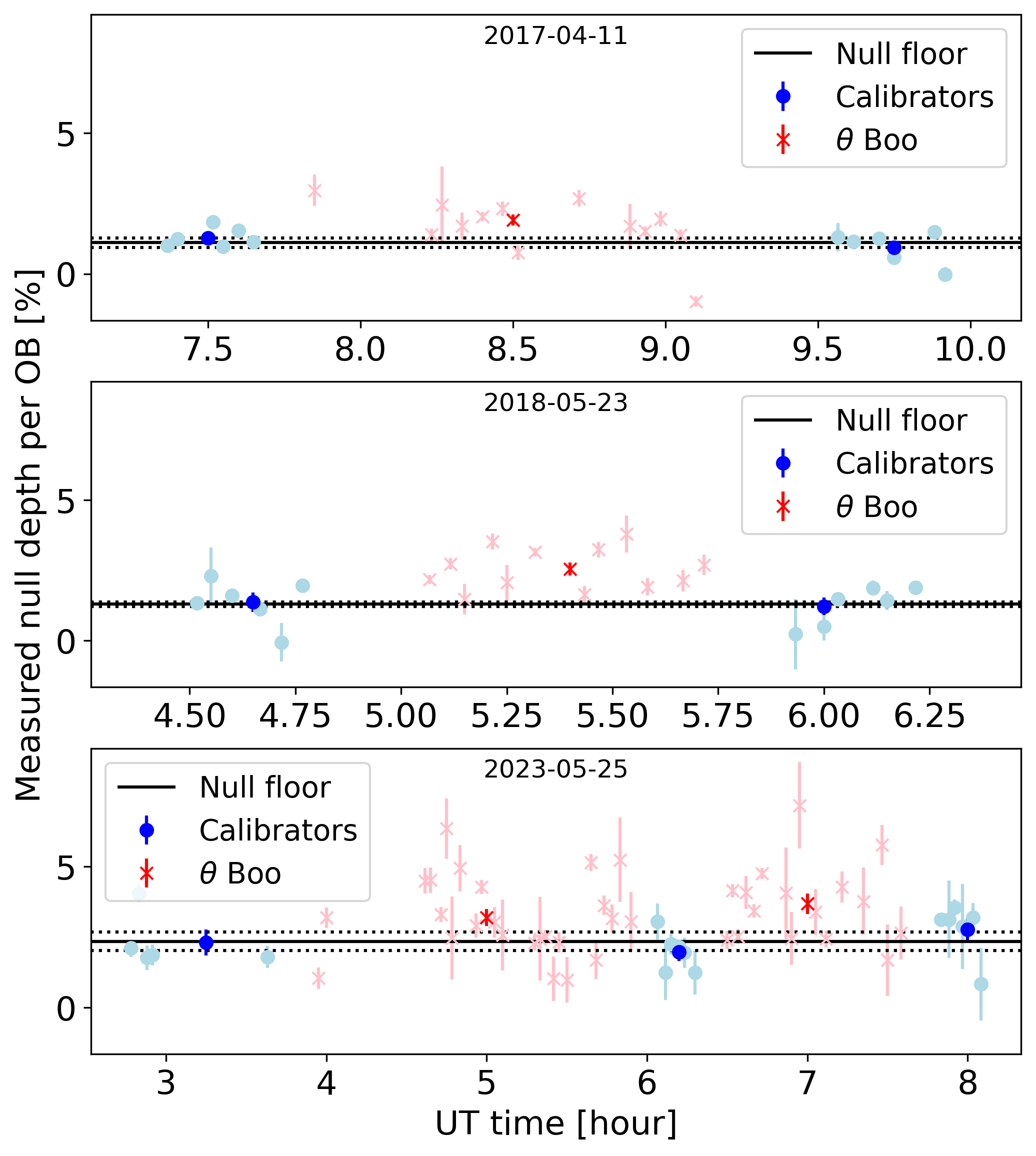}
        \includegraphics[width=0.49\linewidth]{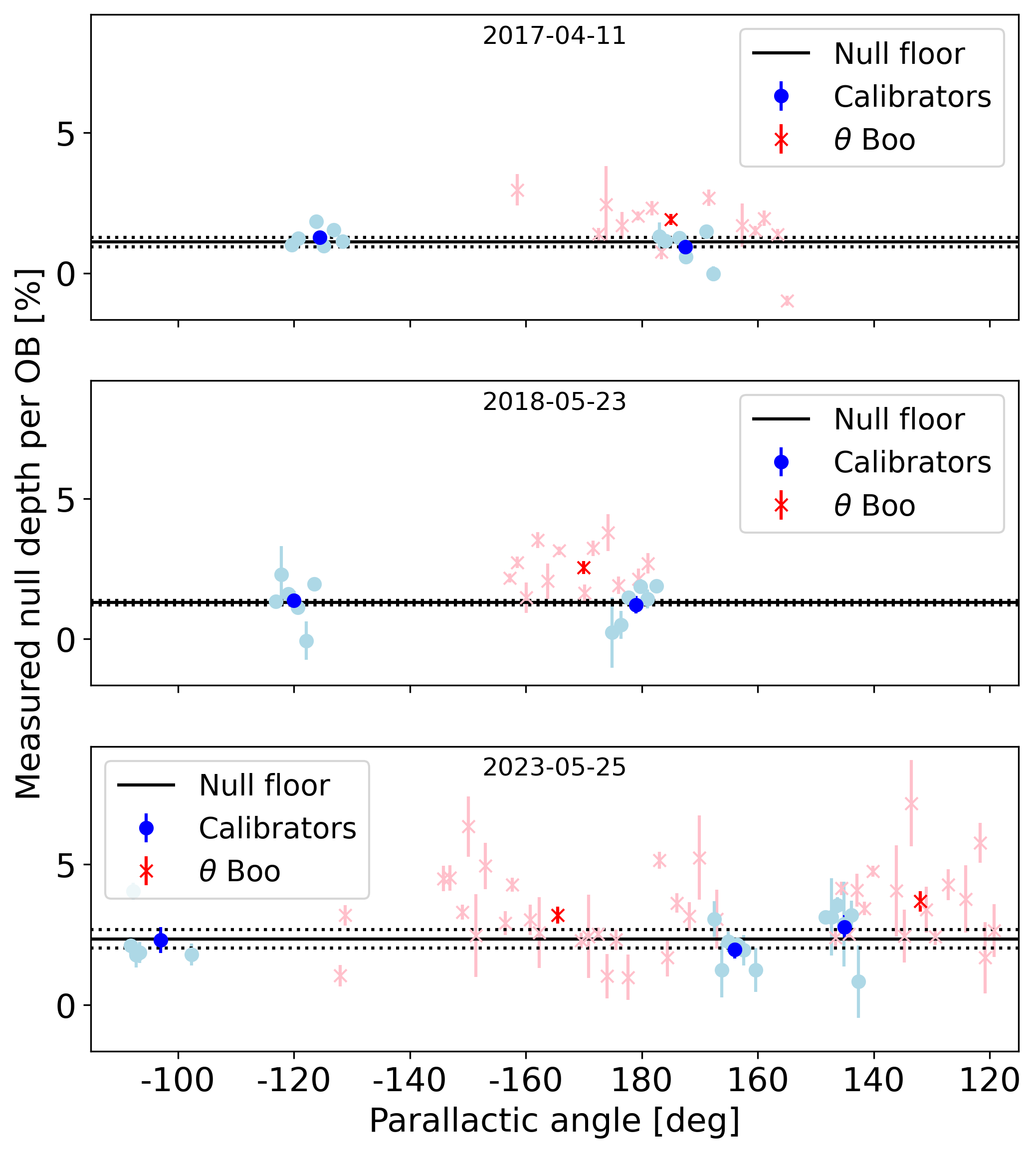}
        \captionof{figure}{Nulls with respect to UT (left) and  parallactic angle (right) for 32\,pixels aperture radius. (top) 2017 April 11, (middle) 2018 May 23, (bottom) 2023 May 25.}
    \end{minipage}
    \vfill
    
\newpage
\quad
\newpage

    \centering
    \begin{minipage}[H]{2\linewidth}
        \includegraphics[width=0.49\linewidth]{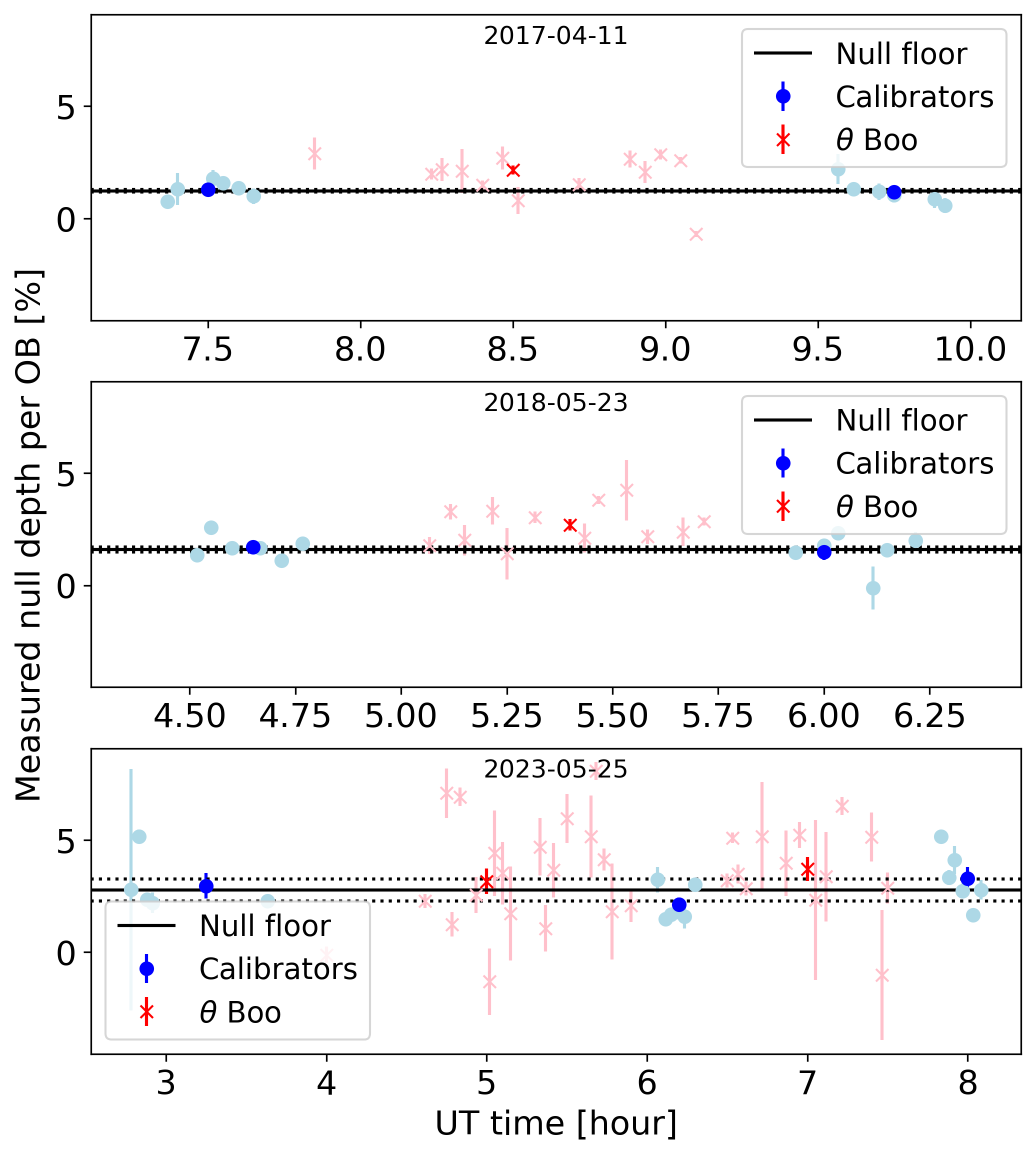}
        \includegraphics[width=0.49\linewidth]{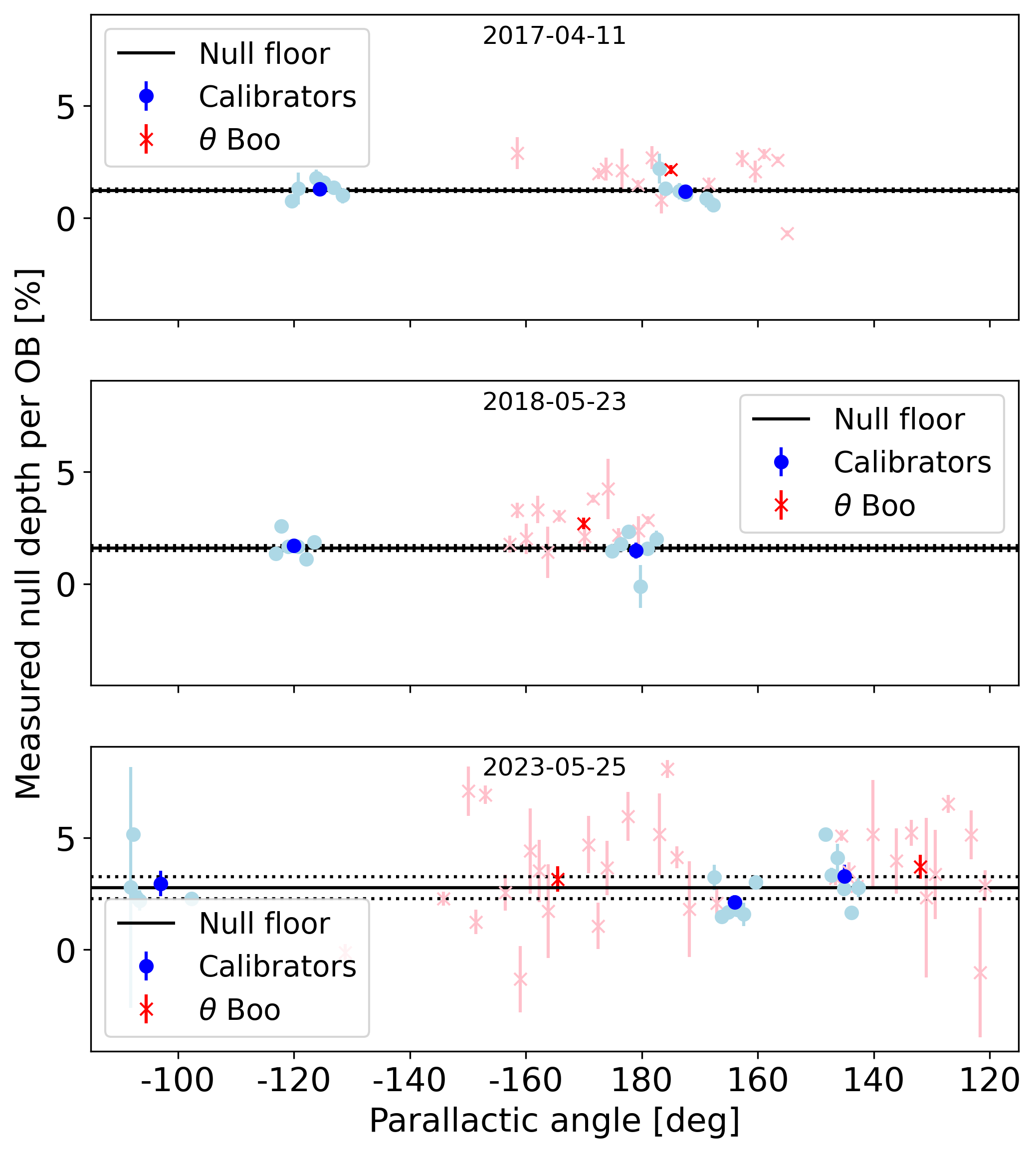}
        \captionof{figure}{Nulls with respect to UT (left) and  parallactic angle (right) for 40\,pixels aperture radius. (top) 2017 April 11, (middle) 2018 May 23, (bottom) 2023 May 25.}
    \end{minipage}

\newpage
\quad
\newpage

\section{Corner plots}\label{ap:corner plots}
    \vfill
    \centering
    \begin{minipage}[H]{2\linewidth}
    \centering
        \hfill
        \includegraphics[width=0.41\linewidth]{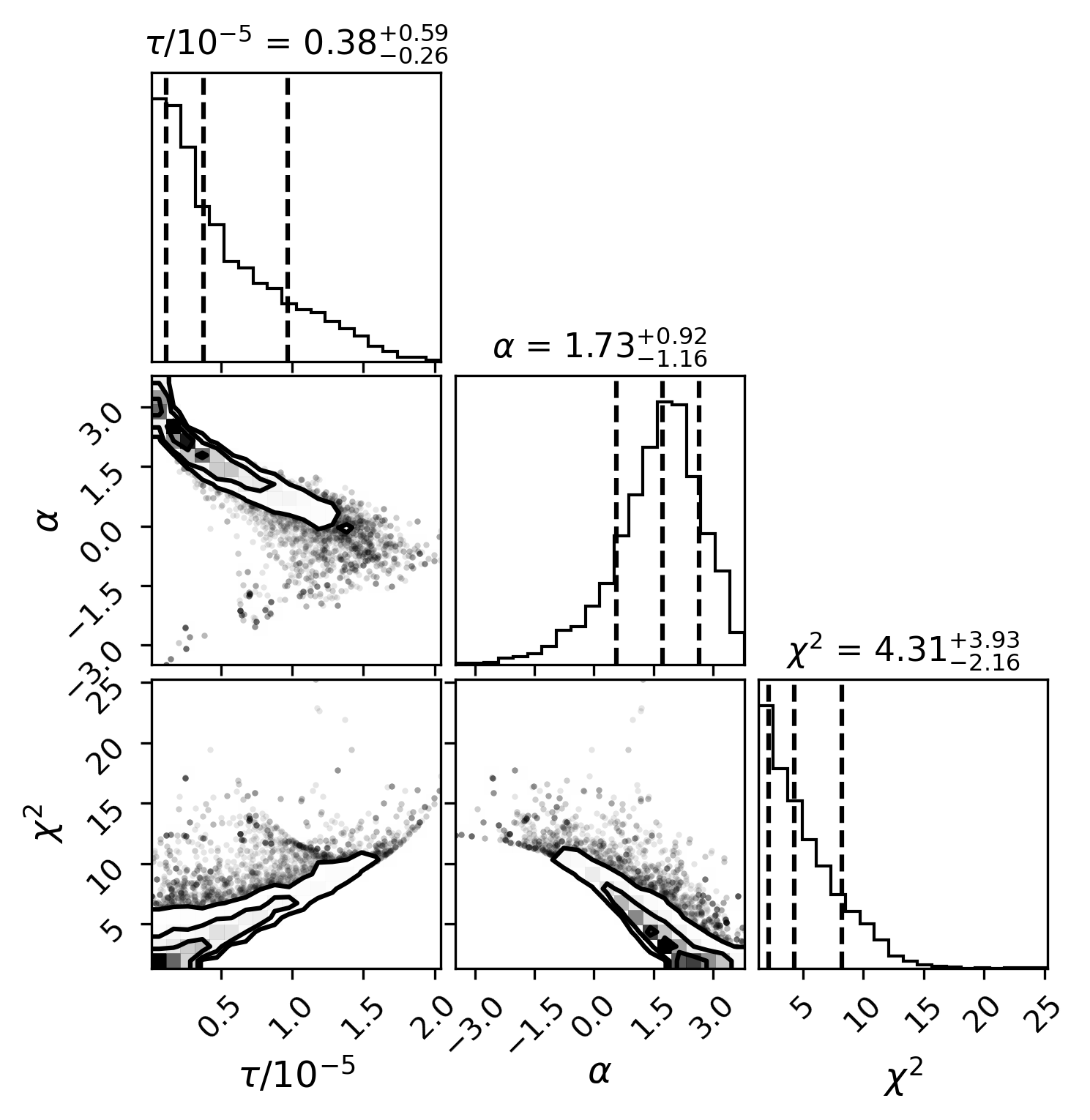}
        \hfill
        \includegraphics[width=0.41\linewidth]{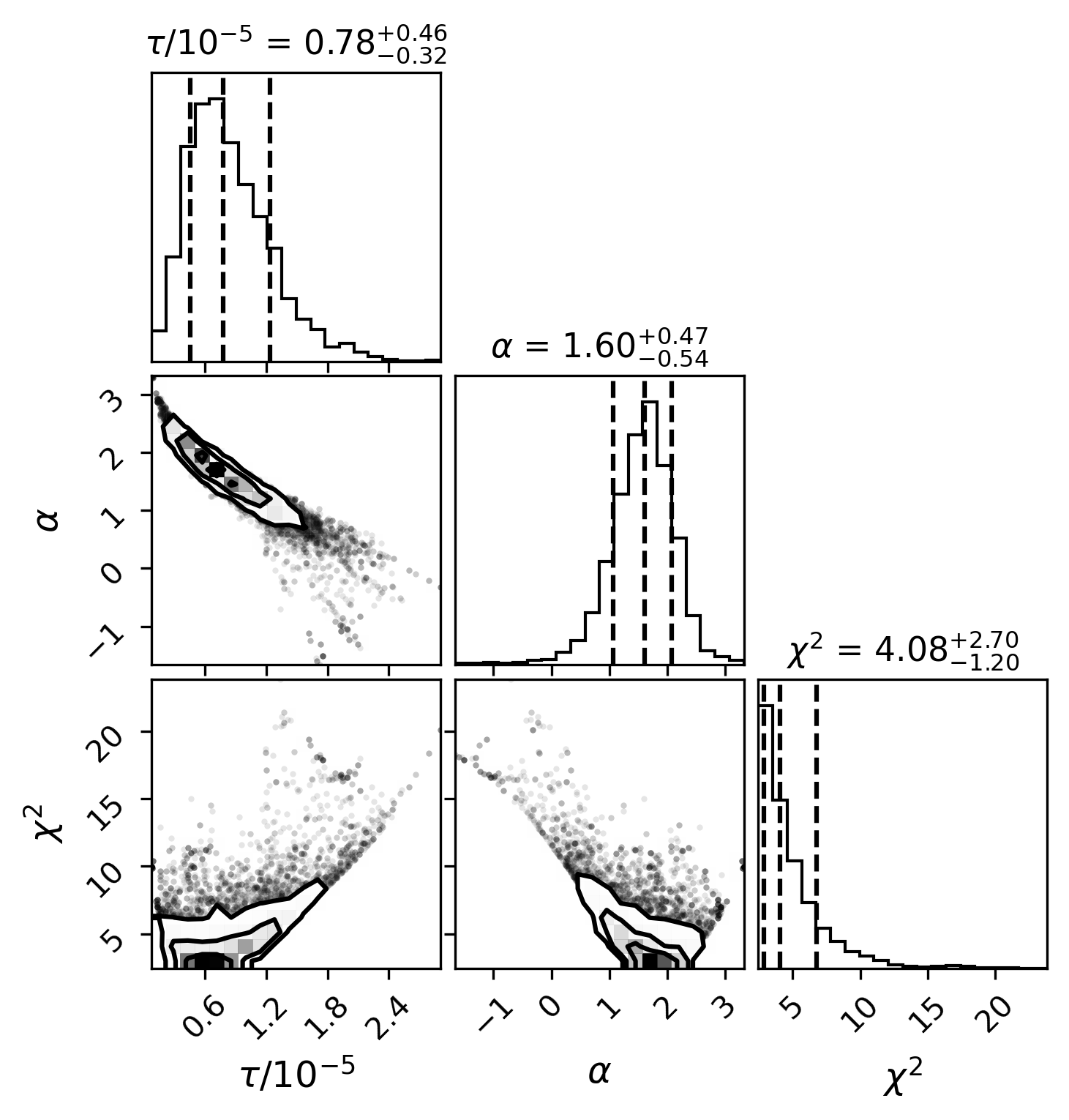}
        \hfill

        \centering
        \hfill
        \includegraphics[width=0.41\linewidth]{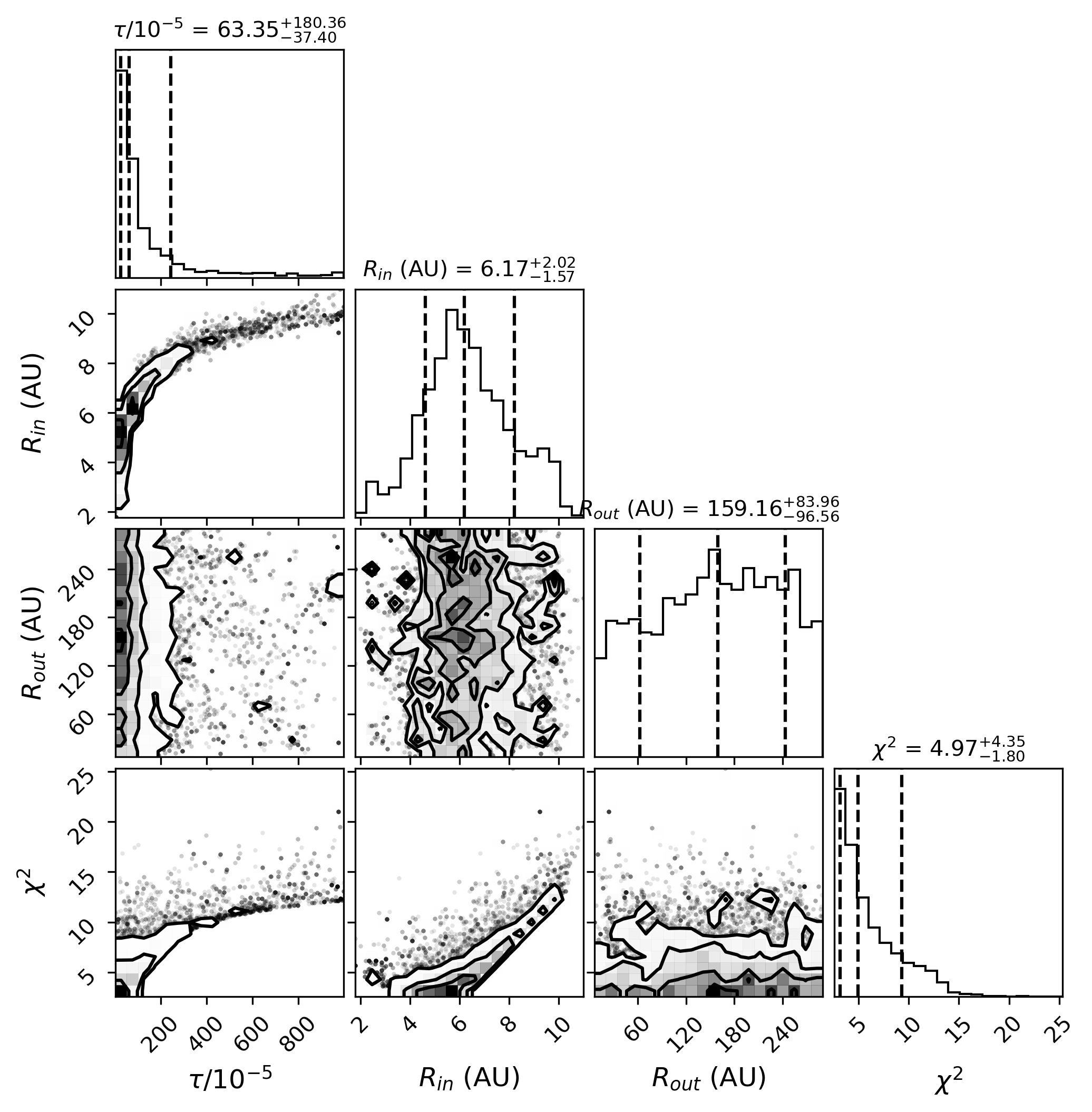}
        \hfill
        \includegraphics[width=0.41\linewidth]{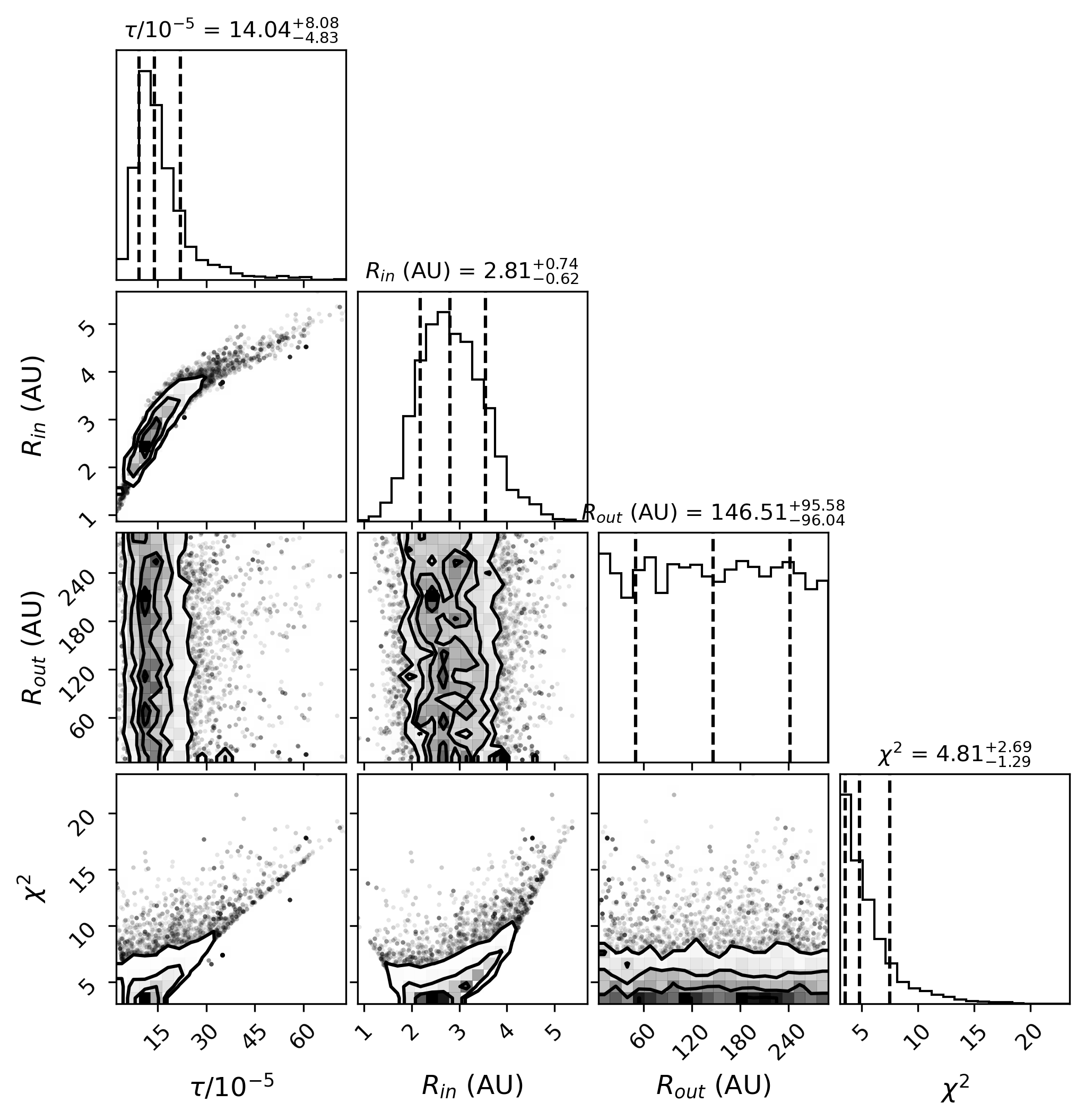}
        \hfill

        \centering
        \hfill
        \includegraphics[width=0.41\linewidth]{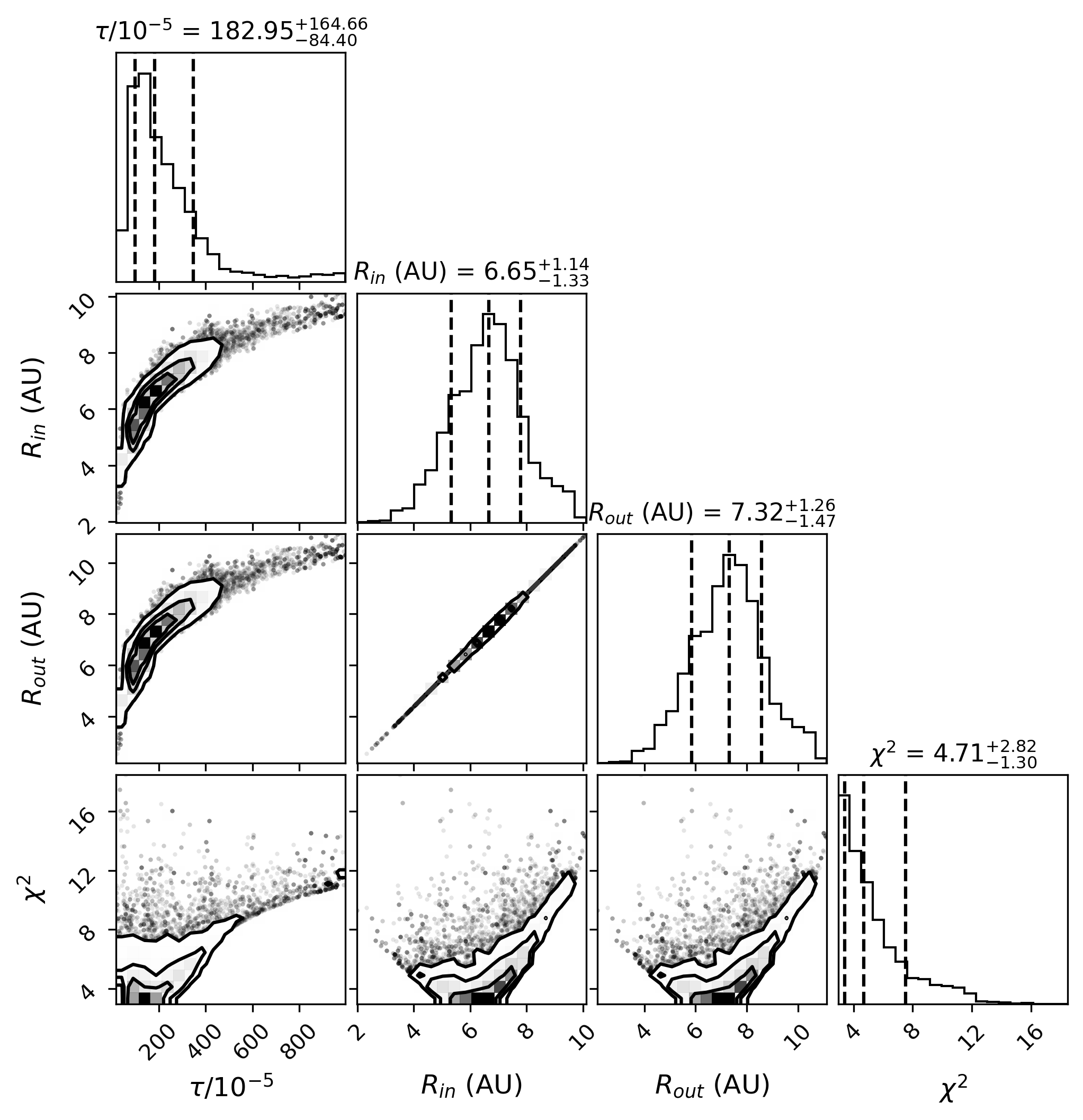}
        \hfill
        \includegraphics[width=0.41\linewidth]{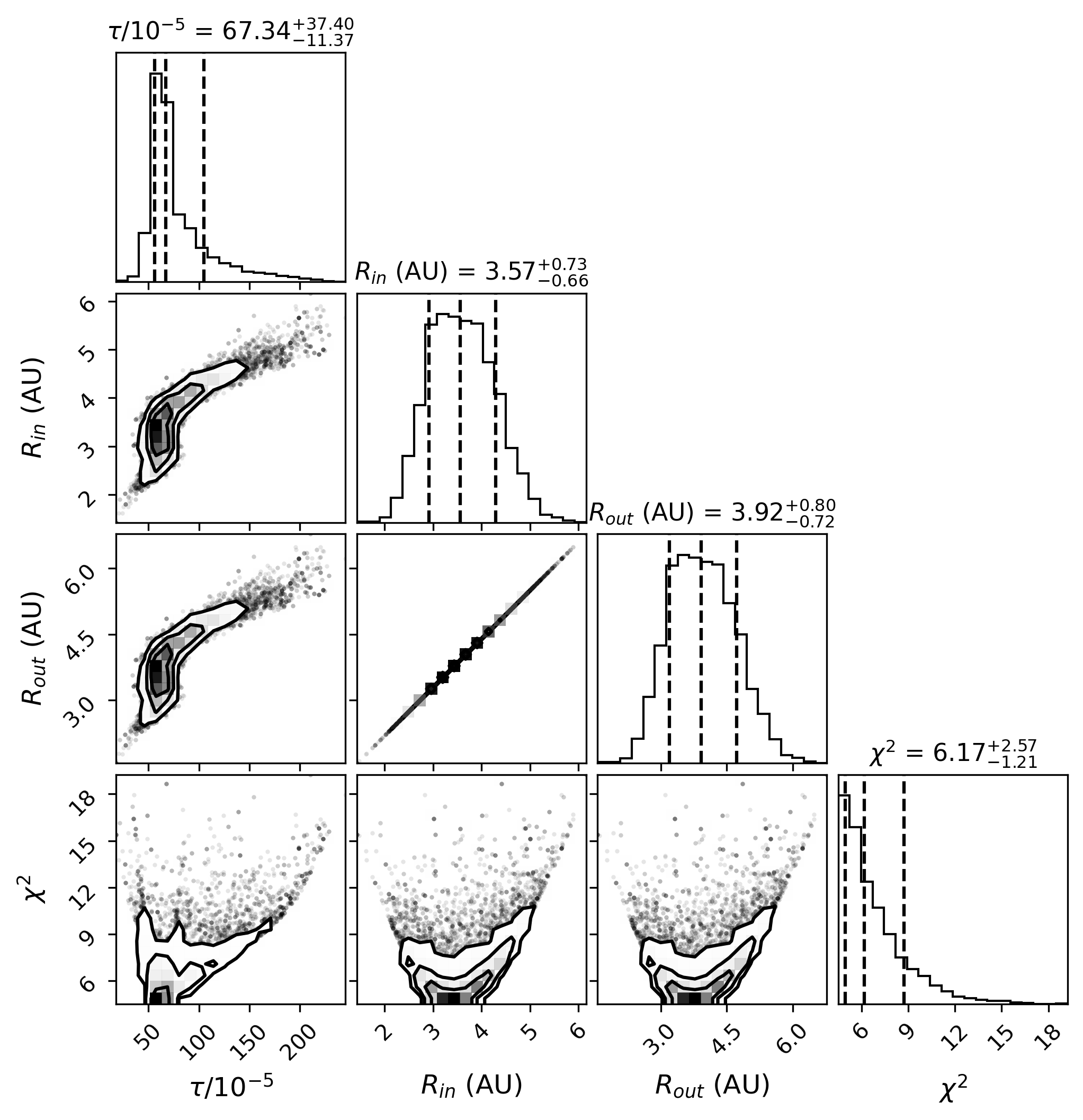}
        \hfill
        \captionof{figure}{Corner plots showing the MCMC posteriors for the model fitting of the 2017 (left) and 2018 (right) datasets. The posteriors are given for a ``disk'' (top), ``wide ring'' (middle), and ``thin ring'' (bottom) geometry.}
        \label{fig:corner pancake wide ring}
    \end{minipage}
    \vfill

    \newpage
    \quad
    \newpage

    \vfill
    \centering
    \begin{minipage}[H]{2\linewidth}
        \centering
        \hfill
        \includegraphics[width=0.41\linewidth]{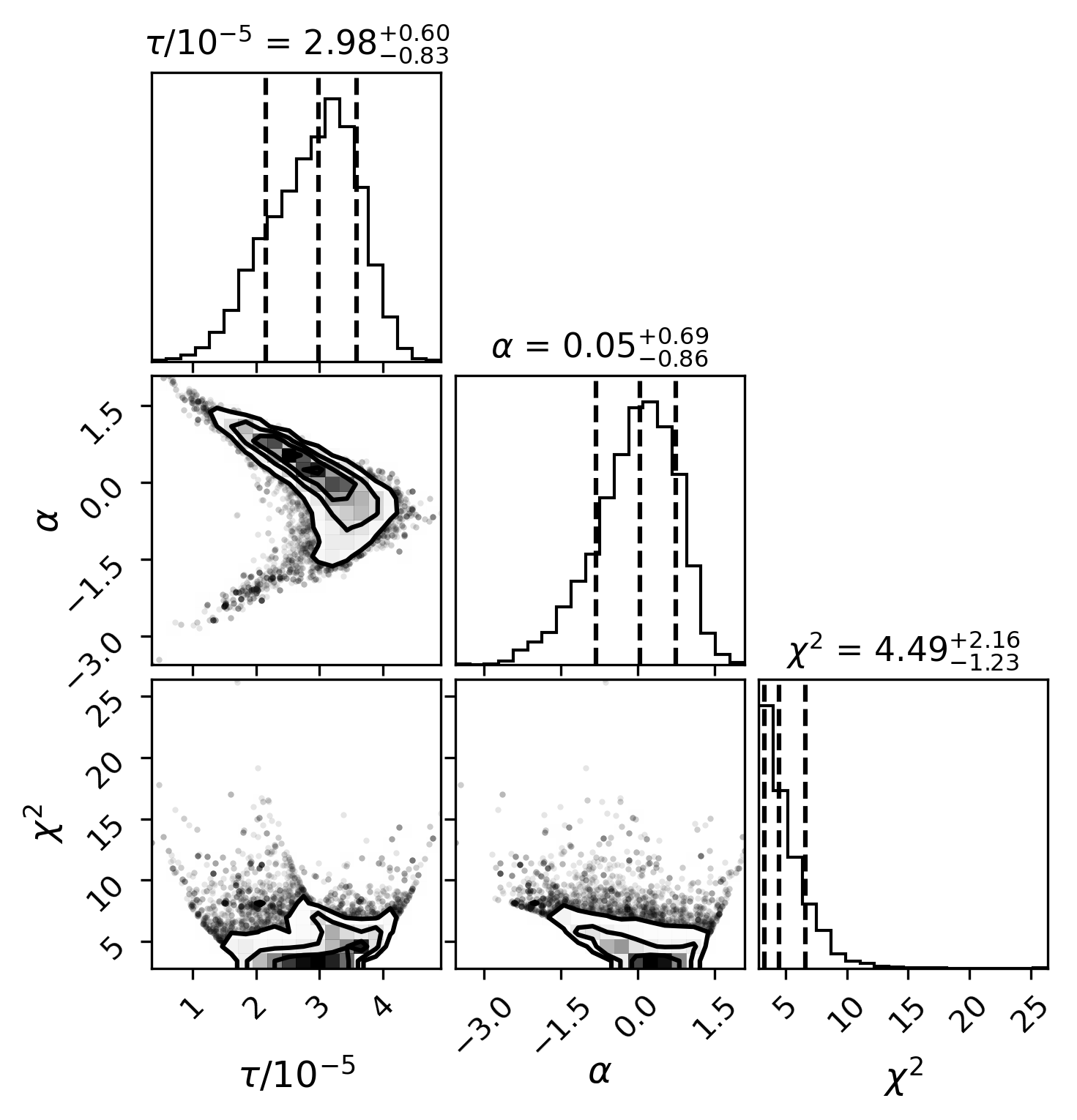}
        \hfill
        \includegraphics[width=0.41\linewidth]{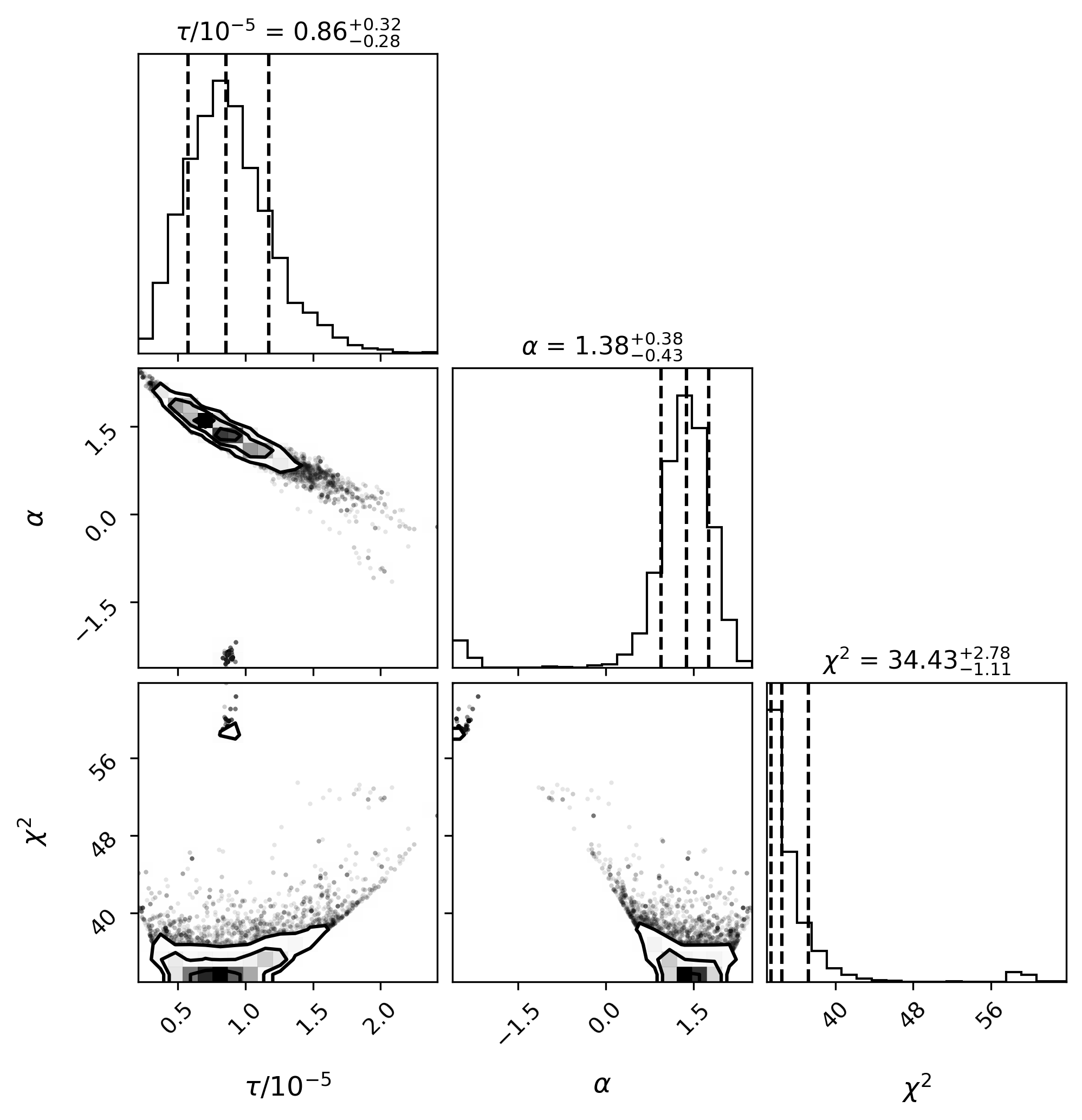}
        \hfill

        \centering
        \hfill
        \includegraphics[width=0.41\linewidth]{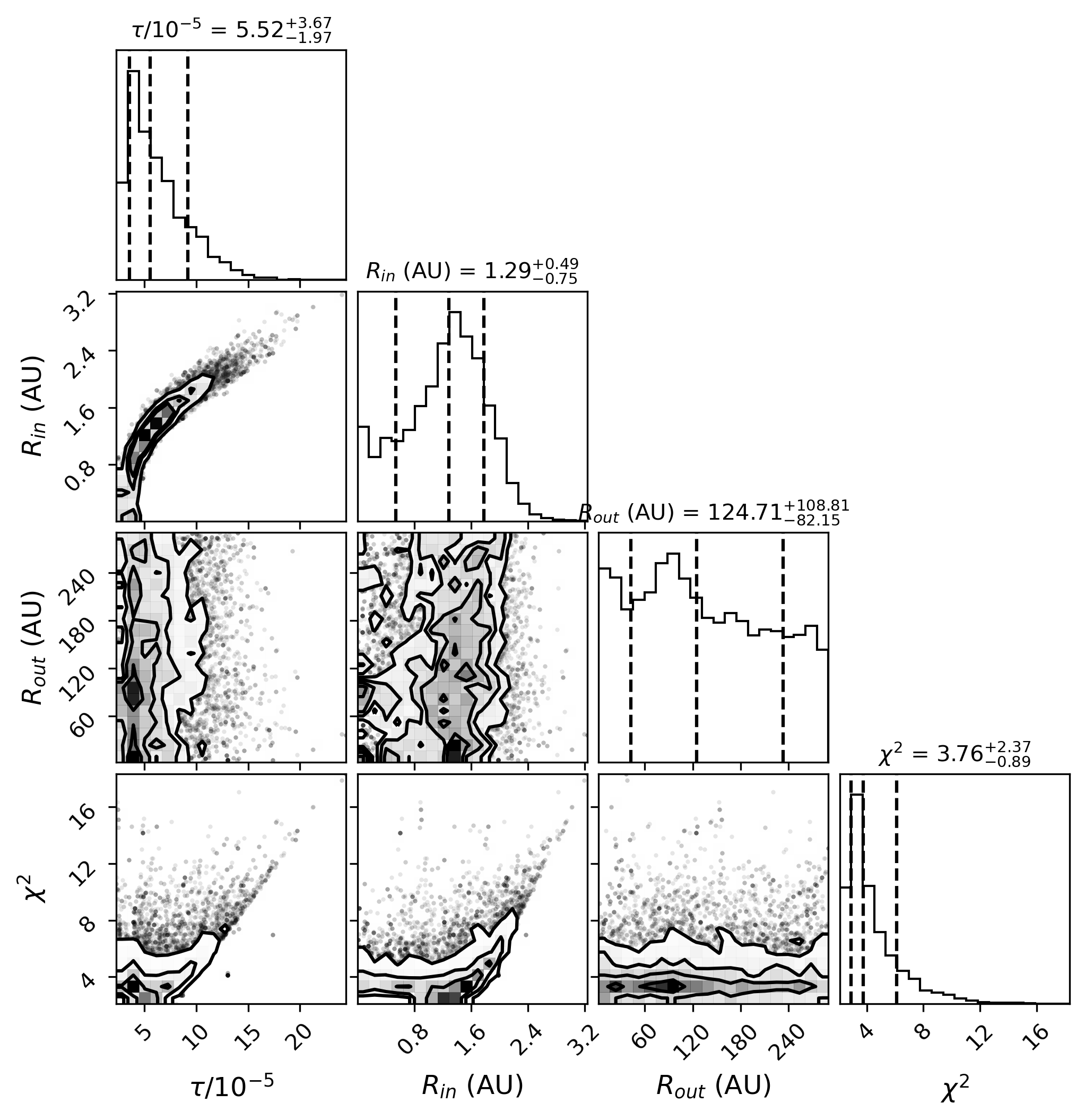}
        \hfill
        \includegraphics[width=0.41\linewidth]{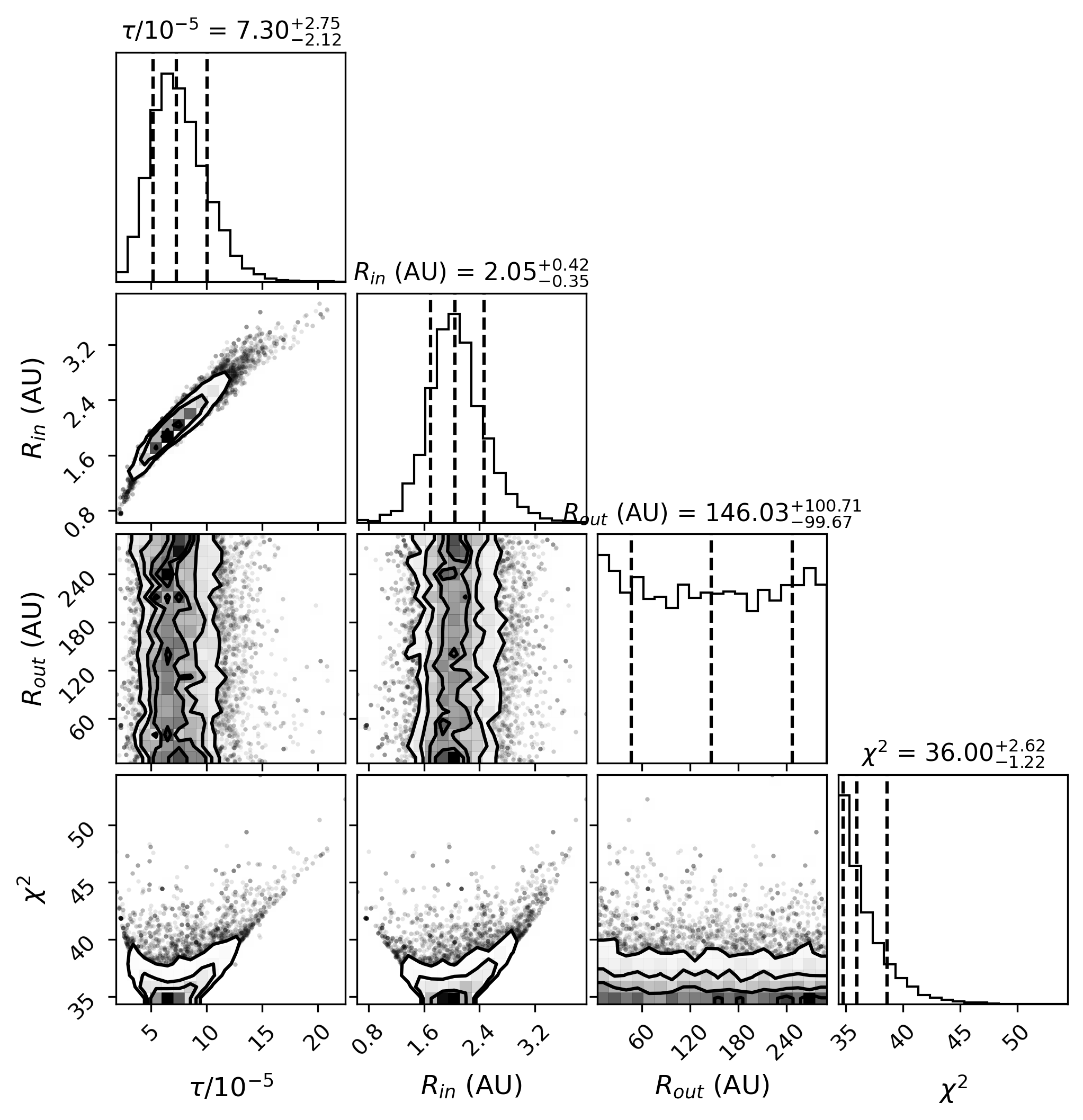}
        \hfill

        \centering
        \hfill
        \includegraphics[width=0.41\linewidth]{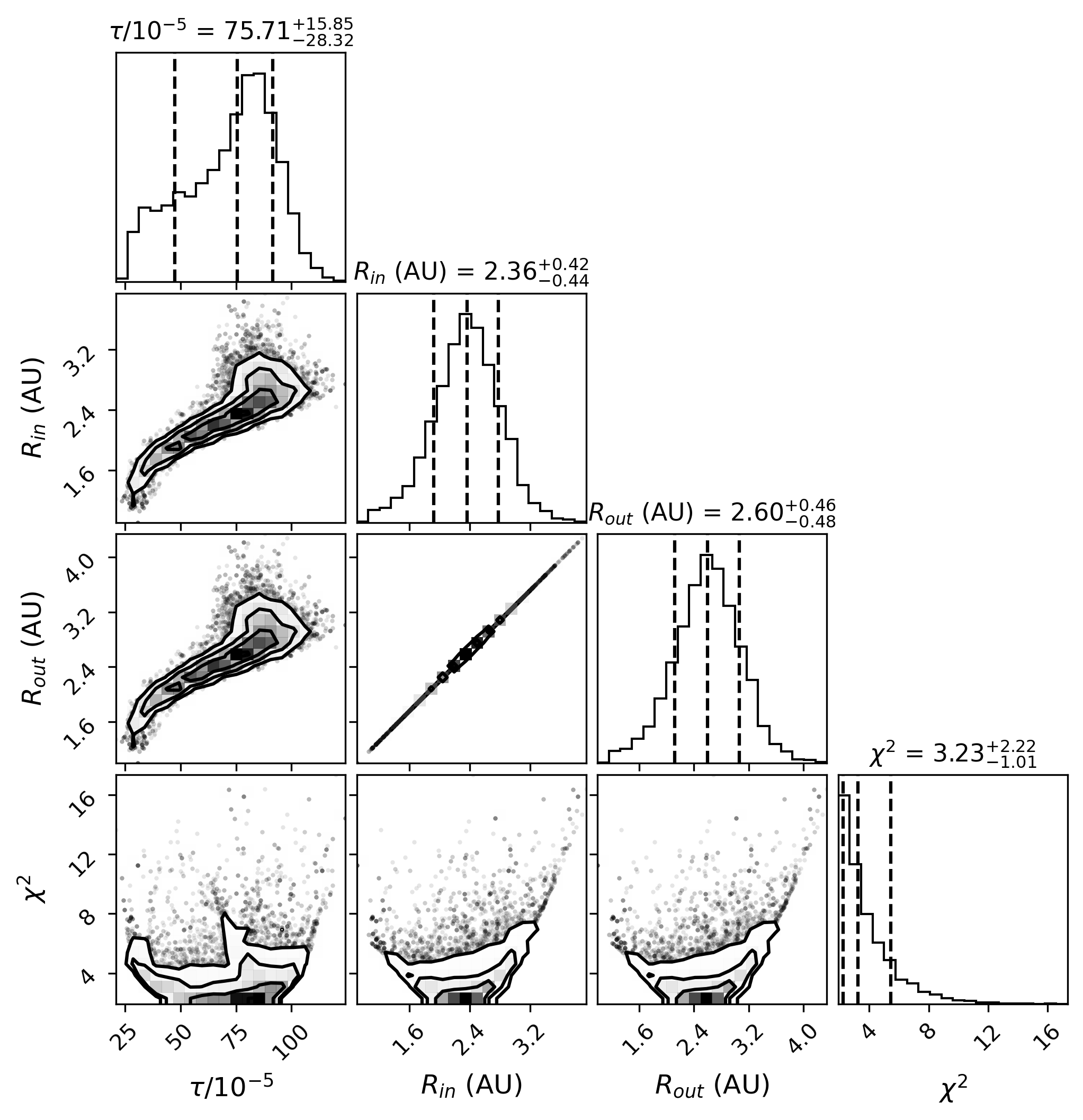}
        \hfill
        \includegraphics[width=0.41\linewidth]{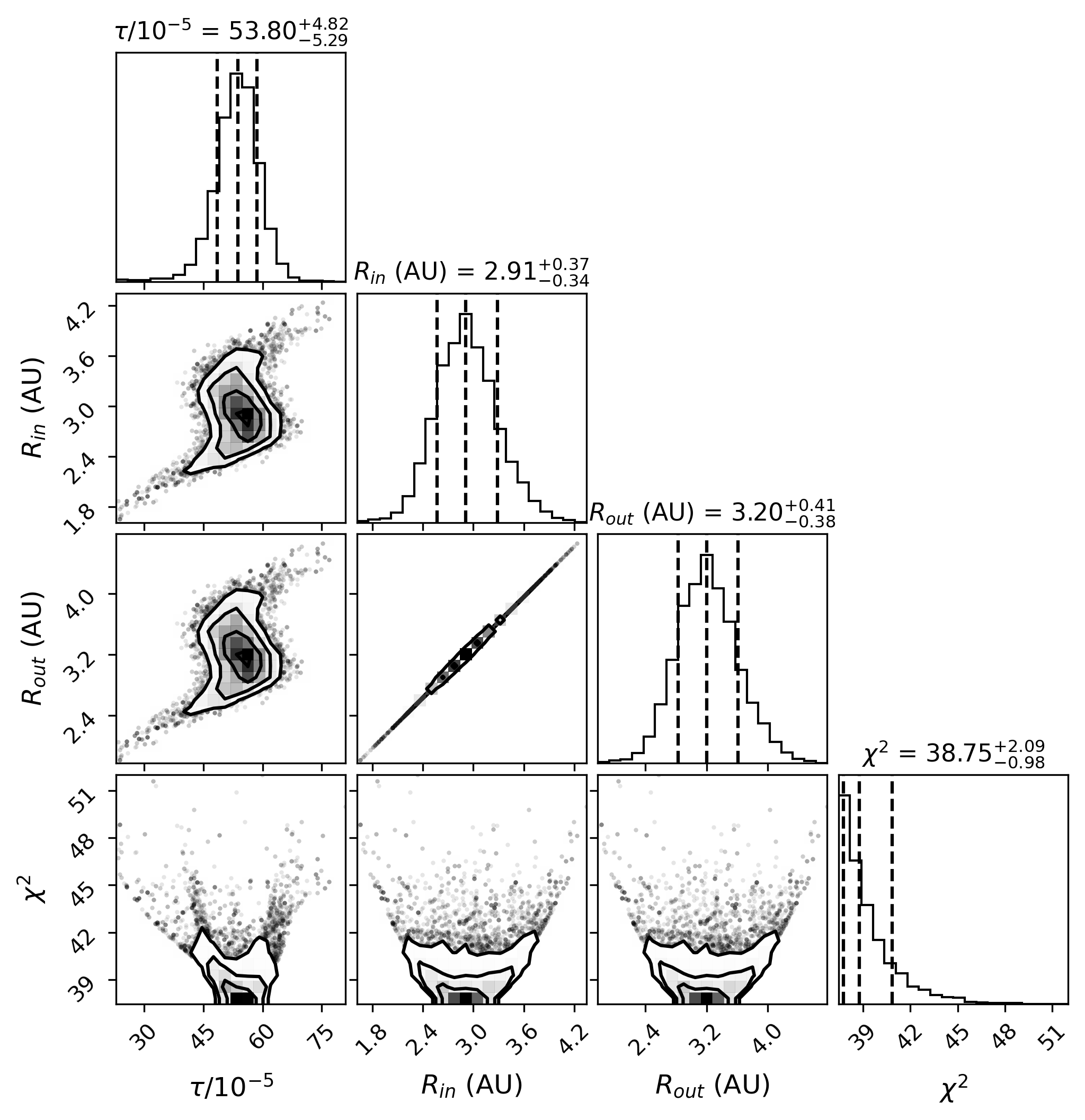}
        \hfill
        
        \captionof{figure}{Corner plots showing the MCMC posteriors for the model fitting of the 2023 (left) and the combined 2017/2018/2023 (right) datasets. The posteriors are given for a ``disk'' (top), ``wide ring'' (middle), and ``thin ring'' (bottom) geometry.}
        \label{fig:corner plots thin ring}
    \end{minipage}
    \vfill

\end{appendix}

\end{document}